\documentclass[aps,reprint,superscriptaddress]{revtex4-2}

\usepackage{soul}
\usepackage{amsmath}
\usepackage{amssymb}
\usepackage{graphicx}
 \usepackage{epstopdf}
\usepackage{float}

\usepackage{lipsum, babel}
\usepackage{placeins}
\usepackage{mathtools}
\usepackage{bm}
\usepackage{array,multirow}

\usepackage{xcolor}
\usepackage{booktabs}

\usepackage{chngcntr}

\usepackage[normalem]{ulem}

\usepackage{xr}
\begin{document}

\title{Controls that expedite first passage times in disordered systems}

\author{Marc H\"oll}\thanks{Corresponding authors: Marc Höll marc-holl@outlook.com, Alon Nissan alonzo.nissan@gmail.com}
\affiliation{Department of Physics, Institute of Nanotechnology and Advanced Materials, Bar-Ilan University, Ramat Gan, 52900, Israel}
\author{Alon Nissan}\thanks{Corresponding authors: Marc Höll marc-holl@outlook.com, Alon Nissan alonzo.nissan@gmail.com}
\affiliation{Institute of Environmental Engineering, ETH Zurich, Zurich, Switzerland}
\author{Brian Berkowitz}
\affiliation{Department of Earth and Planetary Sciences, Weizmann Institute of Science, Rehovot, 7610001, Israel}
\author{Eli Barkai}
\affiliation{Department of Physics, Institute of Nanotechnology and Advanced Materials, Bar-Ilan University, Ramat Gan, 52900, Israel}

\begin{abstract}
First passage time statistics in disordered systems exhibiting scale invariance are studied widely. In particular, long trapping times in energy or entropic traps are fat-tailed distributed, which slow the overall transport process. We study the statistical properties of the first passage time of biased processes in different models, and employ the big jump principle that shows the dominance of the maximum trapping time on the first passage time. We demonstrate that the removal of this maximum significantly expedites transport. As the disorder increases, the system enters a phase where the removal shows a dramatic effect. Our results show how we may speed up transport in strongly disordered systems exploiting scale invariance. In contrast to the disordered systems studied here, the removal principle has essentially no effect in homogeneous systems; this indicates that improving the conductance of a poorly conducting system is, theoretically, relatively easy as compared to a homogeneous system.
\end{abstract}

\maketitle

\section{Introduction}
Tracer pathways, retention, and migration patterns in disordered environments are typically seen to be similar over a large range of length and time scales, being observed ubiquitously in nature, e.g. for the motion of a tracer in porous media \cite{berkowitz2006modeling,berkowitz1997anomalous,berkowitz1998theory,Nissan2019,nissan2018inertial,dullien2012porous}, a colloidal particle in a glassy system \cite{bouchaud1990anomalous,monthus1996models,berthier2011theoretical}, a charge carrier in a strongly disordered amorphous conductor \cite{scher1975anomalous,montroll1973random}, or an  ion channel on the membrane of the cell \cite{fox2021aging,weigel2011ergodic}. In many cases, the probability density function of the transition times (also called sojourn times) is heavy-tailed $\psi(t) \sim A t^{-(1+\alpha)}$ with some amplitude $A$ and the scaling exponent $\alpha>0$ \cite{klages2008anomalous,metzler2000random,metzler2014anomalous,hofling2013anomalous}.  A critical consequence is that extremely long trapping times occur in deep traps, narrow passages, or at major obstacles, hence, the motion slows down dramatically. We present a simple but effective concept that overcomes this slow down. One only needs to remove the single maximum trapping time $\tau_\text{max}$ along the tracer path to gain a surprisingly great effect on the transport behavior. This effect is a consequence of the so-called big jump principle, which has been studied extensively \cite{chistyakov1964theorem,derrida1994non,filiasi2013condensation,vezzani2019single}. We claim that removal of the maximum trapping time along the trajectory will result in a significantly faster, expedited process, more importantly, we will quantify the gain to be achieved. As a reference procedure, we remove the maximum trapping time from each individual particle trajectory. Other techniques will be also discussed, e.g., the removal of only a few very large maxima, or of deep traps from an energy landscape. In all of these techniques, we
eliminate (directly or indirectly) from all or only some trajectories their associated maximum transition time. \\ \\


\indent 
We demonstrate this removal technique for one of the most well-studied observables in stochastic dynamics, namely the time a tracer is advected through a system of length $L$, known as the first passage time $t_f$ \cite{majumdar2010universal,bel2006random,redner2001guide,rangarajan2000anomalous,condamin2007first,condamin2007first2}. A broad class of well-known transport models is considered, e.g. with rigorous theory for the unidirectional transport on a lattice serving as a toy model, via the continuous time random walk \cite{montroll1965random,kutner2017continuous,metzler2000random,montroll1965random,barkai2000continuous,berkowitz1997anomalous,berkowitz1998theory,berkowitz2006modeling,nissan2018inertial,margolin2004continuous,dentz2004time,dentz2005exact,dentz2008transport,cairoli2015anomalous,burioni2014scaling,scalas2006application,weeks1998anomalous,weeks1996anomalous,albers2013subdiffusive} and quenched trap models \cite{bouchaud1990anomalous,monthus1996models,berthier2011theoretical,burov2011time,akimoto2016universal,burov2007occupation,bertin2003subdiffusion,akimoto2018non,burov2012weak}, and with extensive numerical analysis of simulated tracer migration in porous media. The study for each model contains three parts: \textbf{A)} We establish the principle of the single long transition time for these different models, which is based on the principle of the single big jump \cite{chistyakov1964theorem,embrechts2013modelling,embrechts1982estimates,rolski2009stochastic,kyprianou2006introductory,kluppelberg1997large,vezzani2019single,wang2019transport,wang2020large,holl2021big}. 
The tracer path is described by trapping events on a coarse grained scale. We may then define the longest trapping time $\tau_\text{max}$ (defined more precisely below). This time is clearly shorter than the total time $t_f$ it takes the particle to cross the system. Still, in scale free systems, as we show below, and for the slowest particles, $\tau_\text{max} \simeq t_f$ (see precise definition below). This is the long transition time principle that we aim to establish, for widely applicable models. The question is now: how can we enhance the transport? \textbf{B)} We remove the maximum transition times from the associated particle trajectories, and find that the distribution of the modified first passage time 
\begin{equation}\label{modifiedfpt}
t_r=t_f-\tau_\text{max}
\end{equation} 
decays much faster compared to the original distribution of $t_f$. The index ``$r$'' stands for \textit{removal}. \textbf{C)} This transport speed-up is further quantified with the measure of gain 
\begin{equation}\label{gquant}
G=\frac{\langle t_r \rangle}{\langle t_f \rangle},
\end{equation}
where a small value indicates fast transport of the modified process. Clearly, $G < 1$, but the question addressed is more qualitative, what is $G$, and does it exhibit a phase-like transition when the strength of disorder is increased? It is clear that for blind removal of a single trapping time, or when considering homogeneous systems, then $G$ will be close to unity for large systems; however, in the case of strongly disordered systems, we find $G\ll 1$. \\ \\

The main practical challenge is to identify the bottlenecks, namely, the large trapping times in the system. Our theoretical analysis does not address this issue in full detail, but we discuss this point in some depth in Sec.~\ref{sec:outlook}. We must distinguish between several cases. Information on specific pathways is now obtainable in many single particle tracking experiments. In these cases, a ``learning session'' can be completed to identify bottlenecks that slow down the first passage time. These bottlenecks, representing deep traps, are quenched and localized; see Sec.~\ref{sec:qtm2}. Removal of some traps, or restart of the process for a particle in the deepest trap, is in principle possible. Here, we depart from the usual paradigm of restart (see discussion below), where no information on the system is given {\it a priori}. In contrast to annealed models like the continuous time random walk, which is effectively considered a mean field model, the traps are not fixed in a particular location in space. Here, the big jump principle is considered after the trajectory is completed, and the question focuses on analyzing the effect of removal of one long waiting time. In some systems, information about trajectories is not available, and more clever ways must be used to identify bottlenecks; one such method is outlined briefly in Sec.~\ref{sec:outlook}. Summarized, two models in this article (simulation of tracer transport in porous media and the quenched trap model) have quenched disorder and the removal of the biggest trapping times at bottlenecks is a priori possible, while for the two other models (unidirectional transport and continuous time random walk) a removal is only possible a posteriori due to the annealed disorder. The study of the latter models is of academic interest as it indicates the significant effect of removal.\\ \\

A somewhat related concept is the restart protocol. A ``restart'' of a stochastic search process may expedite the search time vastly, and hence this strategy has been extensively studied \cite{yin2023restart,evans2011diffusion,evans2011diffusion2,evans2020stochastic,chechkin2018random,campos2015phase,besga2020optimal,tal2020experimental},  with applications in biological processes \cite{reuveni2014role,budnar2019anillin,bel2009simplicity} and computer science \cite{hamlin2019geometry}, among other fields. The basic paradigm  of restart is to consider a non-biased diffusive particle that is returned to its origin at a given rate. Under certain conditions, this repeated return to the origin will minimize the first passage time $t_f$ to a target \cite{pal2017first}. When a bias is present, this strategy is not necessarily useful. Furthermore, the  number of restarts can be large, and to pick up a particle at some position in space and return it to its origin is typically costly. Here, we present a new method to deal with such questions. Our developments exploit the scale-invariance of transport in  disordered system \cite{klages2008anomalous,metzler2000random,metzler2014anomalous,hofling2013anomalous} and the big jump principle \cite{chistyakov1964theorem,embrechts2013modelling,embrechts1982estimates,rolski2009stochastic,kyprianou2006introductory,kluppelberg1997large,vezzani2019single,wang2019transport,wang2020large,holl2021big,derrida1994non,filiasi2013condensation} to find novel effects. Roughly speaking, along the path of a particle advancing in a disordered system, we identify a bottleneck where transport is slowed down. Namely, the particle is trapped and released many times along its path, and then the basic issue is: will the removal of one and only one of these trapping times dramatically reduce the first passage time? Thus, unlike classical restart theory, we do not send the particle back to its origin several times, and exploit the disorder to obtain a dramatic speed-up of the first passage time. \\ \\
\indent

\indent
The article begins with the basic, unidirectional model (Sec.~\ref{sec:odt}). This model already shows the relevant behavior that is also found in the other models. The most complicated, yet possibly realistic, model considered in this article is the simulation of tracer transport presented in Sec.~\ref{sec:sim_pore}, which shows that the removal principle can serve as a powerful tool in application. Thereafter, the continuous time random walk (Sec.~\ref{sec:ctrw2}) and quenched trap (Sec.~\ref{sec:qtm2}) models are examined as two additional theoretical continuations of the basic unidirectional model. Once these four models are extensively analyzed, an outlook for practical applications is presented (Sec.~(\ref{sec:outlook}) and all results are summarized (Sec.~\ref{sec:discussion}).

\section{Unidirectional transport on a lattice}\label{sec:odt}

\begin{figure*}\begin{center}
\includegraphics[width=0.9\textwidth]{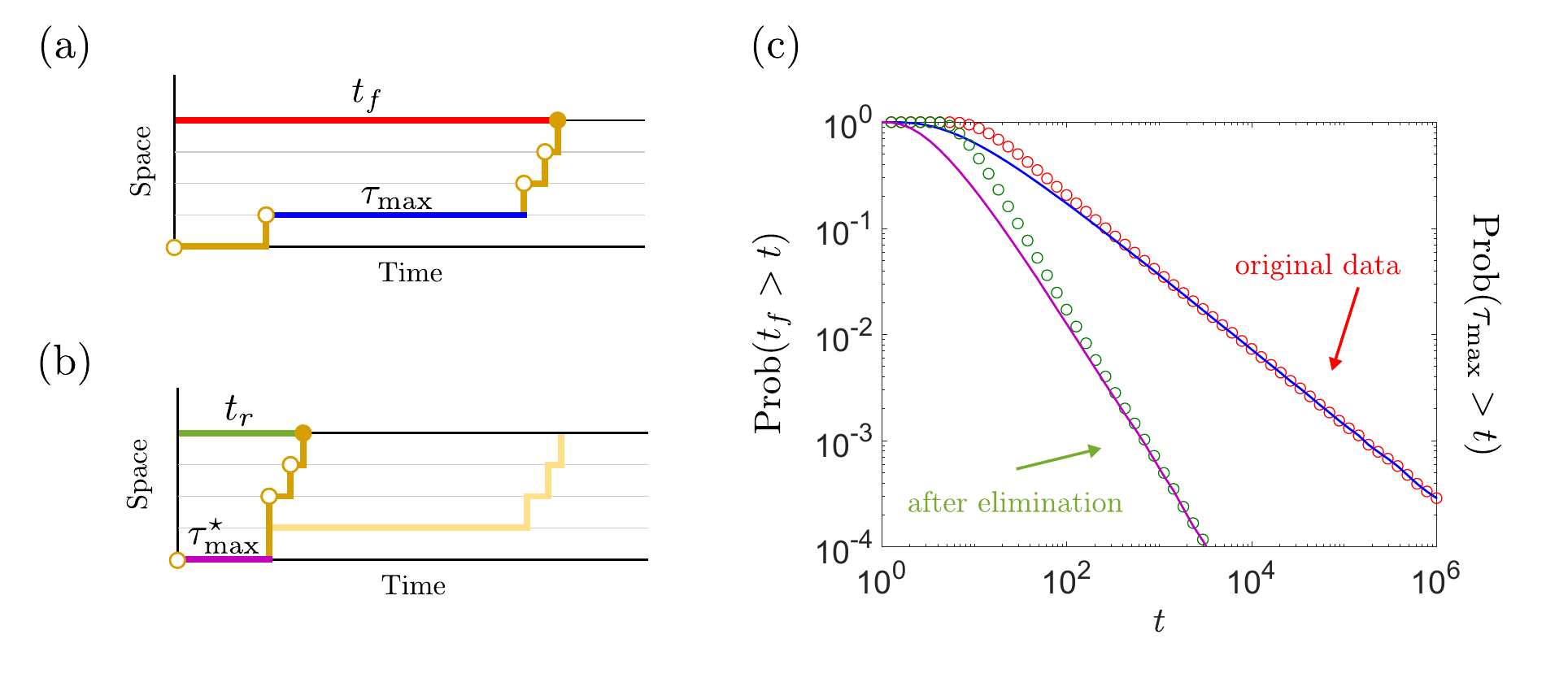}
\caption{(a) Conceptual figure of the unidirectional transport model on a lattice. The first passage time $t_f$ (red) to reach some boundary is the sum of the transition times between lattice points. The maximum transition time $\tau_\text{max}$ (blue) is also marked. (b) Conceptual figure of the same transport model as in (a) but $\tau_\text{max}$ has been removed from the trajectory. This shows the modified first passage time $t_r=t_f-\tau_\text{max}$ (green) and the second maximum $\tau_\text{max}^\star$ (purple). (c) The distributions of $t_f$ (red circles), $\tau_\text{max}$ (blue line), $t_r$ (green circles) and $\tau_\text{max}^\star$ (purple line) obtained from Monte Carlo simulations for the unidirectional model. We used the Pareto distribution with $\alpha=0.7$ for the transition times with $L=5$ and $10^6$ trajectories. As shown, the right tails before and after the elimination of $\tau_\text{max}$ from each trajectory match for the maximum and the first passage time distributions, as predicted by the principle of the single long transition time Eq.~(\ref{bjp_all}) and the relationship after elimination Eq.~(\ref{cap}). Importantly, the elimination decreases the power law of the $t_f$ distribution Eq.~(\ref{bjp_all_2}) to Eq.~(\ref{cap4}). This shows clearly that the elimination of the maximum transition time dramatically modifies the statistics in the tail distribution by orders of magnitude. \label{fig:main_results}}
\end{center}\end{figure*}

We start with a simple model that still can capture some of the complexities in the following, more challenging and realistic, approaches. Consider the transport of particles on a one-dimensional lattice of length $L$ with the lattice points $\{1,2,\ldots,L\}$ and the absorption point (boundary) at $x=L+1$. The movement is in one direction, corresponding to a strongly biased system, i.e., an external, constant large, force that drives the particles to the right. Each particle starts at $x=1$ and is absorbed at 
$x=L+1$; see Fig.~\ref{fig:main_results}(a) for a conceptual figure. Thus, every particle makes exactly $N=L$ jumps before it is absorbed. At each lattice point $x\in \{1,\ldots,L\}$, a particle takes a random transition time $\tau_n$ with $n=x$ before jumping to the nearest neighbor $x+1$. Furthermore, the transition times $\{\tau_n\}_{n=1}^N=\{\tau_1,\ldots,\tau_N\}$ are assumed to be independent and identically distributed random variables. Importantly, the probability density function of these transition times is asymptotically heavy-tailed 
\begin{equation}\label{transitionprob}
\psi(t) \sim At^{-(1+\alpha)}
\end{equation}
for all $n$ with the scaling exponent $\alpha>0$ and the amplitude $A$. The definition of this function is $\psi(t) = - \mathrm{d}/\mathrm{d}t\text{Prob}(\tau_n>t)$ where $\text{Prob}(\tau_n>t) \sim (A/\alpha)  t^{-\alpha}$. Note that if $\alpha>1$, the mean transition time is finite, and otherwise it diverges. It is well-known that physical systems exhibit dynamical transitions, e.g., from normal to anomalous diffusion at $\alpha=1$ \cite{bouchaud1992weak}. A useful example is the Pareto distribution $\psi(t)=\alpha (t_0)^\alpha t^{-(1+\alpha)}$ with $t>t_0$ and otherwise zero. 
\\ \\

There is a vast literature on the physical meaning of Eq.~(\ref{transitionprob}), where it is also standard to use the parameter $\beta$ instead of $\alpha$ \cite{scher1975anomalous,montroll1973random}. For example, in the quenched trap model, $\alpha=T/T_g$ \cite{bouchaud1990anomalous,burov2011time,akimoto2016universal,akimoto2018non,burov2007occupation} where $T_g$ is the glass temperature and $T$ the temperature; see Sec.~\ref{sec:qtm2}. Furthermore, $\alpha$ was recorded in time of flight experiments of charge carriers in disordered systems \cite{scher1975anomalous,montroll1973random}, in contaminant migration in porous media \cite{nissan2018inertial,BERKOWITZ2009}, and for the tracer diffusion process in actin networks \cite{wong2004anomalous,levin2021measurements}, where $\alpha$ is controlled by the size of the tracer in relation to the mesh size of the actin network; it was also observed in the diffusion of channels on the membrane \cite{fox2021aging}. Similar waiting times are found for blinking quantum dots \cite{stefani2009beyond}, tracers in two dimensional rotational flows \cite{solomon1993observation}, and avian predators \cite{vilk2021ergodicity}. To summarize, in many fields of physics, one finds processes described by Eq.~(\ref{transitionprob}), while the unidirectional assumption we used here corresponds to a strong bias acting on the particle (no backward jumps). We will relax this assumption later.

\subsection{Principle of the single long transition time}\label{sec:first_principle}

We now briefly review the principle of the single big jump, which in the context of our work is called the principle of the single long transition time. There are several versions of this principle, and here we consider the simplest case when dealing with independent and identically distributed random variables \cite{chistyakov1964theorem}. We are interested in the first passage time $t_f$ to reach the absorbing site $L+1$. This is the sum of the $N=L$ transition times
\begin{equation}\label{tf_all}
t_f = \sum_{n=1}^N \tau_n.
\end{equation}
When the assumption of unidirectionality is removed, we will have $N\neq L$, see below. The second quantity of interest is the maximum transition time
\begin{equation}\label{max_all}
\tau_\text{max} =\text{max}(\tau_1,\ldots,\tau_N),
\end{equation}
i.e., the longest time a particle transits between two lattice points upon reaching the boundary $L+1$. In Fig.~\ref{fig:main_results}(a), we show both quantities in a conceptual form. 

According to the principle of the single long transition time \cite{chistyakov1964theorem}, we can relate the probabilities of $t_f$ and $\tau_\text{max}$ when both values are large:
\begin{equation}\label{bjp_all}
\text{Prob}(t_f>t) \sim \text{Prob}(\tau_\text{max}>t)
\end{equation}
for large times $t$ and any $N$. In particular, the power law decay is
\begin{equation}\label{bjp_all_2}
\text{Prob}(t_f>t) \sim N \frac{A}{\alpha} t^{-\alpha}
\end{equation}
which is the large $t$ behavior of $N\text{Prob}(\tau_n>t)$. In Fig.~\ref{fig:main_results}(c), we present this principle for transition times following the Pareto distribution with $\alpha=0.7$ and $t_0=1$. The intuitive idea behind Eq.~(\ref{bjp_all}) is that for large $t_f$, the remaining transition times are negligible because $\tau_\text{max}$ is so large that it dominates the statistics indicated by the matching of the tails of the corresponding distributions. We note that in the large $N$ limit, the distributions of the properly rescaled and shifted $t_f$ and $\tau_\text{max}$ converge to an alpha-stable distribution according to the L\'evy central limit theorem \cite{bouchaud1990anomalous} (when $\alpha<2$) and to the Fr\'echet distribution according to the theory of extreme value statistics \cite{majumdar2020extreme}. The tails of these two famed distributions are identical, as the principle of the single long transition time predicts. However, Eq.~(\ref{bjp_all}) is valid for any $N$ which is important for any application with a finite sized system.

\subsection{Elimination of the single long transition time}\label{sec:second_principle}

What is the effect of the elimination of the maximum transition time $\tau_\text{max}$ from the sequence $\{\tau_n\}_{n=1}^N$? We are in particular interested in the modified first passage time
\begin{equation}\label{tfs}
t_r = t_f - \tau_\text{max}
\end{equation}
after removal of the maximum transition time from each particle trajectory. Clearly, this will speed up the transport in the sense that now the time to transverse the system has shortened, but by how much? Before we continue, we address a similar problem. If we remove one transition at random, we have $t_f$ as the sum of $N-1$ instead of $N$ random variables. But this is only a minor change in the statistics, which can be easily seen by Eq.~(\ref{bjp_all_2}). Namely, the power law decay of the sum distribution is still $t^{-\alpha}$. However, as we show now, removing the largest random variable as for $t_r$ in Eq.~(\ref{tfs}), the statistics change dramatically. \\ \\

In the Supplemental Material (SM) Sec.~B, we derive the asymptotic behavior of the distribution of $t_r$. We find 
\begin{equation}\label{cap4}
\boxed{\text{Prob}(t_r>t) \sim \frac{1}{2} N(N-1)\left(\frac{A}{\alpha}\right)^2 t^{-2\alpha}
}
\end{equation}
valid for large times $t$ and any $N$. The remarkable issue is the doubling effect of the exponent $-2\alpha$, i.e. previously we had the $-\alpha$ decay in Eq.~(\ref{bjp_all_2}). Thus, the probability of large $t_r$ is drastically decreased compared to the probability of large $t_f$. In particular, as we will discuss below, if $\alpha<1$, as it is found in many disordered systems, the mean first passage time diverges, but once we eliminate $\tau_\text{max}$ from each trajectory, the mean will diverge only if $\alpha<1/2$. In that sense, we have a dramatic effect upon elimination. In Fig.~\ref{fig:main_results}(c), we present Eq.~(\ref{cap4}), thus showing that the elimination effect is indeed large for $\alpha=0.7$.

\subsubsection{Scale invariance of the single long transition time principle}
Similar to the principle Eq.~(\ref{bjp_all}), we can relate $t_r$ to the maximum transition time after elimination of $\tau_\text{max}$ from each trajectory. So after this elimination, we deal with the $N-1$ transition times $\{\tau_n\}_{n=1,n\neq m}^N=\{\tau_1,\ldots,\tau_{m-1},\tau_{m+1},\ldots,\tau_N\}$ where $\tau_\text{max}=\tau_m$ occurred at the $m$-th step. The step number $m$ is of course random. The maximum transition time after elimination is
\begin{equation}\label{max_all2}
\tau_\text{max}^\star =\text{max}(\tau_1,\ldots,\tau_{m-1},\tau_{m+1},\ldots,\tau_N),
\end{equation}
see Fig.~\ref{fig:main_results}(b) for schematics. The distribution of $\tau_\text{max}^\star$ is known from order statistics \cite{majumdar2020extreme}; we can simply see that its tail behaves as Eq.~(\ref{cap4}). Therefore, we have the asymptotic relationship \begin{equation}\label{cap}
\boxed{\text{Prob}(t_r>t) \sim \text{Prob}(\tau_\text{max}^\star >t)}
\end{equation}
for large times $t$ and any $N$. Thus, the principle of the single long transition time still holds after the elimination of $\tau_\text{max}$, which is an expected effect from scale-free fractal time process. This indicates that continued removal of the second longest transition time will also have a strong effect; see further details below. \\ \\

\subsubsection{Elimination of several long transition times}\label{sec:elimscale}
What happens when we remove not only $\tau_\text{max}$ but also the next longest transition times from the trajectory? To examine this question, we start by ordering the transition times according to their values $\tau_{(1)}<\ldots <\tau_{(N)}$. Obviously, then, $\tau_{(N)}=\tau_\text{max}$ and $\tau_{(1)}=\text{min}(\tau_1,\ldots,\tau_N)$. Then we eliminate the $s$, where $s=1,\ldots,N-1$, longest transition times $\{t_{(N-s+1)},\ldots,t_{(N)}\}$ and are left with $\{\tau_{(1)},\ldots,\tau_{(N-s)}\}$. The first passage time after elimination is  
\begin{equation}
t_r(s) = t_f - \sum_{q=N-s+1}^N \tau_{(q)}.
\end{equation}
In SM Sec.~B, we derive the asymptotic behavior of the distribution of $t_r(s)$. We find 
\begin{equation}\label{newgen}
\text{Prob}(t_r(s)>t ) \sim \frac{N!}{(s+1)!(N-s-1)!}\left(\frac{A}{\alpha}\right)^{s+1}t^{-(s+1)\alpha}
\end{equation}
for large $t$ and any $N$.  Importantly, $\text{Prob}(t_r(s)>t)$ decays with the exponent $-(s+1)\alpha$ much faster than $\text{Prob}(t_f>t)$ with the exponent $-\alpha$. So the elimination of any additional long transition time yields an additional power law decrease by the exponent $-\alpha$. It follows that by removing the $s$ longest transitions times, we can strongly damp the tail of the time of flight distribution.\\ \\

The maximum transition time after elimination is
\begin{equation}
\tau_\text{max}^\star(s) = \tau_{(N-s)}
\end{equation}
for which the distribution is known from order statistics \cite{majumdar2020extreme}, and we see that its tail behaves as Eq.~(\ref{newgen}). Therefore, we have
\begin{equation}\label{secbjp222}
\text{Prob}(t_r(s)>t) \sim \text{Prob}(\tau_\text{max}^\star(s)>t)
\end{equation}
for large $t$ and any $N$. This relationship generalizes the principle Eq.~(\ref{cap}) due to the scale invariance of the transition times Eq.~(\ref{transitionprob}). In Fig.~B.1 in SM Sec.~B, we present the four cases $s=0,\ldots,3$.

\subsubsection{Mean first passage time}

\begin{figure}\begin{center}
\includegraphics[width=1\columnwidth]{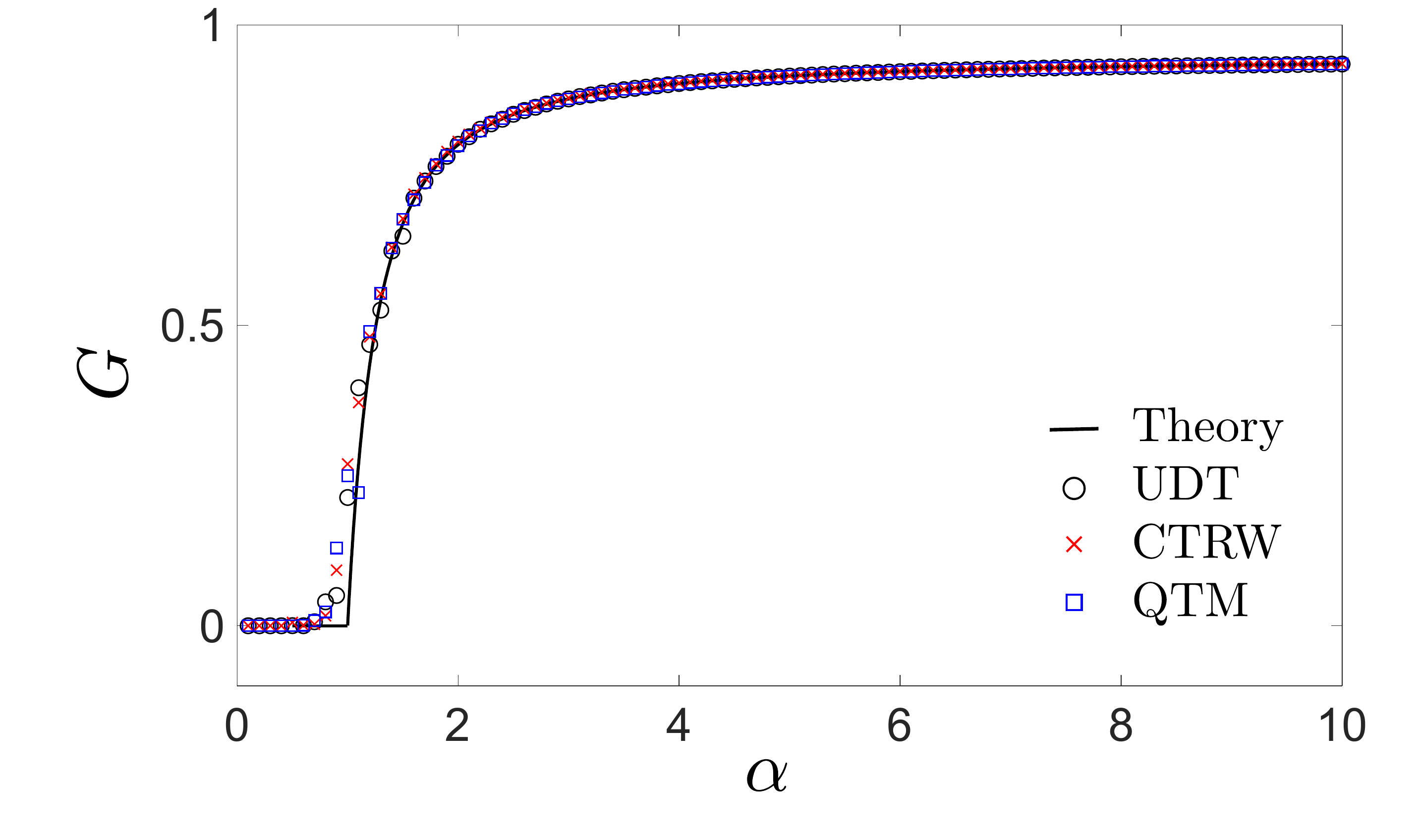}
\caption{The quantifier $G$, which is a measure of the transport improvement Eq.~(\ref{gquant}), versus the parameter $\alpha$, which is a measure of disorder. When $G=0$, the benefit from removal of the maximum transition time is optimal, in contrast with $G=1$. A transition is found at $\alpha=1$, for three models of transport considered in this work, the unidirectional model (UDT) (black circles), the continuous time random walk (CTRW) (red crosses), see Sec.~\ref{sec:ctrw2}, and the quenched trap model (QTM) with strong bias (blue squares), Sec.~\ref{sec:qtm2}. We used Pareto distributed transition times with $t_0=1$. Each point in the plot used $10^5$ trajectories in the Monte-Carlo simulation. For the UDT, we have $N=20$, for CTRW $L=10$ and $p=0.75$ so that $\langle N \rangle = 20$, and for QTM $L=20$. \label{fig:g1_phase}}
\end{center}\end{figure}

The different power law decays of $\text{Prob}(t_f>t)$ in Eq.~(\ref{bjp_all}) and $\text{Prob}(t_r>t)$ in Eq.~(\ref{cap4}) have a dramatic effect on the mean first passage times. The mean $\langle t_f \rangle$ is finite only for $\alpha>1$ but the mean after elimination $\langle t_r \rangle$ for $\alpha>1/2$, as mentioned. We quantify the elimination effect by the ratio
\begin{equation}\label{gquant}
G=\frac{\langle t_r \rangle}{\langle t_f \rangle}
\end{equation}
where $G$ is a measure of gain, in the sense that the smaller its value, the faster is the modified transport process. We must separate the two cases $\alpha<1$ and $\alpha>1$. As an example, we present $G$ for Pareto distributed transition times (see SM Sec.~C)
\begin{equation}\label{gth}
G=
\begin{cases}
0 & \text{ for } 0<\alpha<1,\\
1-(-1)^N \frac{\alpha-1}{\alpha} (N-1)! \frac{\Gamma\left(-N+\frac{1}{\alpha}\right)}{\Gamma\left(\frac{1}{\alpha}\right)} & \text{ for } 1<\alpha.
\end{cases}
\end{equation}
Generally, for power law transition times Eq.~(\ref{transitionprob}), we see a transition in the behavior of $G$ at the critical point $\alpha=1$ similar to a dynamical phase transition. When the disorder becomes stronger $\alpha<1$, the ratio $G$ is zero, which indicates a significant effect upon elimination. In contrast, if we use exponential transition times (see SM Sec.~C), we do not witness a critical transition in the statistical properties of the system. A subtle issue is found when $\alpha<1/2$. Then based on our formulas Eq.~(\ref{bjp_all}) and Eq.~(\ref{cap4}), both means $\langle t_f \rangle$ and $\langle t_r \rangle$ diverge, so we have in Eq.~(\ref{gquant}) the ratio of two infinities. We may still consider the sample means; the ratio is zero as shown in Fig.~\ref{fig:g1_phase}, where we compare the theory of $G$ with Monte Carlo simulations. Note that close to $\alpha=1$, the simulation deviates slightly from the non-analytical prediction of the theory.\\ \\

What is the effect of the elimination in the thermodynamic limit $N\to\infty$? To answer this question, it is useful to eliminate the $s$ longest transition times $\{\tau_{(N-s+1)},\ldots t_{(N)}\}$. We quantify the gain by the ratio $ G(s)=\langle t_r(s) \rangle/\langle t_f \rangle$. In SM Sec.~C, we calculate $G(s)$ exactly for the Pareto and exponentially distributed transition times. To define the limit, we consider a fixed ratio $s=f N$, with the fraction $0<f<1$. We obtain for the example of the Pareto distributed transition times
\begin{equation}\label{impalg}
G(s) \sim 
\begin{cases}
0 & \text{ for }0<\alpha<1, \\
1-f^{1-1/\alpha} & \text{ for }1<\alpha.
\end{cases}
\end{equation}
So even though the system is infinite, a relatively small $f$ leads to a qualitative improvement of the first passage time. Note that if $f=1$, we have $G=0$.  Furthermore, we see that the gain undergoes a phase-like transition when the control parameter is $\alpha$, which as mentioned is proportional, for example, to temperature for the trap model or the size of a bead in actin network diffusion \cite{wong2004anomalous}. \\ \\

Finally, the assumptions used so far, namely prescribing independent and identically distributed transition times and the unidirectional transport, are too limiting. Most physical systems have some kind of correlations among the transition times, and particles performing the stochastic process can move in the reverse direction even in the presence of a driving force that transports the particle towards the boundary. Will the above principles Eq.~(\ref{bjp_all}) and (\ref{cap}) hold in more general models, and will the large gain in elimination quantified by $G$ be generic in other models of transport in disordered systems? Further, in what sense is it plausible to eliminate the long sticking times in more realistic processes?

 \begin{figure*}
	\centering\includegraphics[width=1\textwidth]{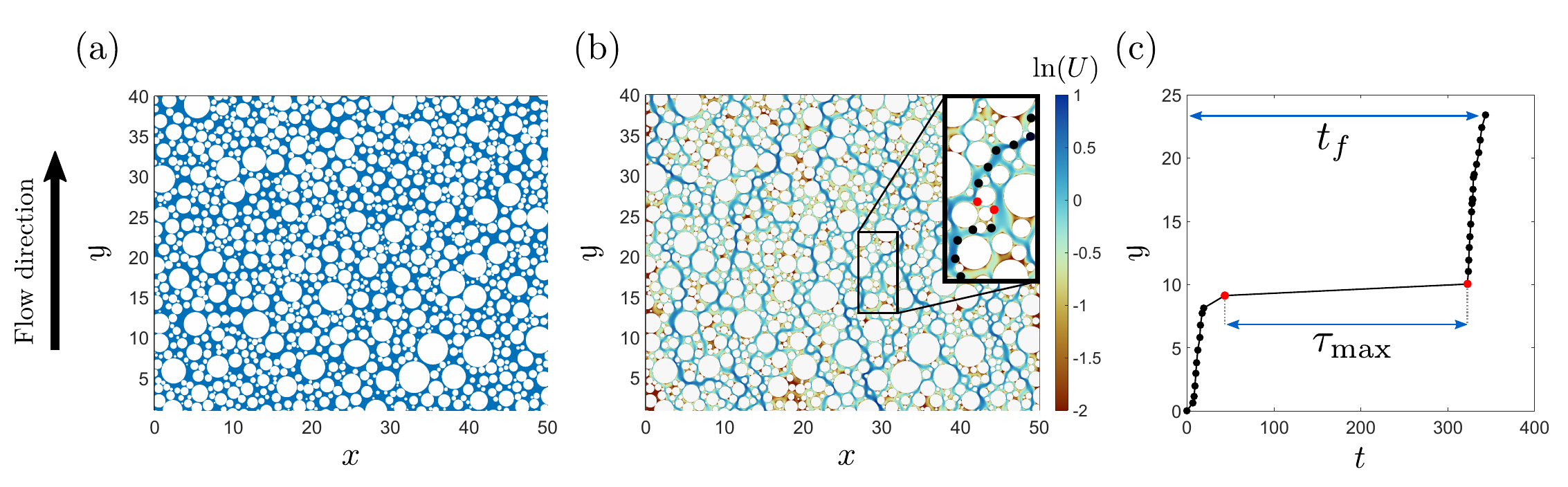}
	\caption{(a) The generated porous medium, with a porosity of 0.35, and mean grain diameter of $\lambda = 1$ [mm]. (b) Fluid velocity field in the generated pore-scale system; the color bar shows the normalized velocity field (divided by the mean value of the field, $\overline{U}$) with a logarithmic scale. The inset shows an example of the location of the longest transition time from a single particle trajectory within a scale of $\lambda$ (see main text). Transitions occur between two successive dots (black marker), where the longest transition time is marked in red. Length unit is [mm], time in [sec], and $\mu$ in [\text{Pa$\cdot$s}]. (c) A single trajectory showing the vertical axis $y(t)$ with $N=30$ transitions. Each time point of a transition is represented by a dot. The $X$-axis is shown in a non-dimensional time domain, i.e. $t = t_\text{dim} \lambda / \overline{U}$ where the dimensional time $t_\text{dim}$ has the unit $\overline{U}/\lambda$, and $\overline{U}$ is the mean velocity. For the presented trajectory, we find $\tau_\text{max} \approx 279 $ and $t_f \approx 343 $ where $t_f$ is the time it takes the particles to transverse the typical distance $L\approx \lambda N$ from bottom to top.} \label{fig1}
\end{figure*}

\section{Simulation of pore-scale transport in a porous medium system}\label{sec:sim_pore}

Over twenty years of field and laboratory experiments \cite{BERKOWITZ2009}, numerical simulations \cite{Nissan2019} and theoretical studies \cite{berkowitz2006modeling} have shown how advective-diffusive tracer motion in hydrogeological systems is characterized by many time scales. In particular, the distribution of trapping times in these systems is very broad, for example for biased transport in porous medium continuous time random walk, with power law sticking times, is a profound model \cite{bel2006random}. In many cases, values of the exponents $\alpha$ are in the regime $1<\alpha<2$. Importantly, since $1<\alpha$, and using the simplified picture used so far, we are still in the phase where $G$ is expected to be finite.

\begin{figure*}\begin{center}
\includegraphics[width=1\textwidth]{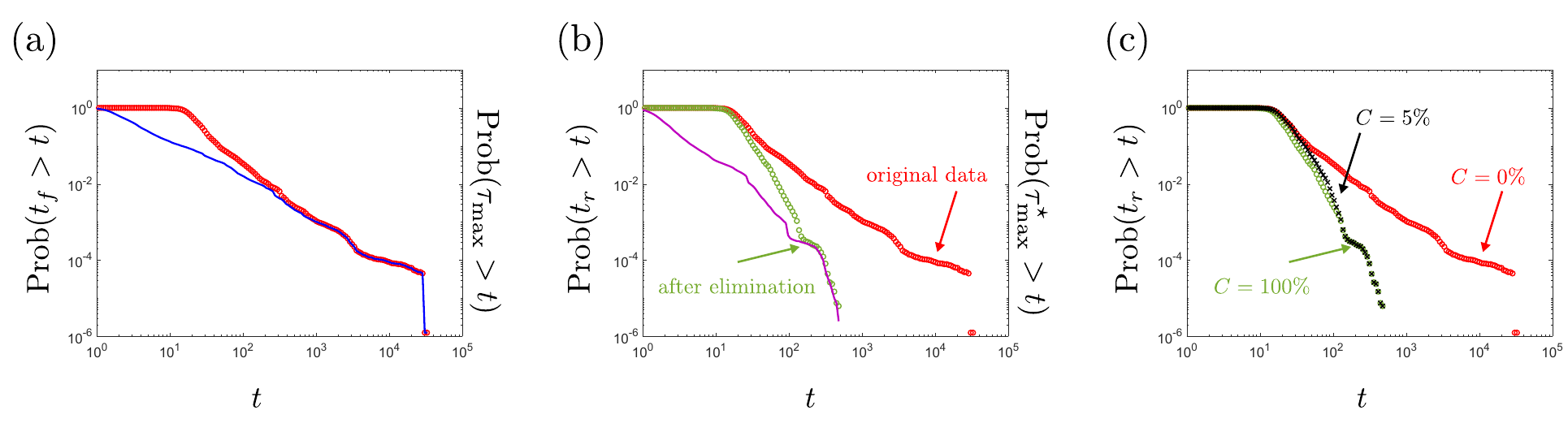}
\caption{(a) The distributions of the first passage time $t_f$ (red circles) and the maximum of the transition times $\tau_\text{max}$ (blue line) for the simulation of transport in a pore-scale system with $N=30$ transition times and $10^6$ particles. The matching of the right tails of the respective distributions shows the single long transition time principle which is non-trivial if compared to the Chistyakov version \cite{chistyakov1964theorem}, as we do not have here a power law tail. (b) The distributions of $t_f^\star$ (green circles) and $\tau_\text{max}^\star$ (purple line), i.e. the after the elimination of the maximum
transition time, for the same system as in (a). We also show the distribution of $t_f$ (red circles) again for comparison. (c) The distribution of the first passage time after eliminating only $C=5\%$ of the largest maxima (black crosses) as explained in the main text, compared with the distributions of the zero cost $C=0\%$ first passage time $t_f$ (red circles) and the costly eliminated $t_f^\star$ (green circles) with $C=100\%$. Clearly, the low level removal $C=5\%$ performs as well as the costly removal $C=100\%$ (the removal of each maximum transition time from its associated trajectory) in the far tails of the distribution. Namely, for the slow movers, which in the context of contaminant spreading  are those who leave long-lasting effects, we can use the cheaper strategy. Note that in our example, the longest transition times we sample are of the order $2\times 10^4$ while after removal we find this to be reduced by a factor of $10$, indicating a large benefit time-wise. \label{fig:alon_results}}
\end{center}\end{figure*}

\subsection{Model}

In this section, we discuss the advection-dominated transport behavior in porous medium. The particles and grains for the porous medium are modelled as hard and impermeable. We have generated a two-dimensional heterogeneous system, by randomly distributing in space solid circular grains from a log-normal distribution, see Fig.~\ref{fig1}(a), with a mean diameter of $\lambda = 1$mm and a standard deviation of $\lambda/2$. The system overall dimensions are: $L_x = 50 \lambda$ and  $L_y = 40 \lambda$, with an average porosity of $\phi = 0.35$. A lognormal distribution can characterize the grain size distribution of different natural soils \cite{Fowler2016}, and therefore was used here as a representative distribution. 
Fluid flow though this system was determined by solving the Stokes equation (using COMSOL Multiphysics$\textsuperscript{\textregistered}$): $\mu \nabla^2 \bf{U} = \nabla p$, where $\bf{U}$ is the pore-scale (local) velocity vector, $p$ the fluid pressure, and $\mu =10^{-3} \text{Pa$\cdot$s}$ the fluid dynamic viscosity of water, coupled with mass conservation, $\nabla U=0$. No-slip boundary conditions were applied on the perimeters of the solid objects (fluid-solid interface) and on the external-vertical boundaries (e.g., impermeable walls). A constant pressure gradient was applied between the bottom and the upper external boundaries of the domain \cite{Nissan2019}. \\ \\

Particle transport was modelled by following an ensemble of particles that move according to the flow field $\textbf{U}$ (Fig.~\ref{fig1}.(b). The spatiotemporal displacement of each particle was determined by a streamline-based method \cite{Bijeljic2011}, which computes the time and distance needed for a particle to exit its current element (within the numerical grid) and arrive to the adjacent element. Particles ($\sim 10^6$) were distributed randomly within a small rectangular strip along the entire (bottom) inlet boundary of the flow domain, as an initial condition, and then moved according to the equation of motion, $d\textbf{x}$ = \textbf{U}[\textbf{x}(t)]dt; where $\textbf{x}$ is the particle location vector and $dt$ is the time step \cite{PereiraNunes2015}. Here, the transport mechanism takes into account only the advective component, while neglecting the occurrence of molecular diffusion. In practice, this scenario is suitable for advection-dominated flow regimes \cite{Bijeljic2011,Nissan2019}.  \\ \\

To relate the transport of particles within the system to the principle of the single long transition time, we first need to characterize the transition times. To do so, we use a standard method \cite{Bijeljic2011} from single particle tracking to define the transition times, using a length scale roughly the size of the grain diameter ($\lambda$); see SM Sec.~D for more information. We use $N = 30$ as a representative number of transitions in the numerical simulations in order to capture the fixed $N$ situation of the previous model. Thus, the typical travel distance of the particles is $L\approx \lambda N$. Smaller values of $N$ were examined (not shown) and showed the same behavior. \\ \\


In Fig.~\ref{fig1}(b) inset, we show an example of a single particle trajectory within the domain, where the locations of transitions are marked in black dots, and the maximum transition time $\tau_\text{max}$ is marked in red. From this example, it can be seen that the maximum transition time occurs when a particle is transported within a narrow pore, perpendicular to the pressure gradient (main flow direction). As a result, the particle velocity magnitude ($\lVert \bf{U} \rVert$) is small (see the color bar in Fig. \ref{fig1}(b), and therefore the transition time $\sim \lambda/\lVert \bf{U} \rVert$ becomes large. The results shown here can be considered typical of other realizations of the domain disorder and with other particle starting positions.

\begin{figure}
	\centering\includegraphics[width = 1\columnwidth]{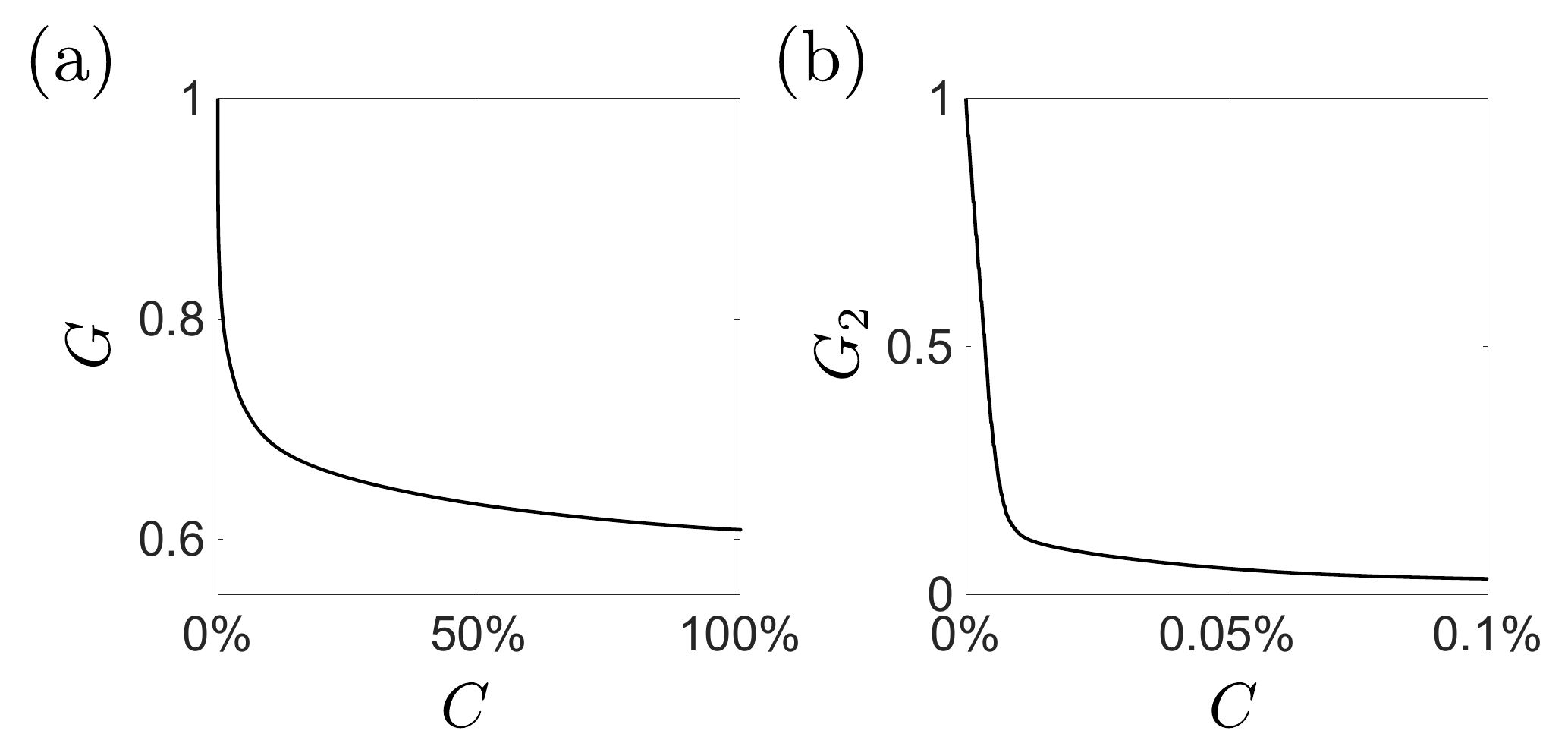}
	\caption{(a) The quantifier $G$ Eq.~(\ref{gquant}) versus the cost factor $C$ for the porous medium simulation presented in Fig.~\ref{fig:alon_results}. As explained in the text, $C$ is the percentage of trajectories where the $\tau_\text{max}$ was removed. (b) The quantifier $G_2$ for the variance versus $C$ for the same simulation.} \label{fig:fig4}
\end{figure}

\subsection{Principle of the single long transition time } 
 
In Fig.~\ref{fig:alon_results}(a), we present the distributions for the first passage time $t_f$ and the maximum transition time $\tau_\text{max}$. As expected from previous works, these distributions are very wide (note the log-log scale). The striking behavior is that we observe excellent matching in the right part of the tails of the distributions, i.e., when $t_f$ and the maximum are large. Thus, we find the principle of the single long transition time in a non-trivial system, as in Eq.~(\ref{bjp_all}). This indicates that the longest first passage times are dominated by the longest transition times, and not by a sequence of many relatively long sticking times, or particles that move against the flow.

An interesting effect is found in Fig.~\ref{fig:alon_results}(a): the tails of both distributions, of $t_f$ and $\tau_\text{max}$, are not smooth, but exhibit step-like structures. The transport is dominated by the disorder configuration of the system. Namely, while the specific pattern of the distributions depends on the details of the porous medium model at hand, the single long transition principle holds also in this fixed background system. It is also clear that the existing theory of the big jump principle for independent and identically distributed random variables \cite{chistyakov1964theorem}, presented in the previous section, needs modifications to treat quenched disorder (fixed in time like the scattering centers). This we will be done later in Sec.~\ref{sec:qtm2} when treating the well-known quenched trap model.

\subsection{Elimination of the single long transition time}\label{sec:cost_function}

Now we eliminate $\tau_\text{max}$ from each trajectory and study the first passage time $t_r = t_f - \tau_\text{max}$. This means that that we follow each complete trajectory, and then remove the maximum transition time from it. Thus, the analysis focuses on numerical validation of the removal principle. Regarding the practical applications, we refer to Sec.~\ref{sec:first_outlook} below as well as Sec.~\ref{sec:outlook}. In Fig.~\ref{fig:alon_results}(b), we present the distribution of $t_r$ (green) and the distribution of $t_f$ before elimination (red). Clearly, we see a dramatic improvement in the transport. In the same figure, we also show $\tau_\text{max}^\star$ after elimination (purple). The figure illustrates the second level of the long transition time principle, namely the tails of the distribution of $t_r$ and of $\tau_\text{max}^\star$ match (like in Fig.~\ref{fig:alon_results}(a) where the global maximum distribution is compared with the distribution of $t_f$), compare also to Eq.~(\ref{cap}). The matching of the two distributions for $t_r$ and $\tau_\text{max}^\star$ is valid even for a relatively complicated structure of the right tail.\\ \\

Practically the most important observation is that the elimination leads to a significant reduction of the first passage times. Viewing the tracers as contamination, clearly, the removal or treatment of the long transition time has a dramatic effect on the cleanup of the system (see the original data in red and the data after removal in green). Comparing the mean first passage times before and after elimination, as in Eq.~(\ref{gquant}), we obtain the value $G\approx 0.6085$; thus the elimination of $\tau_\text{max}$ leads to expedited transport by about $39 \%$. \\ \\

So far, we showed that we gain nearly 40\% increase of efficiency of transport by the elimination method. However, the method we used is costly, as it demands the elimination of the maximum transition time from each trajectory. To move closer to real applications, we address the following protocol. We chose to remove the longest transition times from a finite percentage $0<C<1$ of the trajectories ($C$ is for cost). In general, the idea is that in transport we may discover a few pivotal regions or hot-spots where a critical number of very large $\tau_\text{max}$ occur, and then we need to treat/eliminate only these spots, to expedite the transport. First, we order the maxima according to $\tau_{\text{max},(1)}<\ldots<\tau_{\text{max},(R)}$ where $R$ is the number of trajectories. Then we eliminate the $C\times R$ largest maxima $\{\tau_{\text{max},(R[1-C])}<\ldots<\tau_{\text{max},(R)}\}$ and obtain the modified first passage time $t_f^\star$. The remaining $R(1-C)-1$ trajectories (with low maxima) remain with $t_f$. The total elimination $C=1$ is clearly costly (but as shown very efficient) while $C=0$ is the limit of zero cost but also clearly not useful for our purpose. In Fig.~\ref{fig:alon_results}(c), we compare the distribution of $t_f^\star$ with $C=0.05$ with the two distributions of the original $t_f$ with $C=0$ and $t_f^\star$ with $C=1$. Remarkably, for large first passage times, the distributions of $t_f^\star$ with $C=0.05$ and $1$ are similar. Thus, the far right tail of the distribution of $t_f^\star$, for the case of partial removal of merely $5 \%$ ($C=0.05$) is nearly as efficient as the costly case with $C=1$. Thus, because of the scale-free nature of the process, it is sufficient to use a relatively inexpensive method, and small $C$ performs well. \\ \\ 

To further quantify these observations we plot in Fig.~\ref{fig:fig4}(a) the gain quantifier $G$ versus the percentage treated trajectories; as mentioned, when $C=1$, we obtain $G\approx 0.61$, while clearly if $C=0$ then $G=1$. We find that already for the relatively small value $C\approx 0.1$, $G$ converges almost to the fixed value $0.61$, which means that any additional elimination above $C\approx 0.1$ does not yield further significant gain.\\ \\

Finally, we can quantify the gain also with the variance. In Fig.~\ref{fig:fig4}(b), we show the ratio between the variance of the first passage time after elimination of the longest sticking time, and the variance of the original data set. 
This ratio is denoted as $G_2$, and we perform the elimination as before with some percentage $C$. We find that for a small value $C\approx 0.01 \%$ the quantifier $G_2$ dropped quickly. After that, the convergence is very slow until $G_2 \approx 0.25 \%$ for $C=100\%$. We see from the sharp drop in $G_2$, that the variance of the first passage time is very sensitive to the removal. For the advection-diffusion model in the porous medium under study, this is because the variance is by far more sensitive to the shape of the distribution at its fat tail if compared to the mean. And this is also related to the fact that here the disorder is not too strong, namely the mean of the transition times is finite, for the simulation in Fig.~\ref{fig:alon_results} we found $\langle \tau_n \rangle \approx 0.96$. More specifically, the fact that the ratio of the variances is so small is important, because reducing the variance makes the system more homogeneous, 
and hence predictable (we will discuss this in a future publication). Thus, the transport behavior tends toward Fickian behavior as we remove more maxima. 

\subsection{Bottlenecks in the porous medium model}\label{sec:first_outlook}

\begin{figure}
	\centering\includegraphics[width = 1\columnwidth]{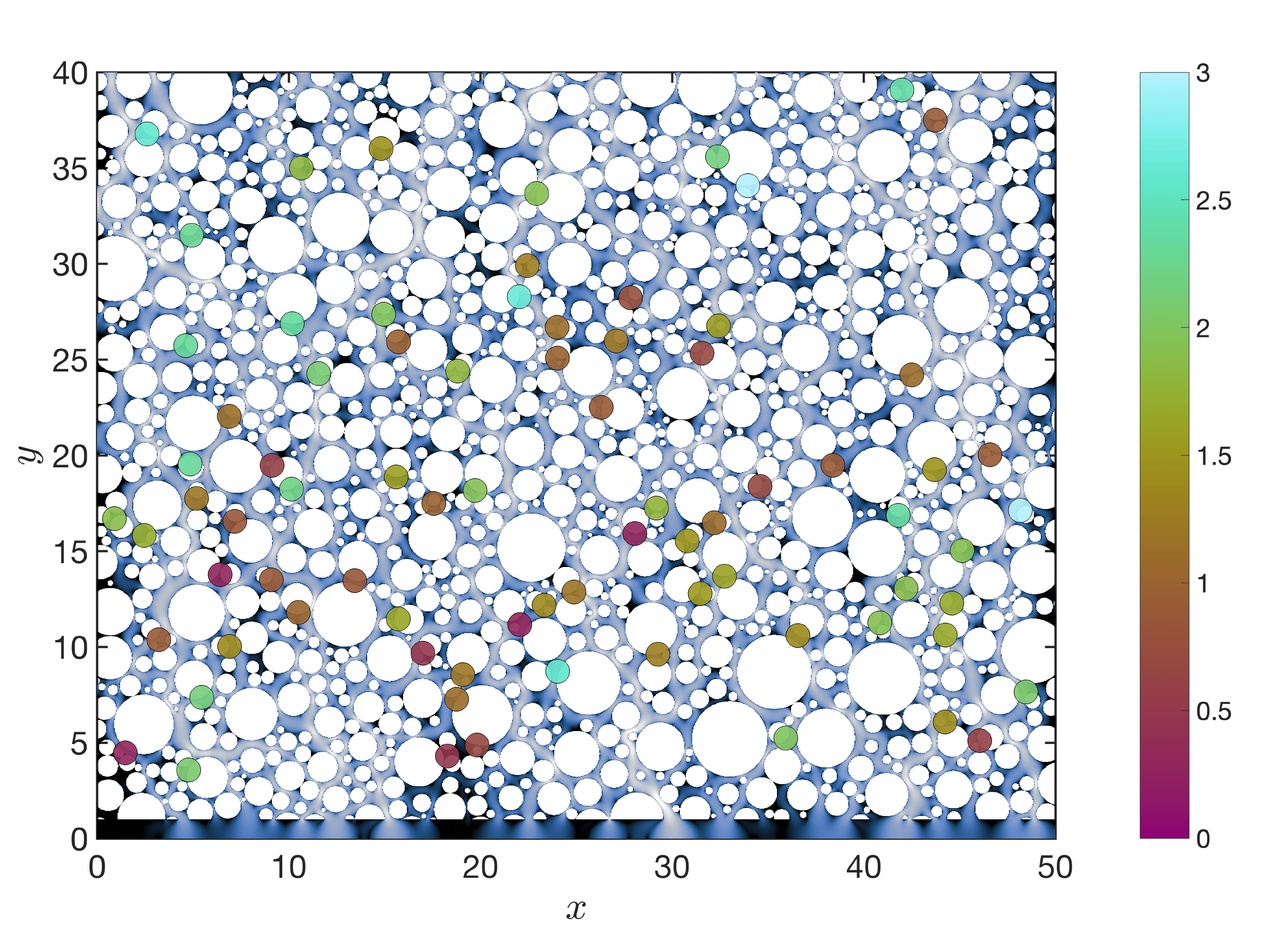}
	\caption{Location of the 1\% largest maximum transition times, i.e. the cost function is $C=0.01$, for the system shown in Fig.~\ref{fig1}. The color bar on the right hande side is displayed in a logarithmic scale, where $0$ corresponds to $1$ particle and, for instance, $3$ represents $10^3$ particles. Clearly, the system exhibits well defined localized spots where the transport is slowed down dramatically.} \label{fig:fig78}
\end{figure}

\begin{figure}
	\centering\includegraphics[width = 1\columnwidth]{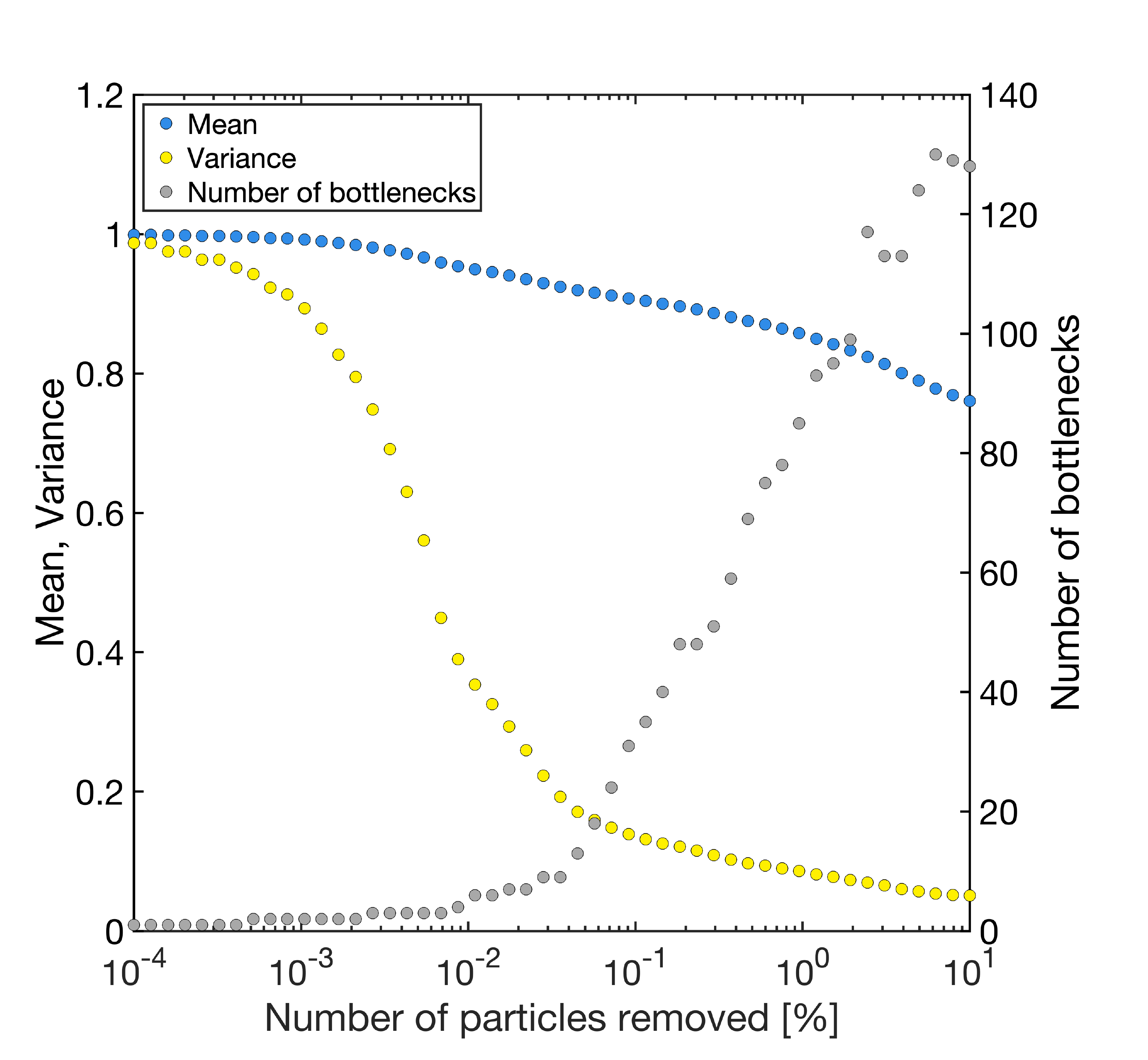}
	\caption{Mean (blue) and variance (yellow) of the first passage time distribution of particles after removing particles with the longest transition times in the deepest bottlenecks (gray) for the system shown in Figs.~\ref{fig1} and~\ref{fig:fig78}. Both moments are normalized with respect to their initial values (i.e., before particles are removed). } \label{fig:fig789}
\end{figure}

As mentioned in the introduction: a natural question is, can we find specific spatial locations where the process is slowed down? The general answer to this question is beyond the scope of this study. Here, we focus on the simulation model described above. After coarse graining, we may define transition times along the paths of individual particles, as already considered. We then, as before, search for the longest transition times, focusing on a certain percentage of the largest maximum sojourn times in the entire system. We can then envision two opposing cases: if the system is homogeneous, these longest (trapping) times will be spread uniformly in the system, while if the system is strongly disordered, the longest times will be distributed non-uniformly.\\ \\

In Fig.~\ref{fig:fig78}, we present the locations for 1\% of the largest maximum sojourn times in our system for a total of $10^6$ particles. It becomes clear that the system shows certain locations where a large number of particles are trapped for extreme times. Thus, the extremes are distributed in a highly non-homogeneous manner throughout the system. This indicates that, at least in principle, we can consider modification of some rather small part of the system and gain a large effect on the reduction in the first passage time. In other words, we envision a situation wherein we use a learning session, with a relatively small number of particles, to gain information from the resulting map  (e.g., Fig.~\ref{fig:fig78}). Identifying the locations of deep bottlenecks (longest trapping times) is key to identifying a removal strategy and an enhancement of the first passage time. At this stage, it is premature to specify how, precisely, a practitioner should treat this problem in practice, in terms of determining how to create the bypass for the slow spots. Moreover, from a theoretical point of view, this is no general theory that can specify exactly when similar effects will occur (beyond the one-dimensional trap model in Sec.~\ref{sec:qtm2} and the model under study). Maps such as those presented in Fig.~\ref{fig:fig78} will in general depend on the bias, the geometry and size of the system, the disorder itself, the coarse graining, the threshold for the number of largest maxima, and the initial conditions. For example, a flux-weighted initial distribution was also investigated but showed no statistical difference to the random distributed initial condition presented here; it is therefore not discussed further. Other initial conditions like point injections, however, are likely to impact the specific results shown in Fig.~\ref{fig:fig78}. This should be more prominent when the bias becomes very strong. Further investigation in this direction is clearly needed. \\ \\

While a detailed study of bottleneck modification is beyond the scope of this study, we provide further analysis of the impact of localized bottlenecks. Fig.~\ref{fig:fig789} shows how the mean and variance of the first passage time distribution are altered when specific particles are removed from the deepest bottlenecks, shown in Fig.~\ref{fig:fig78}. The elimination procedure begins by identifying the particles with the longest transition times and determining the bottlenecks in which they reside (using a fixed-radius, $\lambda$, nearest neighbors clustering algorithm); rather than remove the entire bottleneck from the system (which would then lead to a change in flow patterns and thus a modified overall transport behavior), we extract specific (long transition time) particles that pass through these regions.  \\ \\

In Fig.~\ref{fig:fig789}, the mean and variance (left $y$-axis) are normalized with respect to their initial values (i.e., before particles are removed). The right $y$-axis denotes the number of bottlenecks that have been extracted (shown as grey dots). The $x$-axis indicates the percentage of particles removed from these bottlenecks, and thus the entire system. Clearly, this knowledge of the maximum transition times is attainable only after a learning session. Detecting bottlenecks prior to particle transport necessitates an analysis of the system itself, such as a comparison of trap depths, as demonstrated in the case of the quenched trap model in Sec.~\ref{sec:qtm2}. Fig.~\ref{fig:fig789} shows that, significantly, interception and extraction of even a relatively small number of particles in the deepest bottlenecks triggers notable alteration of the overall transport dynamics. Specifically, addressing just two bottlenecks, which results in the extraction of only $\sim$0.001\% of the particles, yields a notable reduction of $\sim$10\% in the variance of the first passage time. This reduction highlights a substantial transformation in transport characteristics, suggesting a discernible shift towards Fickian behavior. As the proportion of eliminations increases, the pace of transport quickens, leading to a reduction in the mean value of the first passage time. Note that the variance is more strongly affected than the mean when considering the removal of only limited numbers of particles with longest overall transition times. In contrast, the removal of many or all maximum transition times has a large influence also on the mean, as seen in Fig.~\ref{fig:fig4}. Thus, spatial removal of maxima from specific locations and removal of maxima from all paths do not yield the same results.

\section{Continuous time random walk}\label{sec:ctrw2}

We now consider a basic model for anomalous transport, the continuous time random walk \cite{metzler2000random,kutner2017continuous}, which has found application in a vast number of systems. The first passage time problem in this model was studied extensively  \cite{bel2006random,bel2005occupation,rangarajan2000anomalous,condamin2007first,condamin2007first2,balakrishnan1983first,krusemann2015ageing,krusemann2014first,jose2021passage}, starting from the pioneering work of H. Scher and E. Montroll in the context of time of flight of charge carriers in disordered material \cite{scher1975anomalous,montroll1973random}. We study the connection between the first passage time and the maximum transition time, formulating the single long transition time principle, and then investigate the transport enhancement via elimination of the maximum transition time. 

\subsection{Model}

\begin{figure}\begin{center}
\includegraphics[width=1\columnwidth]{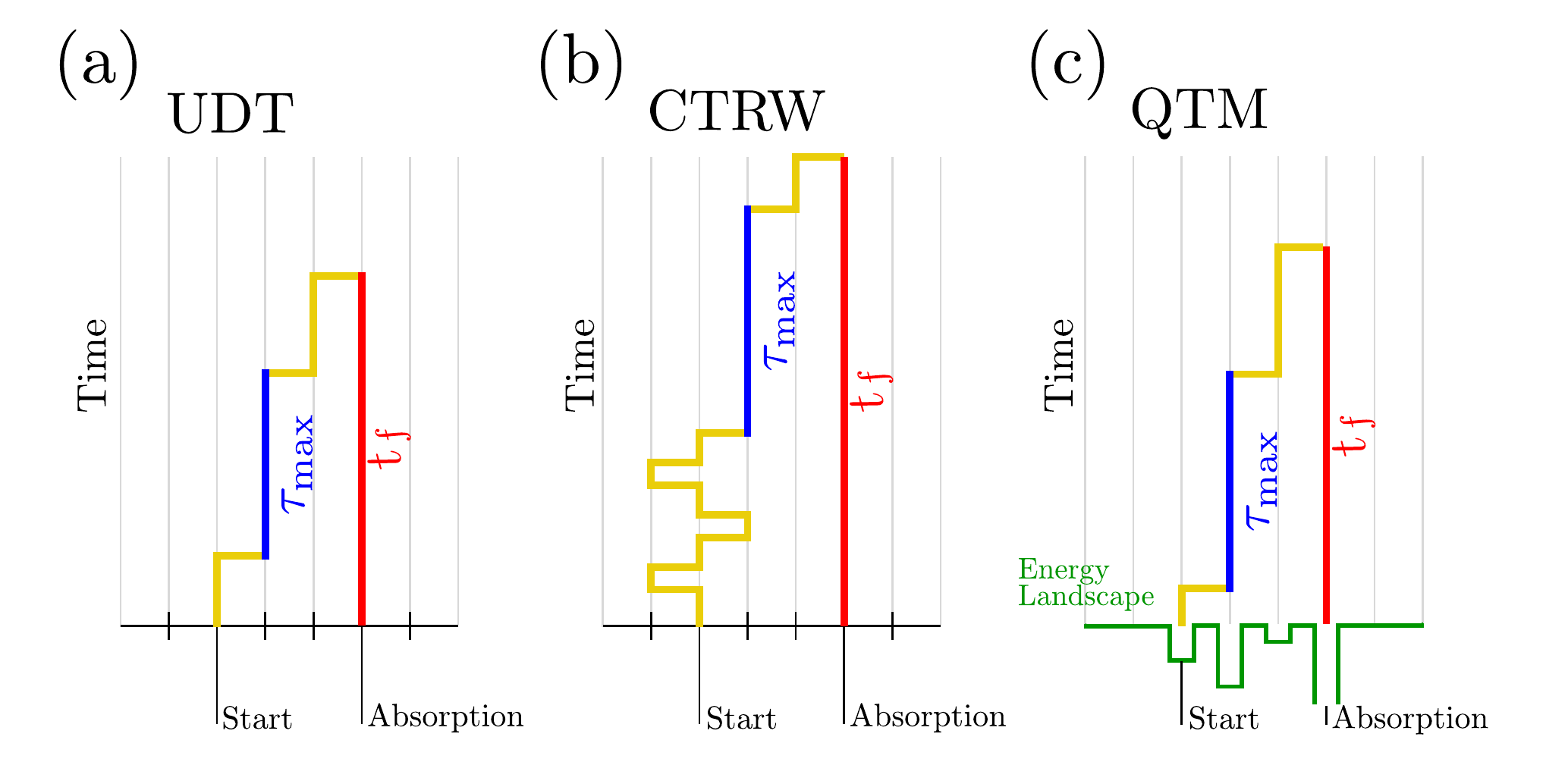}
\caption{Conceptual figure of paths in models considered in this paper. The walks are on a lattice, while the (a) unidirectional transport (UDT) (see also Fig.~\ref{fig:main_results}) and (b) continuous time random walk (CTRW) describe homogeneous processes, in the (c) quenched trap model (QTM) we have a particle in a random, fixed in time, energy landscape.\label{fig:concept_all_models}}
\end{center}\end{figure}

In contrast to the unidirectional model of Sec.~\ref{sec:odt}, jumps to the left and right are permitted, see Fig.~\ref{fig:concept_all_models}. The probability of jumping from some lattice point to the left is $q$ and to the right is $p=1-q$. The difference $p-q$ is related to the driving $F$ stemming from an external force field, in the limit of small $F$ via linear response theory \cite{barkai1998generalized,barkai2000continuous}. In addition, at each lattice point, the particle waits a random time $\tau$ distributed according to the power law distribution Eq.~(\ref{transitionprob}), namely $\psi(t) \sim A t^{-1-\alpha}$, and $\alpha$ was obtained in \cite{metzler2000random}. The particles start at $x=1$. The lattice points are $\{\ldots,-1,0,1,\ldots,L\}$, thus having a semi-infinite lattice, and the absorbing boundary is situated at $L+1$. The first passage time $t_f$ to reach $x=L+1$ is the sum $t_f=\sum_{n=1}^N \tau_n$ and the maximum is $\tau_\text{max}=\text{max}\{\tau_n\}_{n=1}^N$. $N$ which is the number of jumps made by the particle before absorption is random while for the unidirectional model of Sec.~\ref{sec:odt} $N$ is finite namely $N=L$. This is a crucial difference, especially when the fluctuations of $N$ are large. We see that the first passage time $t_f$ in the continuous time random walk is the same as in Eq.~(\ref{tf_all}) and $\tau_\text{max}$ as in Eq.~(\ref{max_all}) but with a random number of transitions $N$. \\ \\

\subsection{Principle of the single long transition time}\label{sec:mainctrwcp}

Because $N$ is random, we have to average over it with a technique called subordination \cite{fogedby1994langevin,bel2005occupation,bel2006random,condamin2007first}. The idea is to consider a discrete time random walk, namely performing jumps every unit of time. In this walk, we use the same bias, initial conditions, and boundary conditions, as for the original model, the continuous time random walk. Let $\phi_\text{dis}(n)$ be the probability that a particle made $N=n$ jumps before its absorption, and the subscript ``dis'' stands for discrete. Obviously, $\phi_\text{dis}(n)$ depends on $p$ and the initial distance $L$ to the absorbing boundary plays a key role. \\ \\

\begin{figure}[H]\begin{center}
\includegraphics[width=1\columnwidth]{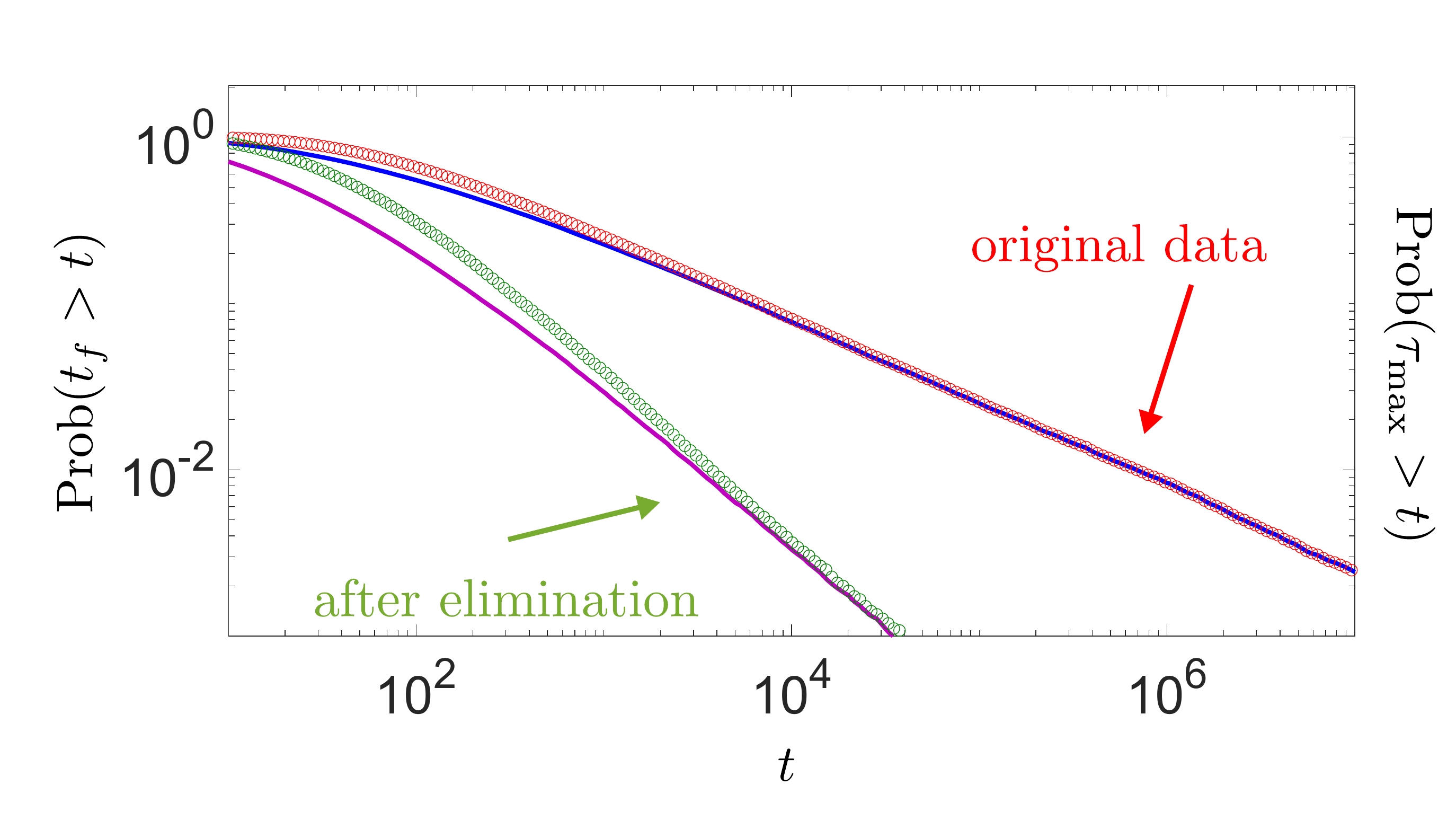}
\caption{The distributions of $t_f $ (red circles), $\tau_\text{max}$ (blue line), $t_r$ (green circles) and $\tau_\text{max}^\star$ (purple line) for the continuous time random walk with $L=5$ and right bias $p=0.8$. Pareto transition times with $t_0=1$ and $\alpha=0.5$ were used for the Monte-Carlo simulations with $10^5$ particles. We find perfect matching between the tails of the two distributions as predicted by Eq.~(\ref{ctrw_bjp_random}) and (\ref{secbjp_ctrw}). We clearly see also here a large benefit from the elimination of the longest transition time $\tau_\text{max}$. \label{fig:ctrw_cp_right_bias}}
\end{center}\end{figure}

The following analysis depends on the mean $\langle N \rangle = \sum_{n=1}^\infty n\phi_\text{dis}(n)$, namely the mean number of steps in a biased discrete time random walk made before absorption. We have to differ  the two cases of finite $\langle N \rangle < \infty$ and the case where $\langle N \rangle =\infty$ diverges, which yields a vastly different behavior for the long transition time principle. We first study the case $p>1/2$ where $\langle N \rangle$ is finite. In this case we are treating a problem, investigated by \cite{kluppelberg1997large} in generality, namely where the large deviations are studied for a random number of random variables. As before, we are interested when the first passage time is large, and find
\begin{equation}\boxed{\begin{split}\label{ctrw_bjp_random}
\text{Prob}(t_f > t) &\sim \text{Prob}(\tau_\text{max}>t) \\
&\sim \sum_{n=1}^\infty n \phi_\text{dis}(n)  \frac{A}{\alpha}t^{-\alpha} \\
&\sim \langle N \rangle \frac{A}{\alpha}t^{-\alpha},
\end{split}}\end{equation}
see SM Sec.~E. We see that the difference to the unidirectional case Eq.~(\ref{bjp_all}) is that we replace $N$ by the mean number of jumps $\langle N \rangle$. More importantly, the principle of single long transition time holds, as before, see Fig.~\ref{fig:ctrw_cp_right_bias}.\\ \\

We find $\phi_\text{dis}(n)$ and then $\langle N \rangle$ which is easy to obtain $\langle N \rangle = L/(p-q)$ (see \cite{redner2001guide,klafter2011first,bel2006random} and also SM Sec.~E). Then Eq.~(\ref{ctrw_bjp_random}) reads 
\begin{equation}\begin{split}\label{ctrwbjexample}
\text{Prob}(t_f>t) &\sim \text{Prob}(\tau_\text{max}>t) \\
&\sim \frac{L}{p-q} \frac{A}{\alpha} t^{-\alpha}.
\end{split}\end{equation}
Notice that when $p\to 1/2$, and hence $q=1-p\to 1/2$ as well, the amplitude diverges. Similar results can be obtained for other models (see SM Sec.~E, F), for example the case when we have a reflecting wall situated possibly far from the initial condition, or when we replace the lattice model with a continuous space version.

\subsubsection{Principle of the extended long transition time}

A very different behavior is found for the non-biased continuous time random walk, namely $p=q=1/2$. Then the mean $\langle N \rangle$ is infinite, which is well-known \cite{redner2001guide}, as it stems from the fact that the walker can drift to the left, in the direction opposing the absorbing boundary. Nevertheless, the random walk is recurrent. The open question is will the single long transition time principle still hold and if so what are the asymptotics? \\ \\

\begin{figure}[H]\begin{center}
\includegraphics[width=1\columnwidth]{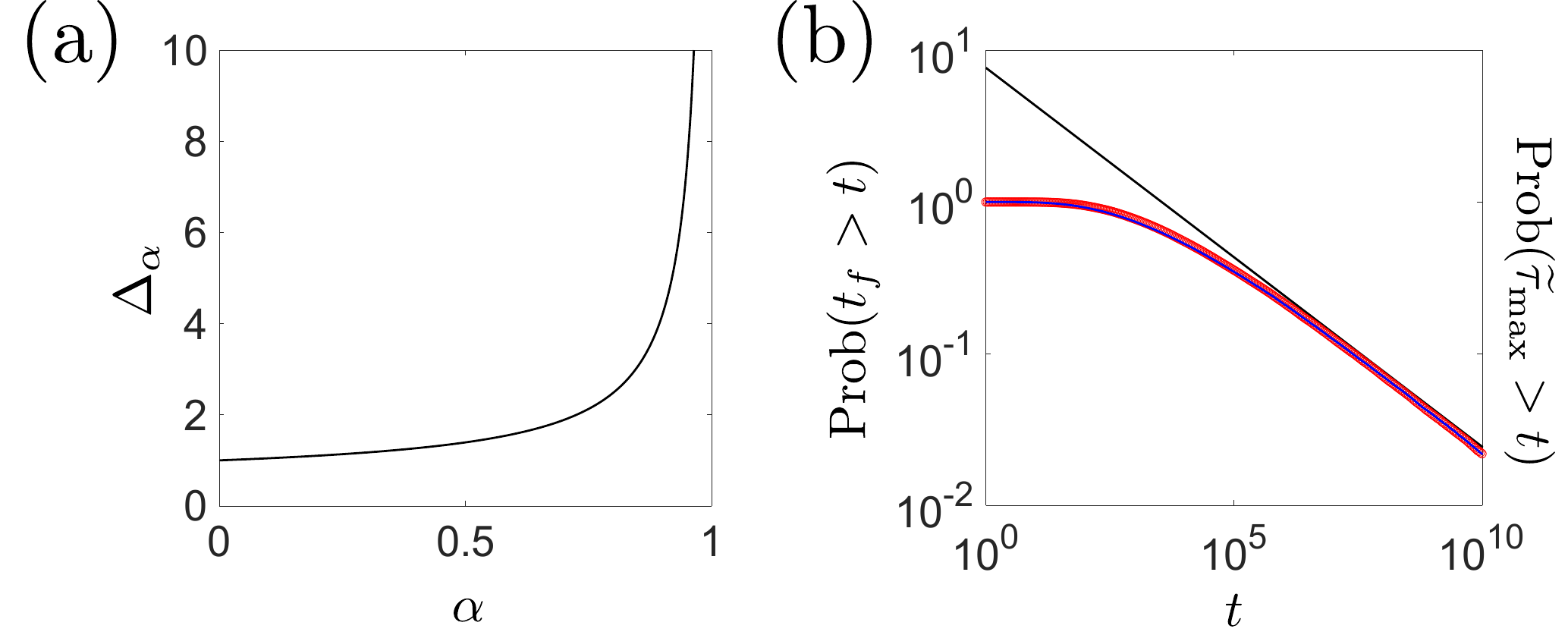}
\caption{(a) The factor $\Delta_\alpha$ given in Eq.~(\ref{stretched3}) to obtain the rescaled transition time $\tilde{\tau}_\text{max}= \Delta_\alpha \tau_\text{max}$. (b) The distributions of $t_f$ (red circles) and $\tilde{\tau}_\text{max}$ (blue line) compared with the theory Eq.~(\ref{stretched2}) (black line) for the continuous time random walk with no bias $p=0.5$ and $L=5$. We used Pareto transition times with $t_0=1$ and $\alpha=0.5$ for the Monte-Carlo simulations with $10^5$ particles. Due to possible very long left excursions, we cut off the simulations once $t_f>10^{14}$. We find perfect agreement with the theory of the extended long transition time principle Eq.~(\ref{stretched2}).\label{fig:ctrw_cp_no_bias}}
\end{center}\end{figure}

To formulate the single long transition time principle we define a rescaled maximum transition time $\tilde{\tau}_\text{max}= \Delta_\alpha \tau_\text{max}$ with the $\alpha$-dependent factor 
\begin{equation}\label{stretched3}
\Delta_\alpha=\left(\frac{2\sqrt{\frac{|\Gamma(-\alpha)|}{\alpha}}}{|\Gamma(-\alpha/2)|}\right)^{2/\alpha}.
\end{equation}
In Fig.~\ref{fig:ctrw_cp_no_bias}(a), this rescaling factor is presented. Clearly $\Delta_\alpha > 1$, hence $\tilde{\tau}_\text{max} > \tau_\text{max}$. For $\alpha\to 0$, we obtain the original maximum $\tilde{\tau}_\text{max} \to \tau_\text{max}$ while for $\alpha \to 1$, the scaling factor $\Delta_\alpha$ diverges. Now this rescaled long transition time $\tilde{\tau}_\text{max}$ is related to the first passage time by the asymptotics
\begin{equation}\boxed{\begin{split}\label{stretched2}
\text{Prob}(t_f > t) &\sim \text{Prob}(\tilde{\tau}_\text{max}>t) \\
&\sim  (\Delta_\alpha)^{\alpha/2} \sqrt{\frac{2A}{\alpha}}t^{-\alpha/2},
\end{split}}\end{equation}
see Fig.~\ref{fig:ctrw_cp_no_bias}(b) and the derivation in SM Sec.~E. The fact that here we rescale $\tau_\text{max}$ with the factor $\Delta_\alpha$ means that the previously discussed long transition time principle Eq.~(\ref{ctrw_bjp_random}) still holds but with a renormalized definition of the maximum transition time. But in contrast to Eq.~(\ref{ctrw_bjp_random}), in Eq.~(\ref{stretched2}) the power law decay is $t^{-\alpha/2}$. The first passage time is always larger than the original $\tau_\text{max}$, that is why we rescaled the latter with $\Delta_\alpha > 1$. Thus, we call Eq.~(\ref{stretched2}) the principle of the extended long transition time. In the limit $\alpha \to 0$, the first passage time scales as the longest transition time without rescaling. On the other hand, in the limit $\alpha \to 1$, we have $\Delta_\alpha \to \infty$ so that the principle breaks down as we discuss now. \\ \\

What happens for $\alpha>1$ for the unbiased case $p=1/2$? The probability distribution $\text{Prob}(t_f > t)$ called survival probability decays like $t^{-1/2}$ which is a well-known result in the theory of diffusion \cite{redner2001guide}. Because the mean transition time is finite, $\alpha$ plays no role in the decay of this survival probability. This is vastly different from the distribution of the maximum $\text{Prob}(\tau_\text{max}>t)$ which decays in the continuous time random walk model like $t^{-\alpha/2}$; see SM Sec.~E. Hence, for $\alpha>1$ and $p=1/2$, there is no principle of the single long transition time nor of the extended version. We end this subsection with the reference to SM Sec.~E for the discussion of the left biased case $p<1/2$.

The asymptotic relationship in Eq.~(\ref{stretched2}) requires renormalization of the maximum transition time. We note that scale invariance and renormalization are indeed related to the models under study; for example, the renormalization group was studied for the quenched trap model~\cite{monthus2003anomalous}. We note that removal of the largest waiting time will modify the exponent of the first passage time, in the sense defined above for see Eq.~(\ref{cap4}). In usual renormalization group treatment, the relevant exponents do not change via coarse graining. The renormalization group approach used to find the transport properties of the system is different from the one presented here.

\subsection{Elimination of the single long transition time}

\begin{figure}\begin{center}
\includegraphics[width=0.9\columnwidth]{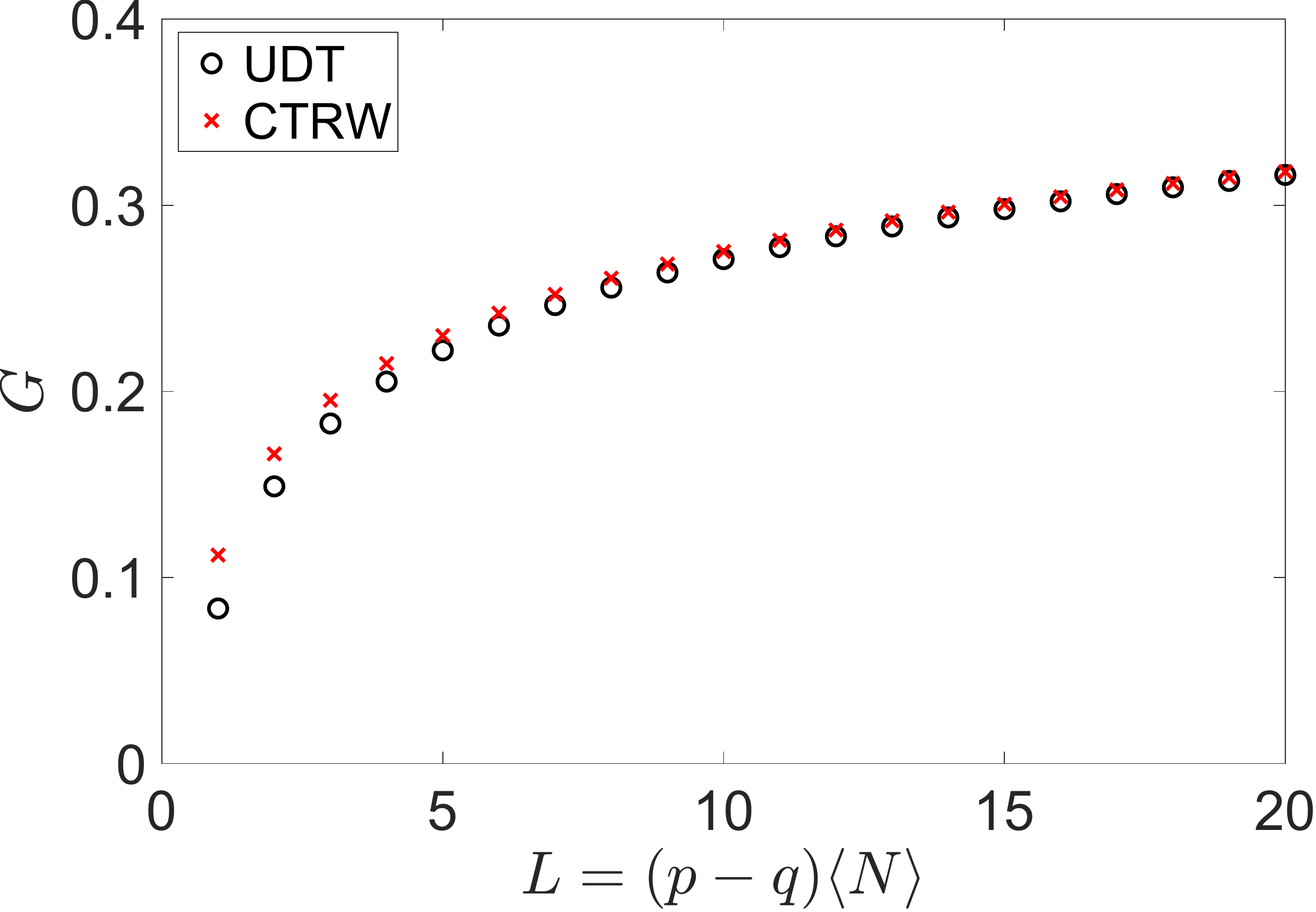}
\caption{The quantifiers $G$ of the unidirectional model Eq.~(\ref{gth}) (black circles) and the continuous time random walk Eq.~(\ref{gctrw}) (red crosses). The transition times follow the Pareto distribution with $\alpha=1.1$ and $t_0=1$. To make the comparison we take $N$ in the unidirectional model equal to $L/(p-q)$, which is the mean number of steps before absorption in the continuous time random walk, namely $\langle N \rangle$. Furthermore, the continuous time random walk has the bias parameter $p=3/4$.\label{fig:gctrw}}
\end{center}\end{figure}

After studying the long transition time principle, we are now ready to make use of it. The fact that tail of distributions of $t_f$ and $\tau_\text{max}$ are related, are a strong indication that the removal of the longest transition times will have a profound effect, which is now quantified. \\ \\

We consider the case $p>1/2$, so the moments of $N$ exist. With the same approach as in Sec.~\ref{sec:mainctrwcp}, we obtain the asymptotic relationship
\begin{equation}\boxed{\begin{split}\label{secbjp_ctrw}
\text{Prob}(t_r > t) &\sim \text{Prob}(\tau_\text{max}^\star>t) \\
&\sim \frac{1}{2}\langle N(N-1) \rangle \left( \frac{A}{\alpha}\right)^2 t^{-2\alpha}
\end{split}}\end{equation}
with $\langle N(N-1) \rangle = \sum_{n=1}^\infty n(n-1)\phi_\text{dis}$. Compared to the result of the unidirectional model Eq.~(\ref{cap4}) and (\ref{cap}), we find again the same power law decay $-2\alpha$ but the $N$-dependent prefactor must be averaged, see Fig.~\ref{fig:ctrw_cp_right_bias}.\\ \\

We measure the gain from the elimination of $\tau_\text{max}$ with the ratio $G=\langle t_r \rangle / \langle t_f \rangle$ as in Eq.~(\ref{gquant}). For the example of the Pareto distributed transition times, it is
\begin{equation}\begin{split}\label{gctrw}
G=\begin{cases}
0 & \text{, } 0<\alpha<1, \\
1-h_\alpha \sum\limits_{n=1}^\infty \phi_\text{dis}(n) (-1)^n n! \Gamma\left(-n+\frac{1}{\alpha}\right) & \text{, } 1<\alpha
\end{cases}
\end{split}\end{equation}
with $h_\alpha=[(p-q)(\alpha-1)]/[\alpha L \Gamma(1/\alpha)]$, see SM Sec.~\ref{sec:cont2}. In Fig.~\ref{fig:gctrw}, we show the quantifier of the elimination effect $G$ versus $L$. The figure also shows $G$ for the unidirectional model of Sec.~\ref{sec:odt}. For small $L$ the two quantifiers are different, but for large enough $L$ they are similar. This similarity is shown in Fig.~\ref{fig:g1_phase} where we plot $G$ versus $\alpha$ of the continuous time random walk. Based on that observation, the thermodynamic limit $L\to\infty$ can be treated similarly as for the unidirectional model. Namely, the removal of the $s=f N$ longest transition times in the continuous time random walk satisfies Eq.~(\ref{impalg}). It is remarkable that the phase like transition in $G$, found when $\alpha$ is varied, is insensitive to model details. \\ \\

{\bf Remark.} In some transport systems, the continuous time random walk exponent $\alpha$ is spatially varying \cite{fedotov2019asymptotic,fedotov2021variable}. Such cases require a separate discussion. In these systems, particles aggregate in regions where roughly speaking $\alpha(x)$ has a minimum. We expect an even larger effect of removal in these systems.

\section{Quenched trap model}\label{sec:qtm2}

In the unidirectional transport model and the continuous time random walk, the transition times are spatially homogeneous, i.e. independent of the lattice points. We drop this simplifying property now, considering the quenched trap model \cite{bouchaud1990anomalous,monthus1996models,berthier2011theoretical,akimoto2018non,burov2007occupation}. Similar to the simulation of the pore-scale system in Sec.~\ref{sec:sim_pore}, in the quenched trap model the disorder is fixed, more specifically the particle is performing a biased random walk with $1/2<p<1$ in a random energy landscape. For the quenched trap model, we deal with energetic traps on the lattice points $x=\{\ldots,-1,0,1,\ldots,L\}$. As before, the particle starts at $x=1$ and the absorption is at $x=L+1$. At each lattice point $x$, an energy trap $E_x$ is located with the distribution $\text{Prob}(E_x>E)=\text{exp}(-E/T_g)$ and $T_g$ is a measure of disorder. At lattice point $x$, where the trap depth is $E_x$, the particle waits the random time $\tau_x$ with the mean $\bar{\tau}_x=t_0\text{exp}(E_x/T)$ which is the well-known Arrhenius time to escape from an energy trap, used in many activation processes. Here, $T$ is the temperature of the system  and $t_0$ is a well-studied timescale for dynamics
in the bottom of the trap \cite{hanggi1990reaction}. According to the basic theory of activation, the distribution of the transition times is exponential $\text{Prob}(\tau_x>t)=\text{exp}(-t/\bar{\tau}_x)$. Averaging $\text{Prob}(\tau_x>t)$ over the disorder gives Eq.~(\ref{transitionprob}) with $\alpha=T/T_g$ \cite{bouchaud1990anomalous}. The idea is that the Arrhenius time is exponential in $E$, and hence even for a thin tailed distribution of the energy, we obtain fat-tailed distributions for the transition times (after averaging over the disorder, which is discussed below). 

\subsection{Principle of the single long transition time}

\subsubsection{Strong bias in a one-dimensional random environment}\label{sec:one_qtm}

\begin{figure}\begin{center}
\includegraphics[width=1\columnwidth]{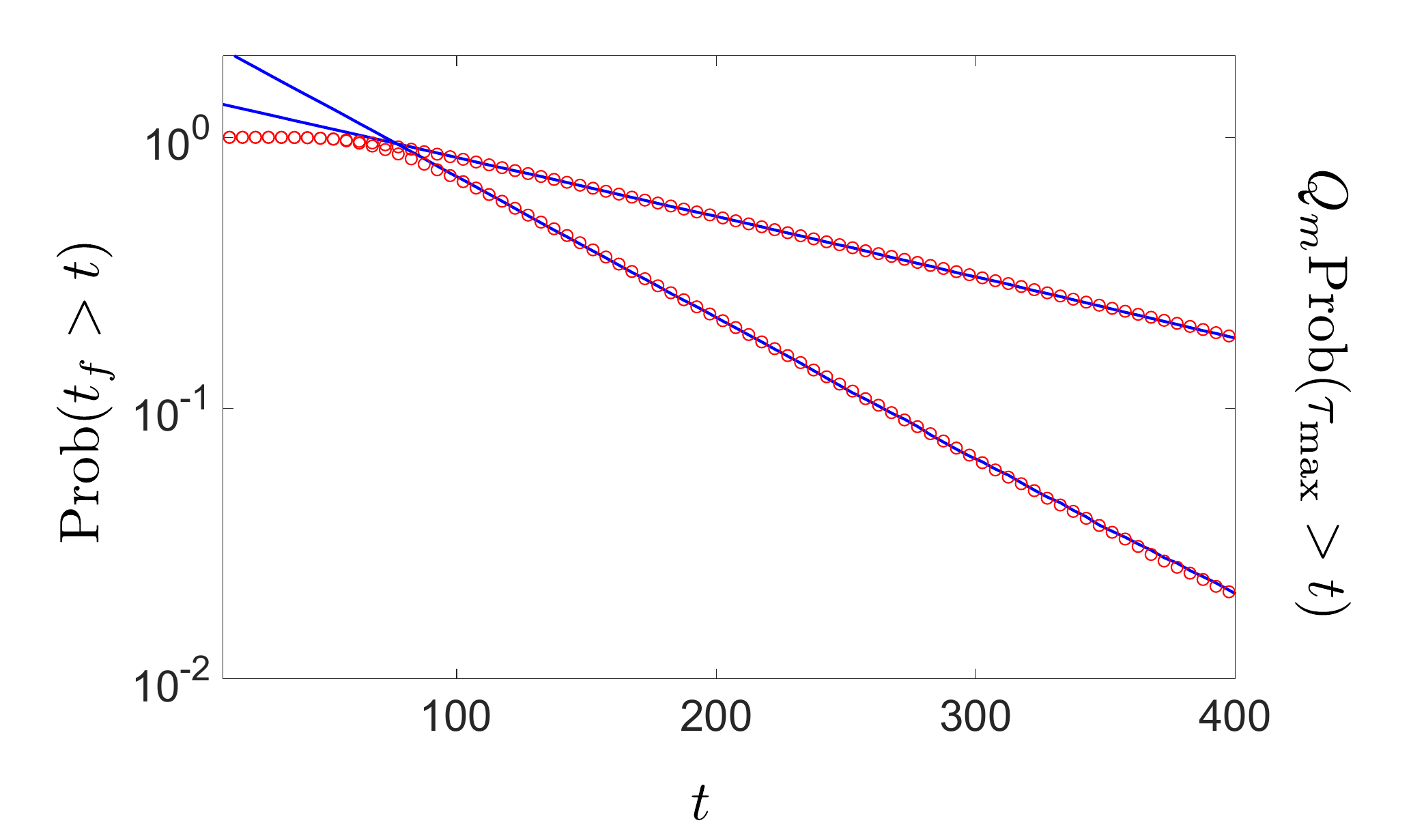}
\caption{The distributions of $t_f$ (red circles) and $\tau_\text{max}$ (blue lines) for the strongly biased quenched trap model with $L=25$. Here, we consider a specific realization of the disorder, corresponding to an experimental situation where no averaging over disorder is made. We find excellent agreement with theory Eq.~(\ref{singe_simple}). The two upper lines represent a unique realization of the disorder with $E_\text{max}\approx 10.61$ and the two lower lines represent another realization with $E_\text{max}\approx 8.86$. We used the parameters $T=T_g=2$ so that $\alpha=1$ and $t_0=1$. To generate the figure, we used $10^5$ trajectories for each disorder. These simulations take into consideration the thermal fluctuations, namely the activation process happens at random times. The fact that different sets of disorder do not produce the same behavior is an indication for non-self averaging for the observables of interest, however, the single long transition time principle clearly holds.
\label{fig:QTM_One_Channel_Unidirect_CP}}
\end{center}\end{figure}

We first consider the strong bias case $p=1$ where the particles move only to the right, namely a constant strong driving force acts on the system. We consider the first passage time without averaging over the disorder, namely we treat a system with a specified realization of the energies $\{E_1, .....E_L\}$ and the absorption at $x=L+1$. This corresponds to a situation when the experiment has one realization of the disordered system. The first passage time $t_f$ is a sum of the microscopic transition times at the traps, as in Eq.~(\ref{tf_all}) with $N=L$. The maximum transition time, in the quenched trap model, is defined as the transition time in the deepest trap $E_\text{max}=\text{max}(E_1,\ldots,E_L)$. Let's say the deepest trap is at the lattice point $x=m$. Then $\tau_\text{max}=\tau_m$. We consider the probability density function of the first passage time $p_{t_f}(t)=\langle \delta(t-[\tau_1+\ldots + \tau_L])\rangle$. Its Laplace transform $\hat{p}_{t_f}(s)=\int_0^\infty p_{t_f}(t)\text{exp}(-st)\mathrm{d}t$ and the maximum probability are
\begin{equation}\begin{split}\label{qtmmethod1}
\hat{p}_{t_f}(s) &= \prod_{x=1}^L (1+\bar{\tau}_x s)^{-1},\\
\text{Prob}(\tau_\text{max}>t)&=  \text{exp} \left(-\frac{t}{\bar{\tau}_m} \right).
\end{split}\end{equation}
Note the equal sign in the second equation due to the assumption that $\tau_\text{max}$ is the transition time from the deepest trap. Asymptotically, when the maximum transition time is very long, as found in strongly disordered systems, then it always happens in the deepest trap. The inverse Laplace transform of the first formula can be calculated exactly $p_{t_f}(t) = \sum_{x=1}^L Q_x p_{\tau_x}(t)$ with $Q_x= \bar{\tau}_x^{L-1} \prod_{y=1,y\neq x}^L [\bar{\tau}_x-\bar{\tau}_y]^{-1}$. From here we find
\begin{equation}\boxed{\begin{split}\label{singe_simple}
\text{Prob}(t_f > t) &\sim Q_m \text{Prob}(\tau_\text{max}>t) \\
&= Q_m  \text{exp} \left(-\frac{t}{\bar{\tau}_m} \right)
\end{split}}\end{equation}
with the prefactor $Q_m= \bar{\tau}_m^{L-1} \prod_{x=1,x\neq m}^L [\bar{\tau}_m-\bar{\tau}_x]^{-1}$.  The decay of both distributions for $t_f$ and $\tau_\text{max}$ is exponential, unlike the power laws found previously. The reason is that in each trap we have exponentially distributed trapping times and hence naturally the distribution cannot be fat tailed, as the system is finite. What is remarkable, is that the principle of the single long transition time still holds, in the sense that the exponential decays of the two probabilities are the same though note the prefactor $Q_m$ in Eq.~(\ref{singe_simple}). In Fig.~\ref{fig:QTM_One_Channel_Unidirect_CP} we demonstrate Eq.~(\ref{singe_simple}) and compare its prediction with Monte-Carlo simulations. Eq.~(\ref{singe_simple}) is an indication that the removal of the deepest trap, is going to qualitatively change the statistical properties of the time to cross the system $t_f$.

\subsubsection{Strong bias with average over the disorder}

\begin{figure}[H]\begin{center}
\includegraphics[width=1\columnwidth]{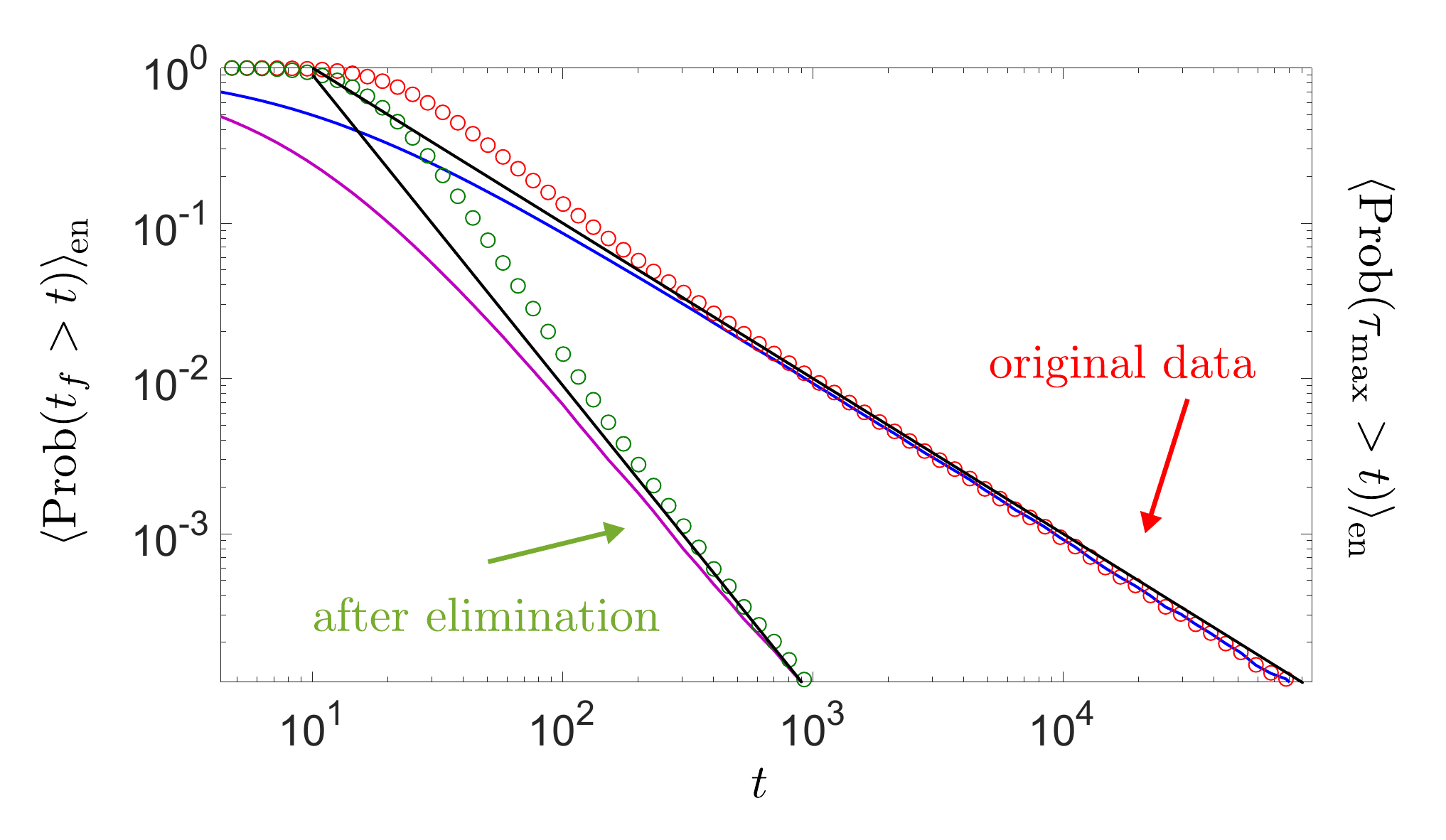}
\caption{The effect of elimination of the deepest trap on transport is studied for the quenched trap model. The plot shows the distributions of $t_f$ (upper red circles), $\tau_\text{max}$ (upper blue line), $t_r$ (lower red circles) and $\tau_\text{max}^\star$ (lower blue line) for the strongly biased quenched trap model with $L=10$. An average over the disorder was performed. We find perfect agreement with the theory, i.e. the principle of the single long transition time Eq.~(\ref{simple_qtm_uni}) (upper black line) and the relationship after elimination Eq.~(\ref{elim_qtm_uni}) (lower black line). We used the parameters $T=T_g=2$ such that $\alpha=1$, $t_0=1$. We used $10^6$ particles in the Monte-Carlo simulations.
\label{fig:QTM_average_Unidirect_CP_elim}}
\end{center}\end{figure}

In the laboratory and also theoretically, averaging over disorder has a profound effect in the sense of modifying statistical laws such as Eq.~(\ref{singe_simple}). As discussed in the review of Bouchaud and Georges \cite{bouchaud1990anomalous}, many channels of disorder may be present. Each particle then encounters a specific realization of disorder, but eventually, the measured quantity is an average. What will be the consequences for the single long transition time principle? \\ \\

The procedure of averaging over disorder of the energy landscape denoted $\langle \circ \rangle_\text{en}$, using Eq. (25), is found in SM Sec.~I. We find the single long transition time principle
\begin{equation}\boxed{\begin{split}\label{simple_qtm_uni}
\langle \text{Prob}(t_f>t)\rangle_\text{en} &\sim \langle \text{Prob}(\tau_\text{max}>t)\rangle_\text{en}\\
&\sim L \Gamma(1+\alpha)(t_0)^\alpha t^{-\alpha}
\end{split}}\end{equation}
with the exponent $\alpha=T/T_g$ valid for large $t$ (see also Fig.~\ref{fig:QTM_average_Unidirect_CP_elim}). This scaling behavior is the same as the large $t$ behavior of $L \langle \text{Prob}(\tau_x>t) \rangle_\text{en}$ where $\langle \text{Prob}(\tau_x>t) \rangle_\text{en}$ is the probability of the transition times after averaging over the disorder. This long transition time principle shows that the theory for the unidirectional model, continuous time random walk and quenched trap model after averaging over the disorder and $p=1$ are similar, i.e., compare Eqs.~(\ref{bjp_all}),(\ref{bjp_all_2}),(\ref{ctrw_bjp_random}) and (\ref{simple_qtm_uni}). The more profound issue is what is the effect of elimination? And what happens when $p\neq 1$, see Eq.~(\ref{bjp_qtm}) below.

\subsubsection{Weak bias with average over the disorder}\label{sec:weakbias}

We now consider the case $1/2<p< 1$, namely the bias is driving the system towards the absorbing boundary $x=L+1$. Unlike the case studied in previous subsection where $p=1$, now the particle can retract. Here, the number of visited traps $K$ is a random integer. Note that the problem of the number of distinct sites visited by a random walker has a long history \cite{dvoretzky1951some,vineyard1963number,biroli2022number}. We also define the total time spent in a trap, the occupation time $\hat{\tau}_x$. This observable is of interest, since if we can eliminate traps, possibly the deepest in our system, we are modifying not a single transition time, since the particle can revisit the trap several times before being absorbed.  The occupation time is $\hat{\tau}_x = \sum_{n_x=1}^{N_x}\tau_x^{(n_x)}$ where $N_x $ is the number of visits of the particle at trap $x$. Further, $N_x=0$ implies that the particle did not visit $x$ before being absorbed (note that $N_1,\ldots, N_L$ are necessarily not equal to zero while $N_{0},N_{-1},\ldots$ can be zero). For each of these $N_x$ visits, the transition time is drawn from the same distribution $\text{Prob}(\tau_x>t)=\text{exp}(-t/\bar{\tau}_x)$. We denote these transition times as $\tau_x^{(n_x)}$ with the visit number $n_x=1,\ldots,N_x$. \\ \\

\begin{figure}[H]\begin{center}
\includegraphics[width=1\columnwidth]{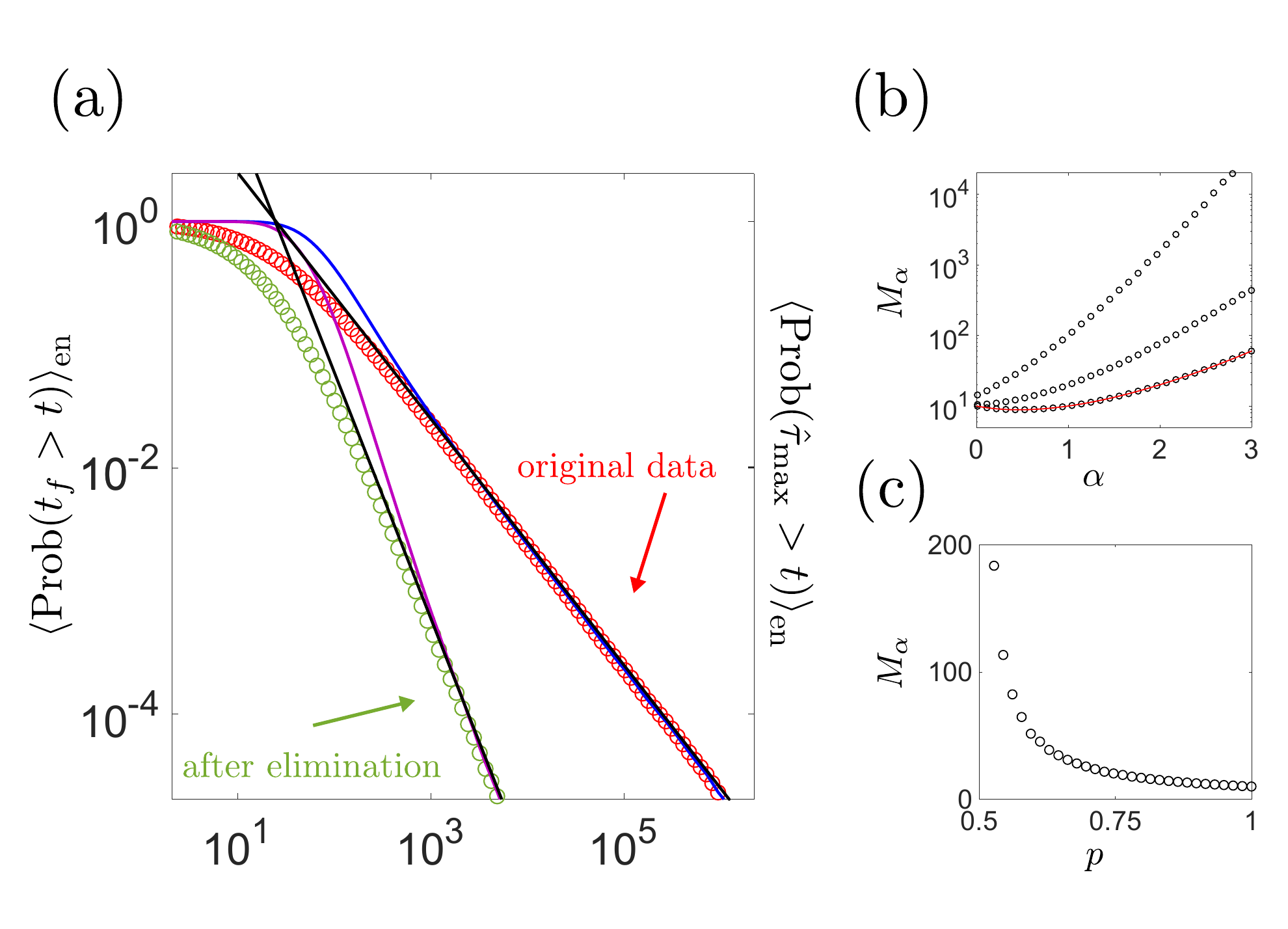}
\caption{(a) The distributions of $t_f$ (red circles), $\hat{\tau}_\text{max}$ (blue line), $t_r$ (green circles) and $\hat{\tau}_\text{max}^\star$ compared with the theories Eq.~(\ref{bjp_qtm}) (upper black line) and Eq.~(\ref{bjp_qtm2223}) (lower black line) for the weakly biased quenched trap model with $L=10$ and $p=0.7$. The functions $M_\alpha$ and $M_\alpha^\star$ were obtained from a discrete time and space random walk, as explained in the main text. We used the parameters $T=T_g=2$ so that $\alpha=1$, $t_0=1$ (as in Fig.~\ref{fig:QTM_One_Channel_Unidirect_CP} and \ref{fig:QTM_average_Unidirect_CP_elim}) and $10^6$ particles. (b) $M_\alpha$ of Eq.~(\ref{malpha}) versus $\alpha$ is plotted for $p=0.55$, $0.75$ and $1$ (three curves with black circles from top to bottom). When $p=1$, we obtain $M_\alpha=L\Gamma(1+\alpha)$ as in Eq.~(\ref{simple_qtm_uni}) which is also shown (red line). We used $L=10$ and $10^4$ particles were simulated to obtain $M_\alpha$. (c) $M_\alpha$ of Eq.~(\ref{malpha}) versus $p$ is plotted for $\alpha=1$ (black circles). We used the same parameters for the simulation as in (b).
\label{fig:QTM_average_rightbias_CP}}
\end{center}\end{figure}

The first passage time is a sum of occupation times $
t_f = \sum_{x=L-K+1}^L \hat{\tau}_x$. Similarly, the occupation time in the deepest visited trap $E_\text{max}=E_m$ with random $x=m$ is denoted $\hat{\tau}_\text{max} = \hat{\tau}_m$. The probability distributions of these two quantities are
\begin{equation}\begin{split}\label{twoeq}
\text{Prob}(t_f>t) &= \sum_{k=1}^\infty \sum_{n_x=0}^\infty \phi(n_x,k)\text{Prob}(t_f>t|n_x,k),\\
\text{Prob}(\hat{\tau}_\text{max}>t) &= \sum_{k=1}^\infty \sum_{n_x=0}^\infty \phi(n_x,k)  \text{Prob}(\hat{\tau}_\text{max}>t|n_x,k).
\end{split}\end{equation}
Here, $\phi(n_x,k)$ is the joint probability that a particle visited $K=k$ traps with $N_x=n_x$ visits at trap $x$. For example, if $p=1$ then clearly $\phi(n_x,k) = \delta_{n_x,1}\delta_{k,L}$ where we use the Kronecker delta. The conditional probabilities on the right-hand side of Eq.~(\ref{twoeq}) are conditioned on the number of visits per trap and of visited traps. The full analysis of Eq.~(\ref{twoeq}) and in particular the derivation of the asymptotic behaviors can be found in SM Sec.~I. For the average over the disorder, we obtain the principle of the long transition time
\begin{equation}\boxed{\begin{split}\label{bjp_qtm}
\langle \text{Prob}(t_f>t)\rangle_\text{en} &\sim \langle \text{Prob}(\hat{\tau}_\text{max}>t)\rangle_\text{en}\\
&\sim (t_0)^\alpha M_\alpha t^{-\alpha}
\end{split}}\end{equation}
with the function
\begin{equation}\begin{split}\label{malpha}
 M_\alpha &=\sum_{x=-\infty}^L\left\langle \frac{\Gamma(N_x+\alpha)}{\Gamma(N_x)} \right\rangle\\
 &=\sum_{x=-\infty}^L\sum_{k=1}^\infty \sum_{n_x=0}^\infty \phi(n_x,k) \frac{\Gamma(n_x+\alpha)}{\Gamma(n_x)}.
\end{split}\end{equation}
While the dimensionless function $M_\alpha$ is non-trivial we see that the long transition time principle holds, in general for the quenched trap model, the far tails of the distribution of the first passage time and the maximum are related. This holds true for any value of $\alpha$ whether one is in the glassy phase $\alpha<1$ or not $\alpha>1$. However, clearly the principle becomes meaningful in practice when $\alpha$ is not too large. In Fig.~\ref{fig:QTM_average_rightbias_CP}(a), we present the simulation for $\alpha=1$ which perfectly matches the long transition time principle Eq.~(\ref{bjp_qtm}). We now explain how to find $M_\alpha$. \\ \\ 

The function $M_\alpha$ can be obtained from the simulation of a discrete time and space random walk with the bias $p$. We generate numerically a trajectory of the discrete time and space random walk which starts at $x=1$ and is absorbed at $x=L+1$. With this trajectory, we count for every simulated trajectory, the number of visits $n_x$ at each lattice point $x<L+1$. Then averaging $\Gamma(n_x+\alpha)/\Gamma(n_x)$ as in Eq.~(\ref{malpha}) gives us $M_\alpha$. In Fig.~\ref{fig:QTM_average_rightbias_CP}(b), we plot $M_\alpha$ versus $\alpha$ and in Fig.~\ref{fig:QTM_average_rightbias_CP}(c), the same function versus $p$. Clearly, for $p\to 0.5$, the value of this parameter blows up, indicating the breakdown of the long transition time principle. The physical reason for this is that when $p=1/2$, the particle can explore in principle a very large number of traps as the motion becomes non-biased (somewhat similar, but far less trivial as the case found with $p=1/2$ for the continuous time random walk).\\ \\

\subsection{Elimination of the deepest trap}

We now study the effect of elimination of the maximum transition time on the statistics of the first passage time for the strongly biased model. The idea is to remove the deepest trap $\text{max}(E_1,\ldots,E_L)$ from the set of traps $\{E_1,\ldots,E_L\}$ and study the effect on the transport.  \\ \\

\subsubsection{Strong bias in a one-dimensional environment}

We apply the methods of order statistics, i.e. we order the traps according to $E_{(1)}<\ldots<E_{(L)}$ and remove $E_{(L)}=\text{max}(E_1,\ldots,E_L)$ from this set. Let $x[E_{(q)}]$ be the lattice point of the $q$-th deepest trap $E_{(q)}$ with $q=1,\ldots,L$. The first passage time and the long transition time both after elimination of $\tau_\text{max}=\tau_m$ (remember that $m$ is the location of the deepest trap $E_m=E_{(L)}$) are $t_r=\sum_{q=1}^{L-1} \tau_{x[E_{(q)}]}=t_f-\tau_{x[E_{(L)}]}$ and $\tau_\text{max}^\star=\tau_{x[E_{(L-1)}]}$. So clearly $\tau_\text{max}^\star$ is the time spent in the trap whose depth is ranked second in the sequence. The Laplace transform $\hat{p}_{t_r}(s)$ of the probability density function $p_{t_r}(t)=\langle \delta(t-[\tau_{x[E_{(1)}]}+\ldots + \tau_{x[E_{(L-1)}]}])\rangle$ and the probability of the maximum in the second deepest trap are
\begin{equation}\begin{split}\label{second_conv}
\hat{p}_{t_r}(s) &= \prod_{q=1}^{L-1} (1+\bar{\tau}_{x[E_{(q)}]} s)^{-1},\\
\text{Prob}(\tau_\text{max}^\star>t)&= \text{exp} \left(- \frac{t}{\bar{\tau}_{x[E_{(L-1)}]}}\right)
\end{split}\end{equation}
We can analyze Eq.~(\ref{second_conv}) for the one-dimensional random environment (i.e. one channel of energy traps) just as Eq.~(\ref{qtmmethod1}), thus, finding after the removal
\begin{equation}\boxed{\begin{split}\label{singe_simple_elim}
\text{Prob}(t_r > t) &\sim Q_{x[E_{(L-1)}]} \text{Prob}(\tau_\text{max}^\star>t) \\
&= Q_{x[E_{(L-1)}]} \text{exp} \left(- \frac{t}{\bar{\tau}_{x[E_{(L-1)}]}}\right)
\end{split}}\end{equation}
with the prefactor $Q_{x[E_{(L-1)}]}= (\bar{\tau}_{x[E_{(L-1)}]})^{L-1} \prod_{x=1,x\neq x[E_{(L-1)}]}^{L-1} [\bar{\tau}_{x[E_{(L-1)}]}-\bar{\tau}_x]^{-1}$. The exponential decay of both distributions is the same, namely the second deepest trap with the rate $-1/\bar{\tau}_{x[E_{(L-1)}]}$ takes control, which is faster than the decay $-1/\bar{\tau}_{x[E_{(L)}]}$ found previously without the elimination in Eq.~(\ref{singe_simple}) because $E_{(L-1)}<E_{(L)}$. Thus, removing the deepest trap yields a gain depending on the particular values of the energies $E_{(L-1)}$ and $E_{(L)}$. Recall that the Arrhenius times are related to the energies $\bar{\tau}_{x[E_{(L-1)}]}=t_0 \text{exp}(E_{(L-1)}/T)$ and $\bar{\tau}_{x[E_{(L)}]}=t_0 \text{exp}(E_{(L)}/T)$, so the times in Eq.~(\ref{singe_simple}) and (\ref{singe_simple_elim}) are mapped to the energies as usual. \\ \\

We now consider the two examples in Fig.~\ref{fig:QTM_One_Channel_Unidirect_CP}, where the energy landscape was generated with $\alpha=1$. We find before elimination $\text{Prob}(t_f>t) \propto \text{exp}(-t/201)$ and $\text{exp}(-t/84)$ while after elimination $\text{Prob}(t_r>t) \propto \text{exp}(-t/18) $ in the first example and $\propto \text{exp}(-t/11)$ in the second. The gain is clearly enormous, and if we would consider $\alpha<1$, we expect an even larger typical gain. However, obviously, since we did not average over disorder, this result is specific for a realization of disorder. To quantify the effect we consider below the ensemble averages. The measure of gain is $G=\bar{t}_r/\bar{t}_f$ with the averages $\bar{t}_f = \sum_{x=1}^L \bar{\tau}_x$ and $\bar{t}_r=\sum_{q=1}^{L-1} \bar{\tau}_{x[E_{(q)}]}$. For the two examples of Fig.~\ref{fig:QTM_One_Channel_Unidirect_CP}, we obtain $G=0.25$ and $0.46$. Note that while $G$ is a measure of gain based on the mean, the above discussion on the exponential decay focuses on large times. Both $G$ and the exponential tails show remarkable sensitivity after the removal. However, $G$ is roughly speaking a statistical measure of typical events, while the tails are naturally sensitive to the longest transition times.

\subsubsection{Strong bias with average over the disorder}
Averaging Eq.~(\ref{second_conv}) over the disorder yields
\begin{equation}\boxed{\begin{split}\label{elim_qtm_uni}
\langle \text{Prob}(t_r > t) \rangle_\text{en} &\sim \langle \text{Prob}(\tau_\text{max}^\star>t) \rangle_\text{en} \\
&\sim L(L-1) \Gamma(1+2\alpha) \frac{(t_0)^{2\alpha}}{2}t^{-2\alpha}.
\end{split}}
\end{equation}
See the SM Sec.~J for the full details of the calculation. Comparison with the long transition time principle Eq.~(\ref{simple_qtm_uni}) shows again the drastic improvement by our method. While the power law decay of Eq.~(\ref{simple_qtm_uni}) is $t^{-\alpha}$, in Eq.~(\ref{elim_qtm_uni}) it is doubled to $t^{-2\alpha}$. We previously found this doubling effect also for the unidirectional model and the continuous time random walk. \\ \\

The measure of gain is $G=\langle \bar{t}_r\rangle_\text{en}/\langle \bar{t}_f\rangle_\text{en}$ and we find exactly
\begin{equation}\begin{split}\label{g_qtm}
G =\begin{cases}
0 & \text{ for } 0<\alpha<1,\\
1-(-1)^L \frac{\alpha-1}{\alpha} (L-1)! \frac{\Gamma\left(-L+\frac{1}{\alpha}\right)}{\Gamma\left(\frac{1}{\alpha}\right)} & \text{ for } 1<\alpha,
\end{cases}
\end{split}\end{equation}
see SM Sec.~J. In Fig.~\ref{fig:g1_phase}, we plot $G$ versus $\alpha$ for $L=20$ and compare this analytical prediction with the simulation, showing excellent agreement without fitting. The behavior is the same for the unidirectional model Eq.~(\ref{gth}) with $L=N$ and Pareto distributed transition times. Thus, in the thermodynamic limit $L\to\infty$, i.e. removing the $s=f L$ deepest traps, we obtain Eq.~(\ref{impalg}).

\subsubsection{Weak bias with average over the disorder}
We consider the weak bias case of Sec.~\ref{sec:weakbias}, namely $1/2<p<1$. For the average over the disorder, we remove the deepest visited trap of each particle. The two probability distributions of the first passage time and of the maximum occupation time, both after the removal, behave as
\begin{equation}\boxed{\begin{split}\label{bjp_qtm2223}
\langle \text{Prob}(t_r>t)\rangle_\text{en} &\sim \langle \text{Prob}(\hat{\tau}_\text{max}^\star>t)\rangle_\text{en}\\
&\sim \frac{(t_0)^{2\alpha}}{2} M_\alpha^\star t^{-2\alpha}
\end{split}}\end{equation}
with the function
\begin{equation}\begin{split}\label{malpha2232}
 M_\alpha^\star &=\sum_{x=-\infty}^L\left\langle(K-1) \frac{\Gamma(N_x+2\alpha)}{\Gamma(N_x)} \right\rangle\\
 &=\sum_{x=-\infty}^L\sum_{k=1}^\infty \sum_{n_x=0}^\infty \phi(n_x,k) (k-1) \frac{\Gamma(n_x+2\alpha)}{\Gamma(n_x)},
\end{split}\end{equation}
see SM Sec.~J. Both probabilities are related and experience a doubling effect in the power law exponent; compare to Eq.~(\ref{bjp_qtm}). We obtain $M_\alpha^\star$ from a discrete space and discrete time random walk (similar to the method of finding $M_\alpha$).  For the case $p=1$, the function is $M_\alpha^\star=L(L-1)\Gamma(1+2\alpha)$ which gives Eq.~(\ref{elim_qtm_uni}). In Fig.~\ref{fig:QTM_average_rightbias_CP}.(a), the two distributions of $t_r$ and $\hat{\tau}_\text{max}^\star$ are demonstrated, and we find full accord between theory and simulation. \\ \\

The gain quantifier $G=\langle \bar{t}_r\rangle_\text{en}/\langle \bar{t}_f\rangle_\text{en}$ has the numerical value $0.151$ for the simulation of Fig.~\ref{fig:QTM_average_rightbias_CP}, showing once again that the elimination of the maximum contribution, here the deepest trap, yields a large gain stemming from doubly decreased power law decay of the probability (compared to the original statistics), especially for $\alpha=1$ as used in this example.

\section{Outlook for Practical Applications}\label{sec:outlook}

Before we summarize our results in Sec.~\ref{sec:discussion}, we briefly discuss potential implications of our findings for practical applications, focusing on the field of contaminant transport in porous media. We first note that current technology allows for the detailed tracking of single tracers in experimental systems; this is used extensively in single molecule tracking in the cell environment. Once the trajectories are analyzed, for example with video microscopy, the experimentalist can, in principle \cite{munoz2021objective}, pinpoint regions where the transport is particularly slow. \\ \\
\indent
As a second example, consider charge carriers in a wire of length $L$, where the disorder is large and hence conductance  is poor. Assume that along the wire we may add a bypass (a low resistivity segment) of length $\Delta L \ll L$. Further, assume one may place this segment anywhere along the wire. The basic question is whether or not the transport will be dramatically improved for a particular choice of location of the bypass. 
If so, one has a method of expediting transport and identifying bottlenecks and their statistical properties. The challenge for the theory would be to predict the magnitude of the effect. Note that in this paper, we investigate improvement of the first passage time and not of transport (we do not optimize the current). For the latter goal, one should consider the inverse of the time required for a particle to traverse the system, which in turn means that the measure studied here, Eq.~(\ref{gquant}), may require modification depending on the observable of interest. \\ \\
\indent
We note that in continuous time random walk theory, we need the full path to determine the maximum waiting time. In this model there is no quenched disorder, hence one cannot find a spatial and localized  bottleneck.  This means that for the continuous time random walk, as a mean field model, the improvement of transport is very different as compared to models of quenched disorder. One can say that the improvement of transport in the continuous time random walk is costly and in some sense more theoretical, because we need the full trajectory to determine the largest waiting time. However, the continuous time random walk offers deep insights, for example regarding the phase-like transition of $G$, which is important in our case. \\ \\
\indent
Additionally, the goal of expediting transport is not limited to tracers in disordered systems.  In the study of wave transmission in strongly scattered media, it was shown how, with clever interference, one may improve transmission \cite{vellekoop2008universal,aulbach2011control,mosk2012controlling} by controlling the many degrees of freedom in the incident waves. Clearly, the basic idea we use is vastly different, and a comparison is meaningless, but we mention this point because the goal of expediting transport is certainly not new.

\subsection{Spreading of Chemical Contamination}

In the studies of porous media in earth sciences and chemical engineering, our results might be generalized to continuum-level (effective medium) treatment of transport in large-scale, heterogeneous porous media, to address critical problems in groundwater quality remediation and management. 
In such cases, it is relevant to consider efforts to expedite chemical transport, to reduce residence times of chemical constituents in a system \cite{edery2014origins,bianchi2011spatial}. \\ \\
\indent
The foundation for a promising real-world technique
(of our theoretical removal approach) was laid in our
analysis of the pore-scale transport in a porous medium
(Sec. III). We demonstrated that removing even a small
portion of the longest transition times is sufficient to expedite the transport behavior significantly. This economic strategy provides the possibility of treating only a few critical regions in the system, namely those regions where
a sufficient number of very large maximum transition
times occur, which are mainly responsible for the slow
down of the process. In the context of our pore-scale transport simulation, we found — based on analysis of particle trajectories as shown in Fig.~\ref{fig1}(b) — that the locations of the longest transition times occur preferentially in a small number of regions within the flow domain. Of course, this depends on the coarse graining applied in the statistical analysis, and a separate, rigorous study of this coarse grain method and of the spatial correlations among the
maxima may prove fruitful. \\ \\
\indent
Our findings regarding tracer transport at the pore-scale, discussed in the preceding paragraph, can be transferred to continuum-level treatment of transport in heterogeneous porous systems. At this level, in strongly heterogeneous porous geological formations, preferential pathways transmit the bulk of the tracer mass \cite{edery2014origins,bianchi2011spatial}. Small numbers of locations along these pathways are characterized by very low permeability properties — corresponding to low velocities and long transitions times — compared to the entire system. In such cases, the longest transitions times will occur in this limited number of locations, suggesting that the overall chemical transport behavior could be expedited in practice by removing or otherwise clogging and bypassing these locations.

\section{Discussion}\label{sec:discussion}

For several frameworks of transport in disordered systems, we studied how the first passage time is strongly modified when we remove the maximum transition times from their associated trajectories. The study contains three parts: \textbf{A)} We established the principle of the single long transition time for different models. This principle states that the first passage time $t_f$, when it is long, is dominated by a single element being the maximum transition time $\tau_\text{max}$; this holds for some of the most well-studied models of transport in strongly disordered systems. For systems where this principle holds, the removal of largest transition times will clearly enhance the transport, which is what we studied next. \textbf{B)} We eliminated the maximum transition times from their trajectories, and found that the distribution of the modified first passage time $t_r=t_f-\tau_\text{max}$ decays much faster compared to the original first passage time $t_f$. \textbf{C)} This transport speed-up was further quantified with the measure of gain $G=\langle t_r \rangle /\langle t_f \rangle$. We explain our results for these three points now. \\ \\

\textbf{A)} The reference setting is unidirectional transport, for which the principle of the single long transition time is well-known \cite{chistyakov1964theorem}. We recapped it in Eq.~(\ref{bjp_all}) (see also Fig.~\ref{fig:main_results}(c)). The $t_f$ distribution (and the $\tau_\text{max}$ distribution) decays algebraically Eq.~(\ref{bjp_all_2}) where the power law exponent is the measure of the disorder $\alpha$ Eq.~(\ref{transitionprob}). Our goal was then to show that basic results hold more generally for well-known and applicable models of transport: the continuous time random walk and the quenched trap model averaged over the disorder. The main difference is that now the paths are no longer unidirectional and the number of jumps may fluctuate. Still, the principle holds: namely the first passage time distribution is the same as the distribution of the largest trapping time, and both decay as a power law with an exponent $\alpha$ (see Eq.~(\ref{transitionprob})). For the continuous time random walk, see Eq.~(\ref{ctrw_bjp_random}), (\ref{ctrwbjexample}) and Fig.~\ref{fig:ctrw_cp_right_bias}. For the quenched trap model, see 
Eq.~(\ref{simple_qtm_uni}), (\ref{bjp_qtm}) and Fig.~\ref{fig:QTM_average_Unidirect_CP_elim}, \ref{fig:QTM_average_rightbias_CP}.(a). Here, $t_f$ is related to the occupation time $\hat{\tau}_\text{max}$, which is the total time of multiple visits in the deepest trap.  \\ \\

We also discovered three vastly different situations, and so we encountered important modifications of the basic long transition time principle. 
(i) For the pore-scale transport simulation, the principle of the single long transition time is presented in Fig.~\ref{fig:alon_results}(a). We observed a striking matching between the distributions of $t_f$ and $\tau_\text{max}$ for large values, although both tails follow a complicated structure. This means that the matching between the distributions of the two observables does not have to follow a power law decay (as found in the other simpler models, i.e. unidirectional transport, continuous time random walk and averaged over disorder quenched model). This is important, because in physical realizations of disordered systems, mean field approaches (like the continuous time random walk) or average over disorder (like those carried out in the quenched model) are not always relevant. As a consequence, delineating a relation between the maximum transition time and the first passage time is non-trivial. (ii) For the continuous time random walk with no bias ($p=1/2$), we obtained Eq.~(\ref{stretched2}) (see also Fig.~\ref{fig:ctrw_cp_no_bias}(b)). Unlike the other cases, here the mean number of jumps before reaching the boundary diverges because the bias is zero. The $t_f$ distribution decay is $\alpha/2$ when $\alpha<1$, and $t_f$ is related to the rescaled maximum $\Delta_\alpha \tau_\text{max}$ (with $\Delta_\alpha > 1$ in Eq.~(\ref{stretched3})). This case is vastly different from other examples because $\Delta_\alpha$ is not equal unity; moreover, as $\Delta_\alpha$ diverges when $\alpha\to 1$. And finally if $\alpha>1$, the principle does not hold at all, showing a qualitative transition in the statistics when $\alpha=1$. (iii) The last example, is the quenched trap model in a one-dimensional random environment Eq.~(\ref{singe_simple}) under strong bias ($p=1$) and for a single realization of disorder. The principle follows an exponential decay with the rate depending solely on the deepest trap. Thus, different realizations of the environment lead to different decays (see Fig.~\ref{fig:QTM_One_Channel_Unidirect_CP}). Remarkably, the distributions of $t_f$ and $\tau_\text{max}$ are asymptotically the same; the distribution is thus by itself random in the sense that it depends on the disorder, although the principle of the long transition time holds. As mentioned previously, the decay of the quenched trap model distributions becomes algebraic when we average over the disorder. Nevertheless, the common trait of all these different examples is that a large value of $t_f$ is dominated by a single element of the trajectory, which is the maximum transition time. This motivated us to eliminate the latter.\\ \\

\textbf{B)} For the three models (unidirectional transport, continuous time random walk, and averaged over disorder quenched trap model) where the $t_f$ distribution decays algebraically with the measure of disorder $\alpha$, the distribution of $t_r$ decays twice as slowly, namely with exponent $2\alpha$. This implies a statistically significant shortened travel time, which was found for all these models; see the reference case of the unidirectional model Eq.~(\ref{cap4}), (\ref{cap}) and Fig.~\ref{fig:main_results}(c), the continuous time random walk Eq.~(\ref{secbjp_ctrw}) and Fig.~\ref{fig:ctrw_cp_right_bias}, and the quenched trap model averaged over the disorder Eq.~(\ref{elim_qtm_uni}), (\ref{bjp_qtm2223}) and  Fig.~\ref{fig:QTM_average_Unidirect_CP_elim}, \ref{fig:QTM_average_rightbias_CP}a. We further discovered that $t_r$ scales asymptotically as the second longest transition time $\tau_\text{max}^\star$, or for the quenched trap model, as the second longest occupation time $\hat{\tau}_\text{max}$. This asymptotic equivalence can be seen as a second level principle of the single long transition time (see Sec.~\ref{sec:elimscale}). The power law exponent switch from $\alpha$ to $2 \alpha$ means that while the mean first passage time diverges $\langle t_f \rangle$ in strongly disordered systems, when $\alpha<1$, after the removal we obtain a finite mean $\langle t_r \rangle$, when $\alpha>1/2$, indicating a large impact of the removal.\\ \\

For the pore-scale simulation, the distribution of $t_r$ presented in Fig.~\ref{fig:alon_results}(b) exhibits a dramatic speed-up compared to the $t_f$ statistics. Furthermore, the distribution matches that of $\tau_\text{max}^\star$ (the second longest transition time per trajectory) even with such a complicated pattern. Finally, the quenched trap model in a one-dimensional random environment under the influence of a strong bias is presented in Eq.~(\ref{singe_simple_elim}). After removal of the deepest trap, the statistics $t_r$ decay exponentially, with the rate depending on the second deepest trap. Thus, we found for all of these examples a strong shortening of the overall travel time in terms of the probability statistics, which implies a significant speed-up of the process.  \\ \\

\textbf{C)} The long transition time principles we found here offer a statement on the relation of the \textit{tails} of the distribution of the $t_f$ and $\tau_\text{max}$. We further quantified the effect using $G$, which is the ratio of the mean first passage times, i.e., $G=\langle t_r \rangle / \langle t_f \rangle$. The influence of the removal undergoes a transition at the critical point $\alpha=1$, see Fig.~\ref{fig:g1_phase}. The stronger the disorder in terms of $\alpha$, the stronger the gain. When $\alpha<1$, $G=0$ indicates the most radical speed-up. We derived $G$ rigorously for the unidirectional transport Eq.~(\ref{gth}), the continuous time random walk Eq.~(\ref{gctrw}), and the quenched trap model averaged over the disorder Eq.~(\ref{g_qtm}). For large systems $L$, the behaviors of $G$ are universal (Fig.~\ref{fig:g1_phase}). The thermodynamic limit ($L\to\infty$) is presented in Eq.~(\ref{impalg}).  \\ \\

We also studied the efficiency of the method and discussed different approaches of the removal of the maxima. For example, in the quenched trap model, we removed the deepest trap, namely we identified a specific location in space that, when eliminated, dramatically enhanced the transport. We also studied the option of removing only a fraction of longest sticking times, from a fraction of the trajectories (this will eventually reduce the resources needed to expedite transport). For example, for the pore-scale system, the removal of even a small portion $C$ of maximum transition times (namely the largest ones among all maxima) is sufficient to expedite the transport behavior significantly (see Fig.~\ref{fig:alon_results}(c)). The quantifier $G$ versus $C$ is presented in Fig.~\ref{fig:fig4}(a), again showing the efficiency of this low cost removal technique. Finally, we studied numerically the spatial locations that yield the  longest trapping times in the model of porous medium. In this system the disorder is quenched and strong, which allows for the spatial identification of bottlenecks, see Fig.~\ref{fig:fig78}. This in turn implies, mainly for future research, that in principle first passage times statistics could be modified   by local changes in the system.  \\ \\

In summary, we provided an extensive study on the fundamental change of the first passage time statistics
under the removal of the maximum transition times,
which demonstrates a drastic speed-up of the transport process; the first passage time distribution decays much faster, which reduces the transport dispersion significantly. Even the latest tracers leave the system rapidly, but also the average velocity is increased. We illustrated this behavior with the much shorter mean first passage time. Generally, the field of transport in disordered systems, and the transport settings investigated here, have a long history with numerous applications in diverse fields. Our results thus have the potential to open a field of actively expedited transport in disordered systems, with many applications (see Sec.~\ref{sec:outlook}).

\section*{Acknowledgement}
M.H. is funded by the Deutsche Forschungsgemeinschaft (DFG, German Research Foundation), Grant No. 436344834. The support of Israel Science Foundation Grant No. 1614/21 (E.B.) and ViTamins project funded by the Volkswagen Foundation grant AZ 9B192 (B.B.) is acknowledged. A.N. is supported by an ETH Zurich Postdoctoral Fellowship. B.B. holds the Sam Zuckerberg Professorial Chair in Hydrology. 

M.H. and A.N. contributed equally to this work.

\bibliographystyle{prestyle}

\bibliography{references} 

\begin{thebibliography}{10}

\bibitem{berkowitz2006modeling}
B.~Berkowitz, A.~Cortis, M.~Dentz, and H.~Scher, Rev. Geophys. \textbf{44}
  (2006).

\bibitem{berkowitz1997anomalous}
B.~Berkowitz and H.~Scher, Phys. Rev. Lett. \textbf{79}, 4038 (1997).

\bibitem{berkowitz1998theory}
B.~Berkowitz and H.~Scher, Phys. Rev. E \textbf{57}, 5858 (1998).

\bibitem{Nissan2019}
A.~Nissan and B.~Berkowitz, Phys. Rev. E \textbf{99}, 033108 (2019).

\bibitem{nissan2018inertial}
A.~Nissan and B.~Berkowitz, Phys. Rev. Lett. \textbf{120}, 054504 (2018).

\bibitem{dullien2012porous}
F.~A. Dullien, \emph{Porous {M}edia: {F}luid {T}ransport and {P}ore
  {S}tructure} (Academic Press, 2012).

\bibitem{bouchaud1990anomalous}
J.-P. Bouchaud and A.~Georges, Phys. Rep. \textbf{195}, 127 (1990).

\bibitem{monthus1996models}
C.~Monthus and J.-P. Bouchaud, J. Phys. A Math. Theor. \textbf{29}, 3847
  (1996).

\bibitem{berthier2011theoretical}
L.~Berthier and G.~Biroli, Rev. Mod. Phys. \textbf{83}, 587 (2011).

\bibitem{scher1975anomalous}
H.~Scher and E.~W. Montroll, Phys. Rev. B \textbf{12}, 2455 (1975).

\bibitem{montroll1973random}
E.~W. Montroll and H.~Scher, J. Stat. Phys. \textbf{9}, 101 (1973).

\bibitem{fox2021aging}
Z.~R. Fox, E.~Barkai, and D.~Krapf, Nat. Comm. \textbf{12}, 6162 (2021).

\bibitem{weigel2011ergodic}
A.~V. Weigel, B.~Simon, M.~M. Tamkun, and D.~Krapf, Proc. Natl. Acad. Sci.
  \textbf{108}, 6438 (2011).

\bibitem{klages2008anomalous}
R.~Klages, G.~Radons, and I.~M. Sokolov, \emph{Anomalous {T}ransport} (Wiley
  Online Library, Hoboken, New Jersey, 2008).

\bibitem{metzler2000random}
R.~Metzler and J.~Klafter, Phys. Rep. \textbf{339}, 1 (2000).

\bibitem{metzler2014anomalous}
R.~Metzler, J.-H. Jeon, A.~G. Cherstvy, and E.~Barkai, Phys. Chem. Chem. Phys.
  \textbf{16}, 24128 (2014).

\bibitem{hofling2013anomalous}
F.~H{\"o}fling and T.~Franosch, Rep. Prog. Phys. \textbf{76}, 046602 (2013).

\bibitem{chistyakov1964theorem}
V.~P. Chistyakov, Theory Probab. Its Appl. \textbf{9}, 640 (1964).

\bibitem{derrida1994non}
B.~Derrida, On Three Levels: Micro-, Meso-, and Macro-Approaches in Physics
  125--137 (1994).

\bibitem{filiasi2013condensation}
M.~Filiasi, E.~Zarinelli, E.~Vesselli, and M.~Marsili, arXiv:1309.7795  (2013).

\bibitem{vezzani2019single}
A.~Vezzani, E.~Barkai, and R.~Burioni, Phys. Rev. E \textbf{100}, 012108
  (2019).

\bibitem{majumdar2010universal}
S.~N. Majumdar, Physica A \textbf{389}, 4299 (2010).

\bibitem{bel2006random}
G.~Bel and E.~Barkai, Phys. Rev. E \textbf{73}, 016125 (2006).

\bibitem{redner2001guide}
S.~Redner, \emph{A {G}uide to {F}irst-{P}assage {P}rocesses} (Cambridge
  university press, 2001).

\bibitem{rangarajan2000anomalous}
G.~Rangarajan and M.~Ding, Phys. Rev. E \textbf{62}, 120 (2000).

\bibitem{condamin2007first}
S.~Condamin, O.~B{\'e}nichou, and J.~Klafter, Phys. Rev. Lett. \textbf{98},
  250602 (2007).

\bibitem{condamin2007first2}
S.~Condamin, O.~B{\'e}nichou, V.~Tejedor, R.~Voituriez, and J.~Klafter, Nature
  \textbf{450}, 77 (2007).

\bibitem{montroll1965random}
E.~W. Montroll and G.~H. Weiss, J. Math. Phys. \textbf{6}, 167 (1965).

\bibitem{kutner2017continuous}
R.~Kutner and J.~Masoliver, Eur. Phys. J. B \textbf{90}, 1 (2017).

\bibitem{barkai2000continuous}
E.~Barkai, R.~Metzler, and J.~Klafter, Phys. Rev. E \textbf{61}, 132 (2000).

\bibitem{margolin2004continuous}
G.~Margolin and B.~Berkowitz, Phys. A: Stat. Mech. Appl. \textbf{334}, 46
  (2004).

\bibitem{dentz2004time}
M.~Dentz, A.~Cortis, H.~Scher, and B.~Berkowitz, Adv. Water Resour.
  \textbf{27}, 155 (2004).

\bibitem{dentz2005exact}
M.~Dentz and B.~Berkowitz, Phys. Rev. E \textbf{72}, 031110 (2005).

\bibitem{dentz2008transport}
M.~Dentz, H.~Scher, D.~Holder, and B.~Berkowitz, Phys. Rev. E \textbf{78},
  041110 (2008).

\bibitem{cairoli2015anomalous}
A.~Cairoli and A.~Baule, Phys. Rev. Lett. \textbf{115}, 110601 (2015).

\bibitem{burioni2014scaling}
R.~Burioni, G.~Gradenigo, A.~Sarracino, A.~Vezzani, and A.~Vulpiani, Commun.
  Theor. Phys. \textbf{62}, 514 (2014).

\bibitem{scalas2006application}
E.~Scalas, Phys. A: Stat. Mech. Appl. \textbf{362}, 225 (2006).

\bibitem{weeks1998anomalous}
E.~R. Weeks and H.~L. Swinney, Phys. Rev. E \textbf{57}, 4915 (1998).

\bibitem{weeks1996anomalous}
E.~R. Weeks, J.~Urbach, and H.~L. Swinney, Physica D \textbf{97}, 291 (1996).

\bibitem{albers2013subdiffusive}
T.~Albers and G.~Radons, EPL \textbf{102}, 40006 (2013).

\bibitem{burov2011time}
S.~Burov and E.~Barkai, Phys. Rev. Lett. \textbf{106}, 140602 (2011).

\bibitem{akimoto2016universal}
T.~Akimoto, E.~Barkai, and K.~Saito, Phys. Rev. Lett. \textbf{117}, 180602
  (2016).

\bibitem{burov2007occupation}
S.~Burov and E.~Barkai, Phys. Rev. Lett. \textbf{98}, 250601 (2007).

\bibitem{bertin2003subdiffusion}
E.~Bertin and J.-P. Bouchaud, Phys. Rev. E \textbf{67}, 026128 (2003).

\bibitem{akimoto2018non}
T.~Akimoto, E.~Barkai, and K.~Saito, Phys. Rev. E \textbf{97}, 052143 (2018).

\bibitem{burov2012weak}
S.~Burov and E.~Barkai, Phys. Rev. E \textbf{86}, 041137 (2012).

\bibitem{embrechts2013modelling}
P.~Embrechts, C.~Kl{\"u}ppelberg, and T.~Mikosch, \emph{Modelling extremal
  events: for insurance and finance} (Springer Science \& Business Media,
  2013).

\bibitem{embrechts1982estimates}
P.~Embrechts and N.~Veraverbeke, Insur. Math. Econ. \textbf{1}, 55 (1982).

\bibitem{rolski2009stochastic}
T.~Rolski, H.~Schmidli, V.~Schmidt, and J.~L. Teugels, \emph{Stochastic
  {P}rocesses for {I}nsurance and {F}inance} (John Wiley \& Sons, 2009).

\bibitem{kyprianou2006introductory}
A.~E. Kyprianou, \emph{Introductory {L}ectures on {F}luctuations of {L}{\'e}vy
  {P}rocesses with {A}pplications} (Springer Science \& Business Media, 2006).

\bibitem{kluppelberg1997large}
C.~Kl{\"u}ppelberg and T.~Mikosch, J. App. Prob. \textbf{34}, 293 (1997).

\bibitem{wang2019transport}
W.~Wang, A.~Vezzani, R.~Burioni, and E.~Barkai, Phys. Rev. Res. \textbf{1},
  033172 (2019).

\bibitem{wang2020large}
W.~Wang, M.~H{\"o}ll, and E.~Barkai, Phys. Rev. E \textbf{102}, 052115 (2020).

\bibitem{holl2021big}
M.~H{\"o}ll and E.~Barkai, Eur. Phys. J. B \textbf{94}, 1 (2021).

\bibitem{yin2023restart}
R.~Yin and E.~Barkai, Phys. Rev. Lett. \textbf{130}, 050802 (2023).

\bibitem{evans2011diffusion}
M.~R. Evans and S.~N. Majumdar, Physical review letters \textbf{106}, 160601
  (2011).

\bibitem{evans2011diffusion2}
M.~R. Evans and S.~N. Majumdar, Journal of Physics A: Mathematical and
  Theoretical \textbf{44}, 435001 (2011).

\bibitem{evans2020stochastic}
M.~R. Evans, S.~N. Majumdar, and G.~Schehr, Journal of Physics A: Mathematical
  and Theoretical \textbf{53}, 193001 (2020).

\bibitem{chechkin2018random}
A.~Chechkin and I.~Sokolov, Phys. Rev. Lett. \textbf{121}, 050601 (2018).

\bibitem{campos2015phase}
D.~Campos and V.~M{\'e}ndez, Phys. Rev. E \textbf{92}, 062115 (2015).

\bibitem{besga2020optimal}
B.~Besga, A.~Bovon, A.~Petrosyan, S.~N. Majumdar, and S.~Ciliberto, Phys. Rev.
  Res. \textbf{2}, 032029 (2020).

\bibitem{tal2020experimental}
O.~Tal-Friedman, A.~Pal, A.~Sekhon, S.~Reuveni, and Y.~Roichman, J. Phys. Chem.
  Lett. \textbf{11}, 7350 (2020).

\bibitem{reuveni2014role}
S.~Reuveni, M.~Urbakh, and J.~Klafter, Proceedings of the National Academy of
  Sciences \textbf{111}, 4391 (2014).

\bibitem{budnar2019anillin}
S.~Budnar, K.~B. Husain, G.~A. Gomez, M.~Naghibosadat, A.~Varma, S.~Verma,
  N.~A. Hamilton, R.~G. Morris, and A.~S. Yap, Developmental cell \textbf{49},
  894 (2019).

\bibitem{bel2009simplicity}
G.~Bel, B.~Munsky, and I.~Nemenman, Phys. Biol. \textbf{7}, 016003 (2009).

\bibitem{hamlin2019geometry}
P.~Hamlin, W.~J. Thrasher, W.~Keyrouz, and M.~Mascagni, Monte Carlo Methods and
  Applications \textbf{25}, 329 (2019).

\bibitem{pal2017first}
A.~Pal and S.~Reuveni, Phys. Rev. Lett. \textbf{118}, 030603 (2017).

\bibitem{bouchaud1992weak}
J.-P. Bouchaud, J. phys., I \textbf{2}, 1705 (1992).

\bibitem{BERKOWITZ2009}
B.~Berkowitz and H.~Scher, Adv. Water Resour. \textbf{32}, 750 (2009).

\bibitem{wong2004anomalous}
I.~Wong, M.~Gardel, D.~Reichman, E.~R. Weeks, M.~Valentine, A.~Bausch, and
  D.~A. Weitz, Phys. Rev. Lett. \textbf{92}, 178101 (2004).

\bibitem{levin2021measurements}
M.~Levin, G.~Bel, and Y.~Roichman, J. Chem. Phys. \textbf{154}, 144901 (2021).

\bibitem{stefani2009beyond}
F.~D. Stefani, J.~P. Hoogenboom, and E.~Barkai, Phys. Today \textbf{62}, 34
  (2009).

\bibitem{solomon1993observation}
T.~Solomon, E.~R. Weeks, and H.~L. Swinney, Phys. Rev. Lett. \textbf{71}, 3975
  (1993).

\bibitem{vilk2021ergodicity}
O.~Vilk, Y.~Orchan, M.~Charter, N.~Ganot, S.~Toledo, R.~Nathan, and M.~Assaf,
  Phys. Rev. X \textbf{12}, 031005 (2022).

\bibitem{majumdar2020extreme}
S.~N. Majumdar, A.~Pal, and G.~Schehr, Phys. Rep. \textbf{840}, 1 (2020).

\bibitem{Fowler2016}
A.~C. Fowler and B.~Scheu, Proc. Math. Phys. Eng. Sci. \textbf{472}, 20150843
  (2016).

\bibitem{Bijeljic2011}
B.~Bijeljic, P.~Mostaghimi, and M.~J. Blunt, Phys. Rev. Lett. \textbf{107},
  204502 (2011).

\bibitem{PereiraNunes2015}
J.~P. Pereira~Nunes, B.~Bijeljic, and M.~J. Blunt, Transp. Porous Media
  \textbf{109}, 317 (2015).

\bibitem{bel2005occupation}
G.~Bel and E.~Barkai, J. Condens. Matter Phys. \textbf{17}, S4287 (2005).

\bibitem{balakrishnan1983first}
V.~Balakrishnan and M.~Khantha, Pramana \textbf{21}, 187 (1983).

\bibitem{krusemann2015ageing}
H.~Kr{\"u}semann, A.~Godec, and R.~Metzler, J. Phys. A: Math. Theor.
  \textbf{48}, 285001 (2015).

\bibitem{krusemann2014first}
H.~Kr{\"u}semann, A.~Godec, and R.~Metzler, Phys. Rev. E \textbf{89}, 040101
  (2014).

\bibitem{jose2021passage}
S.~Jose, J. Stat. Mech. 113208 (2022).

\bibitem{barkai1998generalized}
E.~Barkai and V.~Fleurov, Phys. Rev. E \textbf{58}, 1296 (1998).

\bibitem{fogedby1994langevin}
H.~C. Fogedby, Phys. Rev. E \textbf{50}, 1657 (1994).

\bibitem{klafter2011first}
J.~Klafter and I.~M. Sokolov, \emph{First {S}teps in {R}andom {W}alks: {F}rom
  {T}ools to {A}pplications} (Oxford University Press, 2011).

\bibitem{monthus2003anomalous}
C.~Monthus, Phys. Rev. E \textbf{68}, 036114 (2003).

\bibitem{fedotov2019asymptotic}
S.~Fedotov and D.~Han, Phys. Rev. Lett. \textbf{123}, 050602 (2019).

\bibitem{fedotov2021variable}
S.~Fedotov, D.~Han, A.~Y. Zubarev, M.~Johnston, and V.~J. Allan, Philos. Trans.
  R. Soc. A \textbf{379}, 20200317 (2021).

\bibitem{hanggi1990reaction}
P.~H{\"a}nggi, P.~Talkner, and M.~Borkovec, Rev. Mod. Phys. \textbf{62}, 251
  (1990).

\bibitem{dvoretzky1951some}
A.~Dvoretzky and P.~Erd{\"o}s, In \emph{Proceedings of the Second Berkeley
  Symposium on Mathematical Statistics and Probability}, vol. 1950, 353--367
  (University of California Press Berkeley and Los Angeles, 1951).

\bibitem{vineyard1963number}
G.~H. Vineyard, J. Math. Phys. \textbf{4}, 1191 (1963).

\bibitem{biroli2022number}
M.~Biroli, F.~Mori, and S.~N. Majumdar, Number of distinct sites visited by a
  resetting random walker (2022).

\bibitem{munoz2021objective}
G.~Mu{\~n}oz-Gil, G.~Volpe, M.~A. Garcia-March, E.~Aghion, A.~Argun, C.~B.
  Hong, T.~Bland, S.~Bo, J.~A. Conejero, N.~Firbas, et~al., Nat. Commun.
  \textbf{12}, 1 (2021).

\bibitem{vellekoop2008universal}
I.~M. Vellekoop and A.~Mosk, Phys. Rev. Lett. \textbf{101}, 120601 (2008).

\bibitem{aulbach2011control}
J.~Aulbach, B.~Gjonaj, P.~M. Johnson, A.~P. Mosk, and A.~Lagendijk, Phys. Rev.
  Lett. \textbf{106}, 103901 (2011).

\bibitem{mosk2012controlling}
A.~P. Mosk, A.~Lagendijk, G.~Lerosey, and M.~Fink, Nat. Photon \textbf{6}, 283
  (2012).

\bibitem{edery2014origins}
Y.~Edery, A.~Guadagnini, H.~Scher, and B.~Berkowitz, Water Resour. Res.
  \textbf{50}, 1490 (2014).

\bibitem{bianchi2011spatial}
M.~Bianchi, C.~Zheng, C.~Wilson, G.~R. Tick, G.~Liu, and S.~M. Gorelick, Water
  Resour. Res. \textbf{47} (2011).

\end{thebibliography}


\begin{thebibliography}{8}%
\makeatletter
\providecommand \@ifxundefined [1]{%
 \@ifx{#1\undefined}
}%
\providecommand \@ifnum [1]{%
 \ifnum #1\expandafter \@firstoftwo
 \else \expandafter \@secondoftwo
 \fi
}%
\providecommand \@ifx [1]{%
 \ifx #1\expandafter \@firstoftwo
 \else \expandafter \@secondoftwo
 \fi
}%
\providecommand \natexlab [1]{#1}%
\providecommand \enquote  [1]{``#1''}%
\providecommand \bibnamefont  [1]{#1}%
\providecommand \bibfnamefont [1]{#1}%
\providecommand \citenamefont [1]{#1}%
\providecommand \href@noop [0]{\@secondoftwo}%
\providecommand \href [0]{\begingroup \@sanitize@url \@href}%
\providecommand \@href[1]{\@@startlink{#1}\@@href}%
\providecommand \@@href[1]{\endgroup#1\@@endlink}%
\providecommand \@sanitize@url [0]{\catcode `\\12\catcode `\$12\catcode
  `\&12\catcode `\#12\catcode `\^12\catcode `\_12\catcode `\%12\relax}%
\providecommand \@@startlink[1]{}%
\providecommand \@@endlink[0]{}%
\providecommand \url  [0]{\begingroup\@sanitize@url \@url }%
\providecommand \@url [1]{\endgroup\@href {#1}{\urlprefix }}%
\providecommand \urlprefix  [0]{URL }%
\providecommand \Eprint [0]{\href }%
\providecommand \doibase [0]{https://doi.org/}%
\providecommand \selectlanguage [0]{\@gobble}%
\providecommand \bibinfo  [0]{\@secondoftwo}%
\providecommand \bibfield  [0]{\@secondoftwo}%
\providecommand \translation [1]{[#1]}%
\providecommand \BibitemOpen [0]{}%
\providecommand \bibitemStop [0]{}%
\providecommand \bibitemNoStop [0]{.\EOS\space}%
\providecommand \EOS [0]{\spacefactor3000\relax}%
\providecommand \BibitemShut  [1]{\csname bibitem#1\endcsname}%
\let\auto@bib@innerbib\@empty
\bibitem [{\citenamefont {Majumdar}\ \emph {et~al.}(2020)\citenamefont
  {Majumdar}, \citenamefont {Pal},\ and\ \citenamefont
  {Schehr}}]{majumdar2020extreme}%
  \BibitemOpen
  \bibfield  {author} {\bibinfo {author} {\bibfnamefont {S.~N.}\ \bibnamefont
  {Majumdar}}, \bibinfo {author} {\bibfnamefont {A.}~\bibnamefont {Pal}},\ and\
  \bibinfo {author} {\bibfnamefont {G.}~\bibnamefont {Schehr}},\ }\bibfield
  {title} {\emph {\bibinfo {title} {Extreme value statistics of correlated
  random variables: {A} pedagogical review}},\ }\href@noop {} {\bibfield
  {journal} {\bibinfo  {journal} {Phys. Rep.}\ }\textbf {\bibinfo {volume}
  {840}},\ \bibinfo {pages} {1} (\bibinfo {year} {2020})}\BibitemShut {NoStop}%
\bibitem [{\citenamefont {R{\'e}nyi}(1953)}]{renyi1953theory}%
  \BibitemOpen
  \bibfield  {author} {\bibinfo {author} {\bibfnamefont {A.}~\bibnamefont
  {R{\'e}nyi}},\ }\bibfield  {title} {\emph {\bibinfo {title} {On the theory of
  order statistics}},\ }\href@noop {} {\bibfield  {journal} {\bibinfo
  {journal} {Acta Math. Hungarica}\ }\textbf {\bibinfo {volume} {4}} (\bibinfo
  {year} {1953})}\BibitemShut {NoStop}%
\bibitem [{\citenamefont {Bijeljic}\ \emph {et~al.}(2011)\citenamefont
  {Bijeljic}, \citenamefont {Mostaghimi},\ and\ \citenamefont
  {Blunt}}]{Bijeljic2011}%
  \BibitemOpen
  \bibfield  {author} {\bibinfo {author} {\bibfnamefont {B.}~\bibnamefont
  {Bijeljic}}, \bibinfo {author} {\bibfnamefont {P.}~\bibnamefont
  {Mostaghimi}},\ and\ \bibinfo {author} {\bibfnamefont {M.~J.}\ \bibnamefont
  {Blunt}},\ }\bibfield  {title} {\emph {\bibinfo {title} {Signature of
  {N}on-{F}ickian {S}olute {T}ransport in {C}omplex {H}eterogeneous {P}orous
  {M}edia}},\ }\href {https://doi.org/10.1103/PhysRevLett.107.204502}
  {\bibfield  {journal} {\bibinfo  {journal} {Phys. Rev. Lett.}\ }\textbf
  {\bibinfo {volume} {107}},\ \bibinfo {pages} {204502} (\bibinfo {year}
  {2011})}\BibitemShut {NoStop}%
\bibitem [{\citenamefont {Morales}\ \emph {et~al.}(2017)\citenamefont
  {Morales}, \citenamefont {Dentz}, \citenamefont {Willmann},\ and\
  \citenamefont {Holzner}}]{Morales2017}%
  \BibitemOpen
  \bibfield  {author} {\bibinfo {author} {\bibfnamefont {V.~L.}\ \bibnamefont
  {Morales}}, \bibinfo {author} {\bibfnamefont {M.}~\bibnamefont {Dentz}},
  \bibinfo {author} {\bibfnamefont {M.}~\bibnamefont {Willmann}},\ and\
  \bibinfo {author} {\bibfnamefont {M.}~\bibnamefont {Holzner}},\ }\bibfield
  {title} {\emph {\bibinfo {title} {Stochastic dynamics of intermittent
  pore-scale particle motion in three-dimensional porous media: {E}xperiments
  and theory}},\ }\href@noop {} {\bibfield  {journal} {\bibinfo  {journal}
  {Geophys. Res. Lett.}\ }\textbf {\bibinfo {volume} {44}},\ \bibinfo {pages}
  {9361} (\bibinfo {year} {2017})}\BibitemShut {NoStop}%
\bibitem [{\citenamefont {Metzler}\ and\ \citenamefont
  {Klafter}(2000)}]{metzler2000random}%
  \BibitemOpen
  \bibfield  {author} {\bibinfo {author} {\bibfnamefont {R.}~\bibnamefont
  {Metzler}}\ and\ \bibinfo {author} {\bibfnamefont {J.}~\bibnamefont
  {Klafter}},\ }\bibfield  {title} {\emph {\bibinfo {title} {The random walk's
  guide to anomalous diffusion: a fractional dynamics approach}},\ }\href@noop
  {} {\bibfield  {journal} {\bibinfo  {journal} {Phys. Rep.}\ }\textbf
  {\bibinfo {volume} {339}},\ \bibinfo {pages} {1} (\bibinfo {year}
  {2000})}\BibitemShut {NoStop}%
\bibitem [{\citenamefont {Klafter}\ and\ \citenamefont
  {Sokolov}(2011)}]{klafter2011first}%
  \BibitemOpen
  \bibfield  {author} {\bibinfo {author} {\bibfnamefont {J.}~\bibnamefont
  {Klafter}}\ and\ \bibinfo {author} {\bibfnamefont {I.~M.}\ \bibnamefont
  {Sokolov}},\ }\href@noop {} {\emph {\bibinfo {title} {First {S}teps in
  {R}andom {W}alks: {F}rom {T}ools to {A}pplications}}}\ (\bibinfo  {publisher}
  {Oxford University Press},\ \bibinfo {year} {2011})\BibitemShut {NoStop}%
\bibitem [{\citenamefont {Redner}(2001)}]{redner2001guide}%
  \BibitemOpen
  \bibfield  {author} {\bibinfo {author} {\bibfnamefont {S.}~\bibnamefont
  {Redner}},\ }\href@noop {} {\emph {\bibinfo {title} {A {G}uide to
  {F}irst-{P}assage {P}rocesses}}}\ (\bibinfo  {publisher} {Cambridge
  university press},\ \bibinfo {year} {2001})\BibitemShut {NoStop}%
\bibitem [{\citenamefont {Bel}\ and\ \citenamefont
  {Barkai}(2006)}]{bel2006random}%
  \BibitemOpen
  \bibfield  {author} {\bibinfo {author} {\bibfnamefont {G.}~\bibnamefont
  {Bel}}\ and\ \bibinfo {author} {\bibfnamefont {E.}~\bibnamefont {Barkai}},\
  }\bibfield  {title} {\emph {\bibinfo {title} {Random walk to a nonergodic
  equilibrium concept}},\ }\href@noop {} {\bibfield  {journal} {\bibinfo
  {journal} {Phys. Rev. E}\ }\textbf {\bibinfo {volume} {73}},\ \bibinfo
  {pages} {016125} (\bibinfo {year} {2006})}\BibitemShut {NoStop}%
\end{thebibliography}%

\end{document}


\title{Supplemental Material}

\author{Marc H\"oll}
\affiliation{Department of Physics, Institute of Nanotechnology and Advanced Materials, Bar-Ilan University, Ramat Gan, 52900, Israel}
\author{Alon Nissan}
\affiliation{Institute of Environmental Engineering, ETH Zurich, Zurich, Switzerland}
\author{Brian Berkowitz}
\affiliation{Department of Earth and Planetary Sciences, Weizmann Institute of Science, Rehovot, 7610001, Israel}
\author{Eli Barkai}
\affiliation{Department of Physics, Institute of Nanotechnology and Advanced Materials, Bar-Ilan University, Ramat Gan, 52900, Israel}

\begin{abstract}
We derive the results from the manuscript in this Supplemental Material (SM); these include rigorous calculations for the unidirectional transport, the continuous time random walk and the quenched trap model. For the pore-scale transport, we explain the technique used to obtain the transition times from each trajectory. 
\end{abstract}

\maketitle

\onecolumngrid
\begin{appendix}

\counterwithin{figure}{section}
\setcounter{equation}{0}
\numberwithin{equation}{section}

\section{Introduction}
Here, we summarize the content of the SM. For the unidirectional transport, we calculate the principle of the single long transition time in SM Sec.~\ref{sec:appfractal}, and the measure of gain $G$ in SM Sec.~\ref{sec:appimpr}. Furthermore, SM Sec.~\ref{sec:ctrw_elim} considers the relationship between the modified first passage time and the transition times upon removal, which is not discussed in the main text but necessary for a more complete picture of the removal method. \\ \\

SM Sec.~\ref{sec:porescaleapp} discusses the method applied in the pore-scale transport simulation to obtain the transition times from the trajectories. \\ \\

For the continuous time random walk, we provide the principle of the single, the principle of the single long transition time is derived in SM Sec.~\ref{sec:ctrw_app} where we also present boundary conditions not discussed in the main text. Also, not in the main text, but an important class, is the case of continuous jumps in SM Sec.~\ref{sec:cont}. The measure $G$ can be found in SM Sec.~\ref{sec:cont2}, and the relationship between the modified first passage time and the transition times upon removal in SM Sec.~\ref{sec:ctrw_elim} (not in the main text) \\ \\

For the quenched trap model, the principle of the single long transition time is presented in SM Sec.~\ref{sec:qtm_app} where we additionally discuss deterministic transition times (not shown in the main text) in contrast to exponential transition times. The elimination of the deepest trap and the relevant statistics are calculated in SM Sec.~\ref{sec:app_removal_qtm}.

\section{Elimination of the longest transition times in the unidirectional transport model}\label{sec:appfractal}

For the unidirectional transport of Sec.~II, we derive the asymptotic behavior of the modified first passage time after removal of the maximum transition time, and its relationship to the second longest transition time, which is presented in the main text in Eq.~(9), (11), (13), and (15). We remind that a particle starts at $x=1$ and moves to the right until $x=L+1$. At each lattice point $x \in \{1,\ldots,L\}$, a particle takes a random transition time $\tau_n$ ($n$ is the step number $n=x$) before jumping to the right neighbour. The $N=L$ transition times are independent and identically distributed following the probability density function
\begin{equation}\label{psit}
\psi(t) \sim A t^{-(1+\alpha)}
\end{equation}
with the scaling exponent $\alpha>0$ and the amplitude $A$. This function is defined as $\psi(t)= - \mathrm{d}/\mathrm{d}t \text{Prob}(\tau_n>t)$. In the following, we will usually use the notation $p_{\tau_n}(t)=\psi(t)$ for the probability density function, also for other random variables. In the index, we write the random variable and $t$ is the value. Further, we will also use the cumulative distribution function $P_{\tau_n}(t)=1-\text{Prob}(\tau_n>t)$.\\ \\

\begin{figure}\begin{center}
\includegraphics[width=0.6\columnwidth]{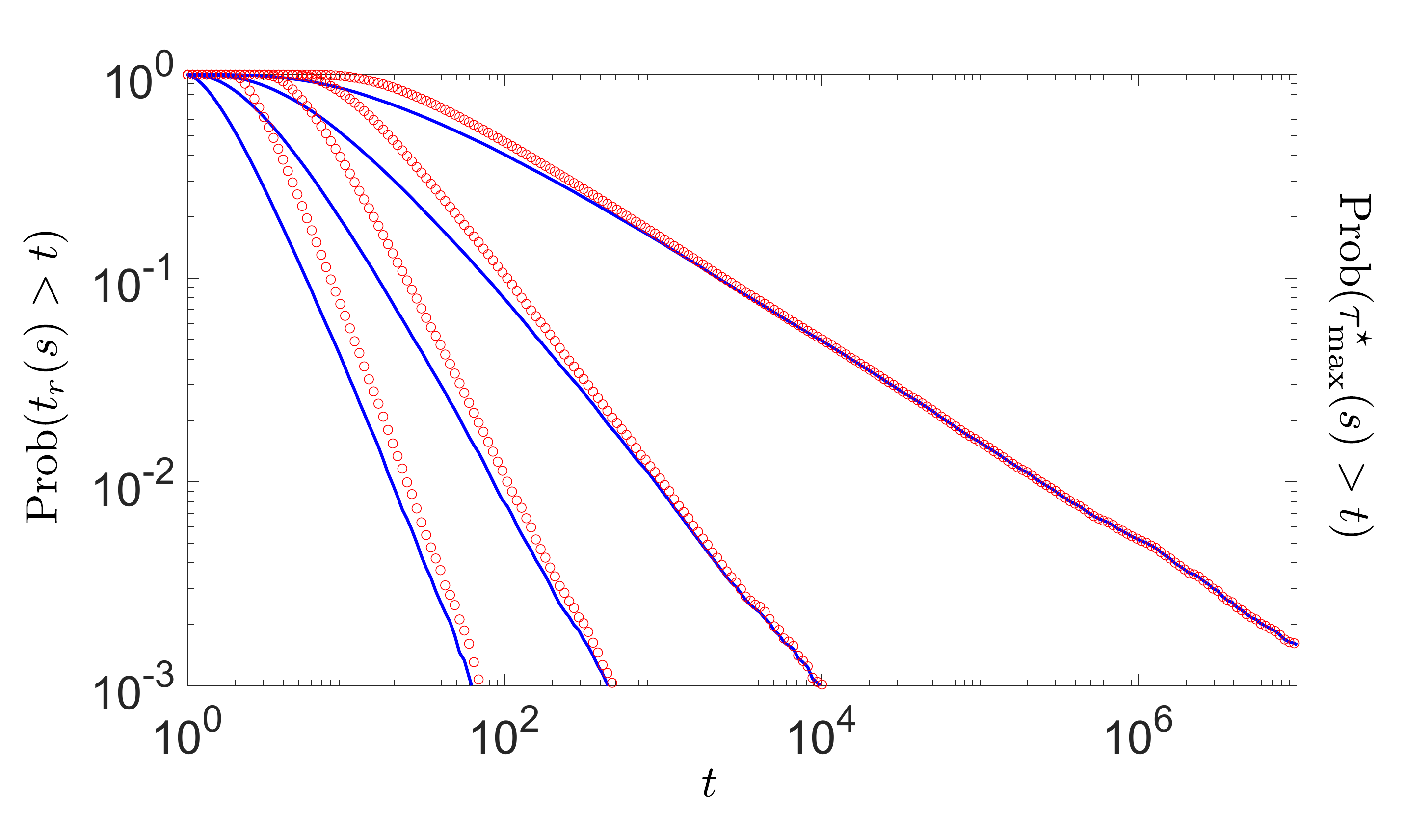}
\caption{Distributions of $t_r(s)$ (red circles) and $\tau_\text{max}^\star(s)$ (black lines) with $s=0,1,2,3$ (from top to bottom) for the unidirectional transport model with $N=5$. We used Pareto transition times with $t_0=1$ and $\alpha=0.5$ and Monte-Carlo simulations with $10^5$ particles. We find the matching predicted by Eq.~(\ref{appremovprinc}) \label{fig:udt_distributionsx}}
\end{center}\end{figure}

So, after $N=L$ steps the particle reaches $L+1$. We have the set of $N$ transition times $(\tau_1,\ldots,\tau_N)$. The idea is to employ order statistics \cite{majumdar2020extreme} and order the set
\begin{equation}
\tau_{(1)}<\ldots<\tau_{(N)}.
\end{equation}
Obviously, it is $\tau_{(N)}=\tau_\text{max}$ and $\tau_{(1)}=\text{min}(\tau_1,\ldots,\tau_N)$. Now, we eliminate the $s=1,\ldots,N-1$ longest transition times.  So we are left with $(\tau_{(1)},\ldots,\tau_{(N-s)})$. The first passage time after elimination of the $s$ largest values is 
\begin{equation}
t_r(s) =\sum_{q=1}^{N-s} \tau_{(q)}
\end{equation}
and the longest transition time after removal is
\begin{equation}
\tau_\text{max}^\star(s)=\tau_{(N-s)}.
\end{equation}
When $s=1$, the two quantities
\begin{equation}
t_r(1)=t_r
\end{equation}
and
\begin{equation}
\tau_\text{max}^\star(1)=\tau_\text{max}
\end{equation}
are mainly considered in the main text. \\ \\
In order to get the asymptotic distributions of both quantities for any $s$, we need the probability density of the $q$-th order statistics $\tau_{(q)}$ \cite{majumdar2020extreme}
\begin{equation}\label{pdf_order}
p_{\tau_{(q)}}(t)=\frac{N!}{(q-1)!(N-q)!} p_{\tau_n}(t)[P_{\tau_n}(t)]^{q-1}[1-P_{\tau_n}(t)]^{N-q}.
\end{equation}
For the maximum after removal, we get the asymptotics
\begin{equation}\label{maxmax3}
p_{\tau_\text{max}^\star(s)}(t) \sim \frac{N!}{s!(N-s-1)!} p_{\tau_n}(t) [1-P_{\tau_n}(t)]^s,
\end{equation}
which is explicitly
\begin{equation}
p_{\tau_\text{max}^\star(s)}(t) \sim \frac{N!}{s!(N-s-1)!} \frac{A^{s+1}}{\alpha^s} t^{-1-(s+1)\alpha}
\end{equation}
and in terms of the probability distribution
\begin{equation}
\text{Prob}(\tau_\text{max}^\star(s) >t) \sim \frac{N!}{(s+1)!(N-s-1)!} \left( \frac{A}{\alpha}\right)^{s+1}t^{-(s+1)\alpha}.
\end{equation}
For the modified first passage time, we first need the joint probability density function of the $N-s$ smallest transition times $(\tau_{(1)},\ldots,\tau_{(N-s)})$. We integrate the joint probability density of all order statistics $N! p_{\tau_1}(t_1)\ldots p_{\tau_N}(t_N)$ over the eliminated transition times $(\tau_{(N-s+1)},\ldots,\tau_{(N)})$ and get
\begin{equation}\begin{split}\label{halfjoint}
p_{\tau_{(1)},\ldots,\tau_{(N-s)}}(t_1,\ldots,t_{N-s})= \frac{N!}{s!}[1-P_{\tau_n}(t_{N-s})]^s\prod_{r=1}^{N-s} p_{\tau_n}(t_r)\Theta(t_r-t_{r-1})
\end{split}\end{equation}
with the Heaviside function $\Theta$. The probability density function of $t_r(s)$ is
\begin{equation}
p_{t_r(s)}(t) = \langle \delta(t-t_r(s)) \rangle,
\end{equation}
which is the generalised convolution given by the $(N-s-1)$-multiple integral
\begin{equation}\begin{split}
p_{t_r(s)}(t) = \int\limits_0^\infty \int\limits_0^\infty \ldots \int\limits_0^\infty \mathrm{d}t_1 \ldots\mathrm{d}t_{N-s-1}\times p_{\tau_{(1)},\ldots,\tau_{(N-s)}}(t_1,\ldots,t_{N-s-1},t-\|\bm{t}\|_{N-s-1}) .
\end{split}\end{equation}
We use the notation $\|\bm{t}\|_{N-s-1}= \sum_{i=1}^{N-s-1} t_i$. Following, we wish to derive the large $t$ behavior of $p_{t_r(s)}(t)$. The $r=N-s$ term is asymptotically
\begin{equation}\begin{split}
p_{\tau_n}(t-\|\bm{t}\|_{N-s-1})[1-P_{\tau_n}(t-\|\bm{t}\|_{N-s-1})]^s \sim p_{\tau_n}(t) [1-P_{\tau_n}(t)]^s
\end{split}\end{equation}
which we can put in front of the integrals. This can be verified by using the Binomial Theorem and considering only the asymptotic behavior. Furthermore, the lower limits of the integrals are set by $\Theta(t_r-t_{r-1})$ in Eq.~(\ref{halfjoint}) for all $r$. Thus, we have asymptotically
\begin{equation}\begin{split}
p_{t_r(s)}(t) \sim \frac{N!}{s!} p_{\tau_n}(t) [1-P_{\tau_n}(t)]^s \int\limits_{t_0}^\infty \int\limits_{t_1}^\infty \ldots \int\limits_{t_{N-s-2}}^\infty\mathrm{d}t_1 \ldots \mathrm{d}t_{N-s-1}  p_{\tau_n}(t_1) \ldots p_{\tau_n}(t_{N-s-1}) .
\end{split}\end{equation}
Without loss of generality, we work with the Pareto probability density function $p_{\tau_n}(t)=\alpha(t_0)^\alpha t^{-1-\alpha}$ with $t\ge t_0$. Generally, one uses the Binomial theorem. Let us start with the first inner integral
\begin{equation}
\int\limits_{t_{N-s-2}}^\infty\mathrm{d}t_{N-s-1} p_{\tau_n}(t_{N-s-1}) = (t_0)^\alpha (t_{N-s-2})^{-\alpha}.
\end{equation}
The second inner integral gives
\begin{equation}\begin{split}
(t_0)^\alpha\int\limits_{t_{N-s-3}}^\infty\mathrm{d}t_{N-s-2} p_{\tau_n}(t_{N-s-2}) (t_{N-s-2})^{-\alpha} = \frac{1}{2} (t_0)^{2\alpha} (t_{N-s-3})^{-2\alpha}.
\end{split}\end{equation}
We can continue similar with the other integrals. Generally, the $k$-th inner integral (with $k=1,\ldots,N-s-1$) is
\begin{equation}\begin{split}
\frac{(t_0)^{k-1}}{(k-1)!} \int\limits_{t_{N-s-k-1}}^\infty\mathrm{d}t_{N-s-k} p_{\tau_n}(t_{N-s-k}) (t_{N-s-k})^{-(k-1)\alpha}= \frac{1}{k!} (t_0)^{k\alpha} (t_{N-s-k-1})^{-k\alpha}.
\end{split}\end{equation}
In particular, the final inner integral (the outer integral) with $k=N-s-1$ is
\begin{equation}\begin{split}
\frac{(t_0)^{N-s-2}}{(N-s-2)!} \int\limits_{t_0}^\infty\mathrm{d}t_1 p_{\tau_n}(t_1)  (t_1)^{-(N-3)\alpha}= \frac{1}{(N-s-1)!} (t_0)^{(N-s-1)\alpha} (t_0)^{-(N-s-1)\alpha}= \frac{1}{(N-s-1)!} .
\end{split}\end{equation}
So, we have finally
\begin{equation}\label{appasymtr}
p_{t_r(s)}(t) \sim \frac{N!}{s!(N-s-1)!} p_{\tau_n}(t) [1-P_{\tau_n}(t)]^s.
\end{equation}
which is the same as Eq.~(\ref{maxmax3}) with $A=\alpha (t_0)^\alpha$. This asymptotic equivalence shows the principle of the single long transition time 
\begin{equation}\label{appremovprinc}
\text{Prob}(t_r(s)>t) \sim \text{Prob}(\tau_\text{max}^\star(s)>t),
\end{equation}
which for $s=1$ is presented in the main text Eq.~(11), and for any $s$ in Eq.~(15). The asymptotic behavior of the $t_r(s)$ distribution Eq.~(\ref{appasymtr}) is presented for $s=1$ in Eq.~(9) and any $s$ in Eq.~(13). In Fig.~\ref{fig:udt_distributionsx}, we present Eq.~(\ref{appremovprinc}) for the four cases $s=0,1,2$ and $3$. When $s=0$, we have the original first passage time and maximum transition time without any removal.

\section{The measure of gain $G$ for the unidirectional transport model}\label{sec:appimpr}
Let us shortly consider the removal of the maximum with $s=1$ (then consider any $s$). When the transition time distribution follows a power law with the exponent $\alpha < 1$ Eq.~(\ref{psit}), then $\langle t_f \rangle$ and $\langle \tau_\text{max} \rangle$ are both infinite. What does this imply for the mean first passage time after elimination
\begin{equation}
\langle t_r\rangle = \langle t_f \rangle - \langle \tau_\text{max} \rangle?
\end{equation}
We discuss this problem here, i.e. we calculate $\langle t_r \rangle$. In general, we consider any $\alpha>0$ and any $s \ge 1$. Exponential transition times are considered in addition. With the results of this section, we provide the derivations for the measure of gain $\langle G \rangle = \langle t_r \rangle / \langle t_f \rangle$ in Eq.~(17) and (18) (the thermodynamic limit $N\to \infty$) presented in the main text.\\ \\

The basis of all following calculations is again that we use the order statistics of the transition times. The mean first passage time after elimination
\begin{equation}
\langle t_r(s) \rangle = \sum_{q=1}^{N-s} \langle \tau_{(q)} \rangle
\end{equation}
and the quantifier of the elimination (Eq.~(16) in the main text) is
\begin{equation}
G(s) = \frac{\left\langle t_r(s) \right\rangle}{\langle t_f \rangle}
\end{equation}
for $s=1,\ldots,N-1$. When $s=0$, it is $t_r(0) =t_f$. When $s=N$, we set $t_r(N)=0$. 

\subsection{The measure of gain $G$ for exponentially distributed transition times}
For the exponentially distributed transition times $p_{\tau_n}(t)=\text{exp}(-t/\langle \tau_n \rangle)/\langle \tau_n \rangle$, the mean order transition time is
\begin{equation}
\langle \tau_{(q)} \rangle = \langle \tau_n \rangle \sum_{j=1}^q \frac{1}{N-j+1},
\end{equation}
see \cite{renyi1953theory}. The mean first passage time after elimination can be calculated as
\begin{equation}\begin{split}\label{mean_exp}
\langle t_r(s) \rangle = \langle \tau_n \rangle \sum_{q=1}^{N-s} \sum_{j=1}^q \frac{1}{N-j+1}= \langle \tau_n \rangle \sum_{j=1}^{N-s} \left( 1-\frac{s}{N-j+1}\right) .
\end{split}\end{equation}
Using the harmonic number $H(K)=\sum_{j=1}^K 1/j$, we write
\begin{equation}
\langle t_r(s) \rangle = \langle \tau_n \rangle \left[N-s-s(H(N)-H(s)) \right].
\end{equation}
Therefore, the improvement quantifier is exactly given by
\begin{equation}
G(s) = 1- \frac{s}{N}[1-H(s)+H(N)].
\end{equation}
Now, we consider the limit $N\to\infty$. For fixed $s$, the large $N$ limit is
\begin{equation}
G(s) \sim 1.
\end{equation}
For a fixed fraction $f$ with $s=f N$, $0<f<1$, the large $N$ limit is
\begin{equation}
G(s) \sim 1-f [1-\text{ln}(f)],
\end{equation}
which is independent of $\langle \tau_n \rangle$.
\subsection{The measure of gain $G$ for Pareto distributed transition times} \label{sec:gforpareto}
For the Pareto distributed transition times $p_{\tau_n}(t) = \alpha (t_0)^\alpha t^{-1-\alpha}$, $t \ge t_0$, the mean order transition time can be calculated as
\begin{equation}
\langle \tau_{(q)} \rangle = \frac{(-1)^q t_0 N! \Gamma(-N+1/\alpha)\Gamma(q)}{(N-q)!(q-1)!\Gamma(-N+q+1/\alpha)}
\end{equation}
for $\alpha > 1/(N-q+1)$. This result has been obtained with Mathematica using the formula $\langle \tau_{(q)} \rangle =\int_0^\infty tp_{\tau_{(q)}}(t)\mathrm{d}t$. From that, we obtain the mean first passage time after elimination
\begin{equation}\begin{split}\label{mean_alg2}
\langle t_r(s) \rangle  = N t_0 \frac{\alpha}{\alpha-1} - \frac{(-1)^{N-s+1}(-1+\alpha s)t_0 N! \Gamma(-N+1/\alpha)}{(-1+\alpha)(-1+s)!\Gamma(1-s+1/\alpha)}
\end{split}\end{equation}
for $\alpha>1/(s+1)$. Thus, we have $G$ taking the ratio of this formula between the general $s$ and $s=0$ case. The latter is $\langle t_r(0) \rangle = \langle t_f \rangle = N t_0 \alpha /(\alpha-1)$ for $\alpha>1$ and infinite otherwise. The quantifier $G$ with $s=1$ is presented in the main text Eq.~(17). \\ \\

Next, we calculate the large $N$ limit of $\langle t_r(s) \rangle$ for $s=f N$ with the fixed fraction $0<f<1$. Each $N$-dependent term of Eq.~(\ref{mean_alg2}) behaves as
\begin{equation}\begin{split}
N! &\sim \sqrt{2\pi}\text{e}^{-N} N^{N+1/2}  ,\\
(s-1)!  &\sim \sqrt{2\pi} \text{e}^{-s}s^{s-1/2}, \\
\Gamma(-N+1/\alpha) &\sim \sqrt{\pi/2}\text{e}^N N^{-N-1/2+1/\alpha}\text{csc}[\pi(-N+1/\alpha)]=(-1)^N\sqrt{\pi/2}\text{e}^N N^{-N-1/2+1/\alpha}\text{csc}[\pi/\alpha], \\
\Gamma(-s+1+1/\alpha)&\sim \sqrt{\pi/2}\text{e}^s s^{-s+1/2+1/\alpha}\text{csc}[\pi(-s+1+1/\alpha)]=(-1)^s \sqrt{\pi/2}\text{e}^s s^{-s+1/2+1/\alpha}\text{csc}[\pi(1+1/\alpha)].
\end{split}\end{equation}
Putting these terms together yields
\begin{equation}
\langle t_r(s)\rangle  \sim N t_0 \frac{\alpha}{\alpha-1} - f^{-1/\alpha} t_0 \frac{1-\alpha f N}{1-\alpha}.
\end{equation}
Since $s\to\infty$ the existence of this mean is ensured for $\alpha>0$. Finally, the improvement quantifier for $\alpha>1$ scales as
\begin{equation}
G(s) \sim 1- f^{1-1/\alpha}
\end{equation}
and for $0<\alpha<1$, we have $G(s)\sim 0$. This result is shown in Eq.~(18) in the main text.

\section{Transition times for the pore-scale transport simulation}\label{sec:porescaleapp}

For the transport simulation in a porous medium of Sec.~III, we obtained the transition times as follows.
To link the advective transport of particles within the random walk framework, one must consider the spatial and temporal transitions of each particle trajectory. To characterize the particle transition times from the transport simulations, we use a length scale of $\lambda$ (the mean grain diameter) as a constant space transition \cite{Bijeljic2011}. Specifically, for each transition, we define a circle of diameter $\lambda$ around the initial position of the particle. The particle is then transported according to the flow field ($\textbf{U}$). Once it leaves its current circle, the center of the circle is moved to the ``new'' particle coordinates.
In doing so, we assume that the velocity field correlations (in space) breaks (on average) after the particle leaves the vicinity of the previous grain \cite{Morales2017}.

\section{Principle of the single long transition time for the continuous time random walk}\label{sec:ctrw_app}

Particles in the continuous time random walk of Sec.~IV start at $x=1$ and are absorbed at $x=L+1$. The probability to jump left is $q$ and right $p=1-q$ with right bias $p>q$. The number of jumps $N$ is random. Each transition time follows the probability density function Eq.~(\ref{psit}). We provide the principle of the single long transition time Eq.~(9) and (20) presented in the main text. We also calculate different examples here.\\ \\

The probability density functions of $t_f=\sum_{n=1}^N \tau_n$ and $\tau_\text{max}=\text{max}(\tau_1,\ldots,\tau_N)$ obey 
\begin{equation}\begin{split}\label{comp}
p_{t_f}(t) &= \sum_{n=1}^N \phi_\text{dis}(n) p_{t_f|N}(t|n), \\
p_{\tau_\text{max}}(t) &= \sum_{n=1}^N \phi_\text{dis}(n) p_{\tau_\text{max}|N}(t|n)
\end{split}\end{equation}
where $\phi_\text{dis}(n)$ is the probability that a particle made $N=n$ jumps before its absorption. The conditional probability density functions on the right hand side consider fixed $N=n$. \\ \\

 Under the assumption of a finite mean number $\langle N \rangle < \infty$, we can utilize the principle of the single long transition time of the unidirectional model (see Eq.~(9), (11) in the main text). We get in generality
\begin{equation}\label{ctrwgenres}
\text{Prob} (t_f > t ) \sim \text{Prob} (\tau_\text{max} > t ) \sim \langle N \rangle \frac{A}{\alpha} t^{-\alpha}
\end{equation}
with the mean $\langle N \rangle = \sum_{n=1}^\infty \phi_\text{dis}(n) n$. This result is presented in Eq.~(19) in the main text.

\subsection{Principle of the single long transition time with finite $\langle N \rangle$ - Second approach} \label{sec:altcpctrwf}
We derive the principle of the single long transition time Eq.~(\ref{ctrwgenres}) where $\langle N \rangle$ is finite with an alternative approach, which we need later for the unbiased situation (when $\langle N \rangle$ is infinite). Here, we use a generating function approach. The generating function is defined as
\begin{equation}\label{gendef5}
\tilde{\phi}_\text{dis}(z) = \sum_{n=1}^\infty \phi_\text{dis}(n) z^n.
\end{equation}
From Eq.~(\ref{comp}), we see that we can relate the Laplace transform of $p_{t_f}(t)$ (defined as $\hat{p}_{t_f}(s)=\mathcal{L}_{t\to s}[p_{t_f}(t)] =\int_0^\infty p_{t_f}(t) \text{exp}(-st)\text{d}t$) and the cumulative distribution function of $\tau_\text{max}$ to the generating function
\begin{equation}\begin{split}\label{ansatzboth}
\hat{p}_{t_f}(s) &= \sum_{n=1}^\infty \phi_\text{dis}(n) [\hat {p}_{\tau_n}(s)]^n,\\
\mathrm{Prob}(\tau_\mathrm{max}\le t) &= \sum_{n=1}^\infty \phi_\text{dis}(n) [\mathrm{Prob}(\tau_n \le t)]^n.
\end{split}\end{equation}
The $\hat{p}_{\tau_n}(s)\approx 1$ expansion of the first formula of Eq.~(\ref{ansatzboth}) is
\begin{equation}
\hat{p}_{t_f}(s) \sim 1+\langle N \rangle (\hat{p}_{\tau_n}(s)-1) + \ldots.
\end{equation}
Inserting the $s\approx 0$ expansion $\hat{p}_{\tau_n}(s)\sim 1+ A\Gamma(-\alpha) s^\alpha$ yields
\begin{equation}
\hat{p}_{t_f}(s) \sim 1 + \langle N \rangle A\Gamma(-\alpha) s^\alpha.
\end{equation}
The small $s$ behavior of the Laplace transform gives the large $t$ behavior of the probability density function, which is known as the Tauberian theorem \cite{metzler2000random,klafter2011first}. Inverse Laplace transform gives
\begin{equation}\label{tfagain}
p_{t_f}(t) \sim \langle N \rangle A t^{-1-\alpha}.
\end{equation}
The $\mathrm{Prob}(\tau_n\le t)\approx 1$ expansion of the second formula of Eq.~(\ref{ansatzboth}) is
\begin{equation}
\mathrm{Prob}(\tau_\mathrm{max}\le t) \sim 1+\langle N \rangle (\mathrm{Prob}(\tau_n\le 1)-1)+\ldots.
\end{equation}
Inserting the large $t$ behavior $\mathrm{Prob}(\tau_n\le t) \approx 1- At^{-\alpha}/\alpha$ and derivation gives
\begin{equation}
p_{\tau_\mathrm{max}}(t) \sim \langle N \rangle A t^{-1-\alpha}
\end{equation}
which is the same asymptotic behavior as Eq.~(\ref{ctrwgenres}). We now calculate $\langle N \rangle$ rigorously. We will also consider $\langle N \rangle = \infty$ later (that is for example the case of no bias).

\subsection{Weak bias $p>q$ with no left boundary condition}\label{sec:weakctrwfirst}
We consider weak bias $p>q$ and calculate $\langle N \rangle$, thus, the principle of the single long transition time Eq.~(\ref{ctrwgenres}). Here, there is no left boundary condition and the lattice points are $x=\{\ldots,0,1,\ldots,L\}$ with start at $x=1$ and absorption at $L+1$. \\ \\ 

We use a generating function approach. The generating function of the mean number of jumps $\tilde{\phi}_\text{dis}(z)$ defined in Eq.~(\ref{gendef5}). For the special case $L=1$ (i.e. absorption at $x=2$) the generating function is
\begin{equation}\label{l1}
\tilde{\phi}_\text{dis}(z)|_{L=1}=\frac{1-\sqrt{1-4pqz^2}}{2qz},
\end{equation}
which can be derived from the renewal equation \cite{redner2001guide,klafter2011first}. For general $L$, we can use the following argument. Let us say we have now $L=2$ (i.e. absorption at $x=3$). Once the particle reaches $x=2$, it is the problem of Eq.~(\ref{l1}) where both the initial starting point and the absorbing boundary are shifted one lattice point to the right. The same picture holds for any $L$. Thus, the general generating function is
\begin{equation}\label{l11}
\tilde{\phi}_\text{dis}(z)=[\tilde{\phi}_\text{dis}(z)|_{L=1}]^L=\left[ \frac{1-\sqrt{1-4pqz^2}}{2qz}\right]^L.
\end{equation}

The $z=0$ expansion for the case $L=1$ of Eq.~(\ref{l1}) is
\begin{equation}\begin{split}
\tilde{\phi}_\text{dis}(z) |_{L=1} &= \sum_{n=1,3,\ldots} \frac{\left(2\frac{n+1}{2}-3\right)!!}{\left( \frac{n+1}{2}\right)!}2^{\frac{n+1}{2}-1}p^{\frac{n+1}{2}}q^{\frac{n+1}{2}-1}z^{2\frac{n+1}{2}-1} \\
&= \sum_{i=1}^\infty \frac{(2i-3)!!}{i!}2^{i-1}p^iq ^{i-1}z^{2i-1}\\
&=pz+p^2q z^3 +2 p^3q^2z^5+5p^4q^3z^7+\ldots
\end{split}\end{equation}
where the step number is $n=2i-1$. Clearly, for the strongly biased transport $p=1$, we have only one term in the expansion corresponding to the case where  the particle is only jumping to the right and absorbed after one step, namely $\phi_\text{dis}(1)=1$ and otherwise zero. From the last formula, we can read $\phi_\text{dis}(n)$ for $L=1$ as
\begin{equation}\begin{split}
\phi_\text{dis}(n)|_{L=1}=
\begin{cases}
\frac{\left(2\frac{n+1}{2}-3\right)!!}{\left( \frac{n+1}{2}\right)!}2^{\frac{n+1}{2}-1}p^{\frac{n+1}{2}}q^{\frac{n+1}{2}-1} & \text{, for } n=1,3,5,\ldots,\\
0 & \text{, for } n=2,4,6,\ldots.
\end{cases}
\end{split}\end{equation}
The mean $\langle N \rangle=\sum_{n=1}^\infty n \phi_\text{dis}(n)$ for $L=1$ is
\begin{equation}\begin{split}
\langle N \rangle|_{L=1} =\sum_{i=1}^\infty (2i-1) \frac{(2i-3)!!}{i!}2^{i-1}p^iq ^{i-1}= \frac{1}{p-q}.
\end{split}\end{equation}
Here, we see that when $p\to 1/2$ this mean diverges, as it is well-known. Therefore, we can conclude the mean jump number for general $L$ as
\begin{equation}\label{meanlpq}
\langle N \rangle = L \langle N \rangle|_{L=1} = \frac{L}{p-q} .
\end{equation}
Another way, we can also read the mean from 
\begin{equation}
\langle N \rangle = \frac{\mathrm{d}\tilde{\phi}_\text{dis}(z)}{\mathrm{d}z}|_{z=1}.
\end{equation}
Thus, using the $z\approx 1$ expansions
\begin{equation}
\hat{\phi}_\text{dis}(z)|_{L=1} \sim 1 +\frac{1}{p-q} (z-1) + \mathcal{O}((z-1)^2)
\end{equation}
and 
\begin{equation}
\hat{\phi}_\text{dis}(z) \sim 1 + \frac{L}{p-q} (z-1) + \mathcal{O}((z-1)^2),
\end{equation}
we find again the mean numbers of jumps in the second terms. With this exact formula for $\langle N \rangle$, we present the principle of the single long transition time in Eq.~(20) in the main text.

\subsection{No bias $p=1/2$ with no left boundary condition}\label{sec:ctrwnobiasapp}

We consider no bias $p=q=1/2$. Here, there is no left boundary and the lattice points are $x=\{\ldots,0,1,\ldots,N\}$ with start at $x=1$ and absorption at $L+1$. Since the bias is zero, the mean number of jumps $\langle N \rangle$ diverges. So the principle of Eq.~(\ref{ctrwgenres}) does not hold. However, we can use again the approach of SM Sec.~\ref{sec:altcpctrwf} to find the modification of the principle. \\ \\

We start with the generating function for $L=1$, which is
\begin{equation}\label{l2}
\tilde{\phi}_\text{dis}(z)|_{L=1}=\frac{1-\sqrt{1-z^2}}{z}
\end{equation}
where we used $p=q=1/2$ in Eq.~(\ref{l1}).  The $z\approx 1$ expansion is
\begin{equation}
\hat{\phi}_\text{dis}(z)|_{L=1} \sim 1 -\sqrt{2(1-z)} +\ldots
\end{equation}
For general $L$, the $z\approx 1$ expansion is
\begin{equation}
\hat{\phi}_\text{dis}(z) \sim 1 -L\sqrt{2(1-z)} +\ldots
\end{equation}
where we used $\tilde{\phi}_N(z)=[\tilde{\phi}_N(z)|_{L=1}]^L$, see Eq.~(\ref{l11}). \\ \\

In the following, we consider $0<\alpha<1$. First, we discuss $\hat{p}_{t_f}(s)$ where $\hat{p}_{t_f}(s)$ given in Eq.~(\ref{ansatzboth}). We put the $s\approx 0$ expansion of $\hat{p}_{\tau_n}(s)$ into the $\hat{p}_{\tau_n}(s) \approx 1$ expansion of $\hat{p}_{t_f}(s)$ and get
\begin{equation}
\hat{p}_{t_f}(s) \sim 1 -L\sqrt{2A|\Gamma(-\alpha)|} s^{\alpha/2}.
\end{equation}
Inverse Laplace transform obtains
\begin{equation}\label{tfnobias}
p_{t_f(t)} \sim \frac{\sqrt{2A|\Gamma(-\alpha)|}}{|\Gamma(-\alpha/2)|}L t^{-1-\alpha/2}
\end{equation}
valid for long times. On the other hand, the large $t$ behavior of $p_{\tau_\text{max}}(t)$ is obtained as follows. We put the large $t$ behavior of $\text{Prob}(\tau_n \le t)$ into the $\text{Prob}(\tau_n \le t) \approx 1$ expansion of $\text{Prob}(\tau_\text{max} \le t)$, so
\begin{equation}
\mathrm{Prob}(\tau_\mathrm{max}\le t)  \sim 1- L\sqrt{2A/\alpha} t^{-\alpha/2}
\end{equation}
with the derivation
\begin{equation}
p_{\tau_\mathrm{max}}(t)\sim \sqrt{\alpha A/2} L t^{-1-\alpha/2}.
\end{equation}
This as the same power law decay as Eq.~(\ref{tfnobias}) but with a different prefactor. We presented this result in Eq.~(22) in the main text with the proper rescaling of the maximum transition time Eq.~(21). 

\subsection{Left bias $p<q$ with no left boundary condition}

We consider a left bias $p<q$. Here, there is no left boundary and the lattice points are $x=\{\ldots,0,1,\ldots,N\}$ with start at $x=1$ and absorption at $L+1$. Since the bias pushes the particles away from the absorbing wall, the mean number of jumps $\langle N \rangle$ diverges. So the principle of Eq.~(\ref{ctrwgenres}) does not hold. However, we can use again the approach of SM Sec.~\ref{sec:altcpctrwf} to find the modification of the principle. \\ \\

We start with the generating function for $L=1$. The $z\approx 1$ expansion is
\begin{equation}
\hat{\phi}_\text{dis}(z)|_{L=1} \sim \frac{p}{q} + \frac{1-\sqrt{1-4pq}}{2q\sqrt{1-4pq}} (z-1) + \mathcal{O}((z-1)^2)
\end{equation}
For general $L$, the $z\approx 1$ expansion is
\begin{equation}
\hat{\phi}_\text{dis}(z) \sim \left(\frac{p}{q}\right)^L + L \frac{1-\sqrt{1-4pq}}{2q\sqrt{1-4pq}} (z-1) + \mathcal{O}((z-1)^2)
\end{equation}
where we used $\tilde{\phi}_N(z)=[\tilde{\phi}_N(z)|_{L=1}]^L$, see Eq.~(\ref{l11}). We use the definition of the generating function, namely $\phi_N(z)_{z=1} = \sum_{n=1}^\infty \phi_N(n)$ hence the leading term $(p/q)^L <1$ is the probability to reach the absorbing boundary $L+1$. In other words, that absorption is not guaranteed due to the left bias $p<q$. \\ \\

In what follows, we consider an ensemble of trajectories that are conditioned on absorption, namely those trajectories that are never absorbed are discarded. We indicate this with ``(abs)''. Following the procedure of SM Sec.~\ref{sec:altcpctrwf}, we have
\begin{equation}
p_{t_f}^\mathrm{(abs)}(t) \sim p_{\tau_\mathrm{max}}^\mathrm{(abs)}(t) \sim \langle N \rangle^\mathrm{(abs)} At^{-1-\alpha}
\end{equation}
with
\begin{equation}
\langle N \rangle^\mathrm{(abs)} =  \frac{1-\sqrt{1-4pq}}{2q\sqrt{1-4pq}}L.
\end{equation}
The mean $\langle N \rangle^\mathrm{(abs)}$ here is the mean with respect to the ensemble of trajectories that were eventually absorbed.

\subsection{No bias $p=q=1/2$ with left reflecting boundary}
We discuss the problem unbiased situation $p=q=1/2$ in the presence of a reflecting boundary at $R<1$. In this case, the mean $\langle N \rangle$ is finite, in contrast to to unbiased case with no right wall ($R\to -\infty$, see SM Sec.~\ref{sec:ctrwnobiasapp}). The problem for $L=1$ was discussed in \cite{bel2006random}, and the generating function for $L=1$ is
\begin{equation}
\tilde{\phi}_\text{dis}(z)|_{R,L=1} =  \frac{z/2}{1-\frac{z^2}{4}\frac{H_+\Lambda_+^{-R-1}+H_-\Lambda_-^{-R-1}}{H_+\Lambda_+^{-R}+H_-\Lambda_-^{-R}}}
\end{equation}
with the $z$-dependent functions
\begin{equation}\begin{split}
\Lambda_\pm &= \frac{1\pm \sqrt{1-z^2}}{2}, \\
B_- &= \frac{1-z^2/2-\Lambda_-}{\Lambda_- - \Lambda_+},\\
B_+ &= 1- B_-.
\end{split}\end{equation}
For any $L>1$, we obtain the generating function as follows. Let's consider first $L=2$. When the particle reaches $x=2$ for the first time, it is now the $L=1$ problem but the reflecting boundary $R$ is moved one lattice point to the left. Similar, for any $L$. So the generating function is
\begin{equation}\begin{split}
\tilde{\phi}_\text{dis}(z)|_{R,L} = \prod_{R^\prime=0}^{L-1}   \tilde{\phi}_\text{dis}(z)|_{R-R^\prime,L=1}=  \prod_{R^\prime=0}^{L-1}\frac{z/2}{1-\frac{z^2}{4}\frac{H_+\Lambda_+^{-R+R^\prime-1}+H_-\Lambda_-^{-R+R^\prime-1}}{H_+\Lambda_+^{-R+R^\prime}+H_-\Lambda_-^{-R+R^\prime}}}.
\end{split}\end{equation}
The generating function $\tilde{\phi}_\text{dis}(z)|_{R-R^\prime,L=1}$ with absorbing boundary $x=L+1=2$ was derived in \cite{bel2006random}. For $L=1$ the $z\approx 1$ expansion of $\tilde{\phi}_\text{dis}(z)|_{R-R^\prime,L=1}$ is
\begin{equation}
 \tilde{\phi}_\text{dis}(z)|_{R-R^\prime,L=1} \sim 1 + (3-2R-2R^\prime)(z-1) +\ldots
\end{equation}
from which we see the mean number of jumps for $L=1$
\begin{equation}
\langle N \rangle|_{R-R^\prime,L=1}= 3-2R-2R^\prime.
\end{equation}
Therefore, the mean for any $L$ is
\begin{equation}
\langle N \rangle|_{R,L} = \sum_{R^\prime=0}^{L-1} \langle N \rangle_{N}|_{R-R^\prime,L=1} = (2+L+2|R|)L.
\end{equation}
Thus, we have the explicit statement the principle of the single long transition time Eq.~(\ref{ctrwgenres}).

\subsection{Bias $p\neq q$ with left reflecting boundary}

We discuss the problem $p\neq q$ (which includes left and right bias) in the presence of a reflecting boundary at $R<1$. In this case, the mean $\langle N \rangle$ is finite. Again, the particle starts at $x=1$ and absorption is at $x=L+1$. The problem for $L=1$ was discussed in \cite{bel2006random}, and the generating function for $L=1$ is
\begin{equation}\begin{split}
 \tilde{\phi}_\text{dis}(z)|_{R,L=1} =\frac{pz}{1-pqz^2\frac{B_+\lambda_+^{-R-1}+B_-\lambda_-^{-R-1}}{B_+\lambda_+^{-R}+B_-\lambda_-^{-R}}}.
\end{split}\end{equation}
with the $z$-dependent functions
\begin{equation}\begin{split}
\lambda_\pm &= \frac{1\pm \sqrt{1-4pqz^2}}{2},\\
B_+ &= \frac{1-(\lambda_-)^2-pz^2}{\lambda_+ - \lambda_-}, \\
B_- &= 1-B_+.
\end{split}\end{equation}
For any $L>1$, the generating function is
\begin{equation}\begin{split}
\tilde{\phi}_\text{dis}(z)|_{R,L} = \prod_{R^\prime=0}^{L-1}   \tilde{\phi}_\text{dis}(z)|_{R-R^\prime,L=1} =  \prod_{R^\prime=0}^{L-1}\frac{pz}{1-pqz^2\frac{B_+\lambda_+^{-R+R^\prime-1}+B_-\lambda_-^{-R+R^\prime-1}}{B_+\lambda_+^{-R+R^\prime}+B_-\lambda_-^{-R+R^\prime}}}.
\end{split}\end{equation}
For $L=1$ the $z\approx 1$ expansion of $ \tilde{\phi}_N(z)|_{R-R^\prime,L=1}$ is
\begin{equation}
 \tilde{\phi}_\text{dis}(z)|_{R-R^\prime,L=1} \sim 1 + \frac{\mathrm{d} \tilde{\phi}_\text{dis}(z)|_{R-R^\prime,L=1}}{\mathrm{d}z}|_{z=1}(z-1) +\ldots
\end{equation}
from which we see the mean number of jumps for $L=1$
\begin{equation}
\langle N \rangle|_{R-R^\prime,L=1}= \frac{\mathrm{d} \tilde{\phi}_N(z)|_{R-R^\prime,L=1}}{\mathrm{d}z}|_{z=1}.
\end{equation}
Therefore, the mean for any $L$ is
\begin{equation}\begin{split}
\langle N \rangle|_{R,L} = \sum_{R^\prime=0}^{L-1} \langle N \rangle|_{R-R^\prime,L=1}=  \sum_{R^\prime=0}^{L-1}\frac{\mathrm{d} \tilde{\phi}_N(z)|_{R-R^\prime,L=1}}{\mathrm{d}z}|_{z=1}=\frac{\mathrm{d}\tilde{\phi}_N(z)|_{R,L}}{\mathrm{d}z}|_{z=1}.
\end{split}\end{equation}
We suggest to use Mathematica, in order to calculate numerical values of $\langle N \rangle|_{R,L}$ for given parameters $p$, $q$, $R$ and $L$. With that, we have the principle of the single long transition times Eq.~(\ref{ctrwgenres}).

\section{Principle of the single long transition time for the continuous time random walk with continuous jumps}\label{sec:cont}

For the continuous time random walk, the principle of the single long transition time Eq.~(19) and (\ref{ctrwgenres}) depends on the mean number of jumps $\langle N \rangle$ (when $\langle N \rangle$ exists). Here, we discuss the case of continuous jumps $\Delta x_n$ which is not presented in the main text but stills serves as an important example. When the transport is biased, the mean jump size is positive $\langle \Delta x_n \rangle>0$ so that $\langle N \rangle$ exists, and then the principle Eq.~(19) and (\ref{ctrwgenres}) still holds. \\ \\

We discus exemplary the unidirectional transport. The particles start at $x=0$ and are absorbed at $x=L$. The first passage at $L$ happens at the $N$-th jump event under the condition $\sum_{n=1}^{N-1} \Delta x_n < L \le \sum_{n=1}^N \Delta x_n$.  The position at the step $N-1$ before passing $L$ is $\sum_{n=1}^{N-1} x_n$. The last step $\Delta x_N$ can be seen as the survival jump size. Thus, the probability of $N=n$ jumps in Laplace space is  $\hat{\phi}_\text{dis}(n)=\mathcal{L}_{L \to s}[\phi_\text{dis}(n)]=[\hat{p}_{\Delta x_n}(s)]^{n-1} [1-\hat{p}_{\Delta x_n}(s)]/s$. For example, for an exponential jump size distribution, it can be solved exactly and we find the mean number $\langle N \rangle = 1+L/\langle \Delta x_n \rangle$, which gives us an explicit expression for the principle Eq.~(19) and (\ref{ctrwgenres}).

\section{Measure of gain $G$ for the weakly biased ($p>1/2$) continuous time random walk}\label{sec:cont2}

The measure of gain $G= \langle t_r \rangle / \langle t_f \rangle$ for the biased continuous time random walk ($p>1/2$) with Pareto distributed transition times is presented in Eq.~(24) in the main text. We obtained this formula because with the help of the two mean first passage times
\begin{equation}\begin{split}
\langle t_f \rangle &= \sum_{n=1}^\infty \phi_\text{dis} (n) \langle t_f |n \rangle, \\
\langle t_r \rangle &= \sum_{n=1}^\infty \phi_\text{dis} (n) \langle t_r |n \rangle.
\end{split}\end{equation}
The conditional means on the right hand side assume fixed $N=n$ number of jumps. We use now Pareto transition times. That means for the first formula $\langle t_f | n \rangle = n t_0 \alpha/(\alpha-1)$ when $\alpha>1$ and therefore $\langle t_f \rangle = \langle N \rangle t_0 \alpha/(\alpha-1)$. For the second formula, we use $\langle t_r | n \rangle$ from Eq.~(\ref{mean_alg2}) (when $s=1$), which is valid for $\alpha>1/2.$ Putting both results together gives $G$, which is Eq.~(24) in the main text, namely
\begin{equation}\begin{split}
G=\begin{cases}
0 & \text{, } 0<\alpha<1, \\
1-h_\alpha \sum\limits_{n=1}^\infty \phi_\text{dis}(n) (-1)^n n! \Gamma\left(-n+\frac{1}{\alpha}\right) & \text{, } 1<\alpha
\end{cases}
\end{split}\end{equation}
with $h_\alpha=[(p-q)(\alpha-1)]/[\alpha L \Gamma(1/\alpha)]$ where we used $\langle N \rangle = L/(p-q)$ Eq.~(\ref{meanlpq}). In order to obtain numerical values of $G$, we have to calculate the sum (for $1<\alpha$). We use the analytical expression of the generating function $\tilde{\phi}_\text{dis}(z)$ Eq.~(\ref{l11}) and obtain numerical values of $\phi_\text{dis}(z)$ Eq.~(\ref{gendef5}) using Mathematica. This allows us to get $G$ numerically.

\section{Elimination principle depending on the transition time distribution after elimination}\label{sec:ctrw_elim}
The principle of the single long transition time of the unidirectional model (Eq.~(\ref{bjp_all}) in the main text) can also be written as
\begin{equation}\label{bbbvcbb}
\text{Prob}(t_f>t) \sim N \text{Prob}(\tau_n>t),
\end{equation}
i.e. the first passage time distribution is asymptotically distributed as $N$ times the transition time distribution. Here, we want to derive this relationship after we removed the maximum transition time for the unidirectional transport and the continuous time random walk. This is not discussed in the main text but necessary for a more complete picture of the mathematical structure of the removal technique. For the relationship between the distributions of $t_r=t_f-\tau_\text{max}$ and $\tau_\text{max}^\star$ (the second maximum), see the main text Eq.~(\ref{cap}) (unidirectional transport) and Eq.~(\ref{secbjp_ctrw}) (continuous time random walk).

\subsection{Unidirectional transport}
For the unidirectional transport of Sec.~\ref{sec:appfractal}, the transition times after elimination of the maximum at step number $m$, i.e. $\tau_\text{max}=\tau_m$ are $(\tau_1,\ldots,\tau_{m-1},\tau_{m+1},\ldots,\tau_N)$. We call an element of this set $\tau_n^\star$. The probability density function is the average
\begin{equation}\label{fixedndis}
p_{\tau_n^\star}(t) = \frac{1}{N-1} \sum_{q=1}^{N-1} p_{\tau_{(q)}}(t)
\end{equation}
with $ p_{\tau_{(q)}}(t)$ from Eq.~(\ref{pdf_order}). The large $t$ behavior is
\begin{equation}\label{largetnew}
p_{\tau_n^\star}(t) \sim N p_{\tau_n}(t) [1-P_{\tau_n}(t)].
\end{equation}
Comparison with the large $t$ behavior of the $t_r$ distribution Eq.~(\ref{appasymtr}) yields
\begin{equation}\label{needi2}
p_{t_r}(t) \sim (N-1) p_{\tau_n^\star}(t),
\end{equation}
i.e. the transition time distribution after elimination is multiplied with the number of jumps $N-1$.

\subsection{Continuous time random walk with finite $\langle N \rangle$}

We consider the continuous time random walk of Sec.~\ref{sec:ctrw_app} with finite moments of $N$, e.g. the weakly biased $p>q$ case. The transition times after elimination of $\tau_\text{max}=\tau_m$ are $(\tau_1,\ldots,\tau_{m-1},\tau_{m+1},\ldots,\tau_N)$. Importantly, $N$ is random. An immediate question is the following; does Eq.~(\ref{needi2}) still hold when we average the prefactor $p_{t_r}(t) \sim \langle N-1 \rangle p_{\tau_n^\star}(t)$?  \\ \\

Let us start with a straightforward definition of $p_{\tau_n^\star}(t)$, namely the average over $N$ of Eq.~(\ref{fixedndis}), namely
\begin{equation}\label{needi3}
p_{\tau_n^\star}(t) = \langle p_{\tau_n^\star|N}(t|N) \rangle \sim \langle N \rangle p_{\tau_n}(t)[1-P_{\tau_n}(t)].
\end{equation}
The mean is $\langle \circ \rangle = \sum_{n=1}^\infty \phi_\text{dis}(n) \circ$ and  $p_{\tau_n^\star|N}(t|N)$ is given in Eq.~(\ref{fixedndis}) and (\ref{largetnew}). How to obtain an estimate of $p_{\tau_n^\star}(t)$ from simulation? Given $R$ trajectories. For each trajectory $r\in\{1,\ldots,R\}$, we have the set of transition times $\bm{\tau}^\star(r)=(\tau_1,\ldots,\tau_{m-1},\tau_{m+1},\ldots,\tau_N)$. We calculate the histogram of $\bm{\tau}^\star(r)$ denoted as $h_{\tau_n^\star}(t;r)$. For large enough $R$, the sample mean $
\frac{1}{R} \sum_{r=1}^R h_{\tau_n^\star}(t;r) \approx p_{\tau_n^\star}(t)$ estimates our first definition of $p_{\tau_n^\star}(t)$. \\ \\

The asymptotic relationship to the $t_r$ distribution is
\begin{equation}
p_{t_r}(t) \sim \frac{\langle N(N-1) \rangle}{ \langle N \rangle} p_{\tau_n^\star}(t),
\end{equation}
which is obtained via averaging Eq.~(\ref{appasymtr}) (with $s=1$) over $N$ and comparing with Eq.~(\ref{needi3}). Thus, the prefactor is not the averaged prefactor of Eq.~(\ref{needi2}).\\ \\

Now we consider a second definition
\begin{equation}\begin{split}\label{needi5}
p_{\tau_n^\star}(t) = \frac{\langle(N-1) p_{\tau_n^\star |N}(t|N) \rangle}{\langle N-1 \rangle} \sim \frac{\langle N(N-1) \rangle }{\langle N -1 \rangle } p_{\tau_n}(t)[1-P_{\tau_n}(t)].
\end{split}\end{equation}
Again, the mean is $\langle \circ \rangle = \sum_{n=1}^\infty \phi_\text{dis}(n) \circ$ and  $p_{\tau_n^\star|N}(t|N)$ is given in Eq.~(\ref{fixedndis}) and (\ref{largetnew}). This second definition weights $p_{\tau_n^\star|N}(t|N)$ with the number of jumps $N-1$ in the numerator and divides by the mean number of jumps $\langle N-1\rangle$. How to obtain an estimate of $p_{\tau_n^\star}(t)$ from simulation? Given $R$ trajectories. For each trajectory $r\in\{1,\ldots,R\}$, we have the set of transition times $\bm{\tau}^\star(r)=(\tau_1,\ldots,\tau_{m-1},\tau_{m+1},\ldots,\tau_N)$. We do not calculate the histogram $h_{\tau_n^\star}(t;r)$ of this set as for the first definition. Instead, we first collect all transition times (from every trajectory) so that we have $\{\bm{\tau}^\star(1),\ldots,\bm{\tau}^\star(R)\}$. And now we calculate the histogram of this set denoted as $h_{\tau_n^\star}(t;1,\ldots,R)$. For large enuogh $R$, this histogram approximates the second definition of the density $h_{\tau_n^\star}(t;1,\ldots,R)\approx p_{\tau_n^\star}(t)$. \\ \\

The asymptotic relationship to the $t_r$ distribution is
\begin{equation}
p_{t_r}(t) \sim \langle N -1 \rangle  p_{\tau_n^\star}(t),
\end{equation}
which is obtained via averaging Eq.~(\ref{appasymtr}) (with $s=1$) over $N$ while using Eq.~(\ref{needi5}). Thus, the prefactor is the averaged prefactor of Eq.~(\ref{needi2}). So the second definition Eq.~(\ref{needi5}) can be seen as way to obtain a straightforward generalisation of the unidrectional transport relationship Eq.~(\ref{needi2}). However, also the first definition Eq.~(\ref{needi3}) can still be useful for data analysis where one needs to estimate asymptotic behaviors.

\section{Principle of the single long transition time for the quenched trap model}\label{sec:qtm_app}

For the quenched trap model, we presented the principle of the single long transition time in the main text in Eq.~(26), (27) and (29). We derive these results here. Generally, the following calculations assume any bias $p>1/2$. The special case of strong bias $p=1$ can be found below in SM Sec.~\ref{sec:strongqtm}.\\ \\

\subsection{Basic setting}

We start the analysis of the transport in a one-dimensional energy landscape. The lattice points are $x\in \{\ldots,0,1,\ldots,L\}$ with $L\ge 1$, there is no left boundary condition. The start is at $x=1$ and $x=L+1$ is the absorbing wall. The bias $p>1/2$ pushes the particles towards $x=L+1$. Each lattice point $x$ has a fixed energetic trap $E_x$, which was coined by the probability density function
\begin{equation}
p_{E}(E_x) = \frac{1}{T_g} \text{e}^{-E_x/T_g}.
\end{equation}
At each trap $x$, the transition time depends on trap depth $E_x$. We consider the probability density function
\begin{equation}\label{randomtrans}
p_{\tau_x}(t) = \frac{1}{\bar{\tau}_x} \text{e}^{-t/\bar{\tau}_x}
\end{equation}
with $\bar{\tau}_x=t_0 \text{exp}(E_x/T)$ and the temperature $T$. A particle starting at $x=1$ has the span $\{L-K+1,\ldots,L\}$ of visited lattice points. The number of visited traps $K \ge L$ is random, and its moments exists (since we assume right bias $p>1/2$). The furthest point visited is $L-K+1\le  1$. \\ \\

\subsection{Occupation time}
Before we discuss the two important quantities, namely the first passage time and the maximum occupation time (see definition below), we have to think about the transition times. A trap can be visited multiple times by the same particle. Each visit takes a new transition time distributed from the same distribution Eq.~(\ref{randomtrans}). The occupation time is the total time of all visits, that is the sum of all transition times of one particle at the same trap
\begin{equation}\label{occdefnvdsfa4}
\hat{\tau}_x=\sum_{n_x=0}^{N_x} \tau_x^{(n_x)}.
\end{equation}
Here, $N_x$ is the number of visits at $x$, $n_x=0,\ldots,N_x$ is the specific number of the visit at $x$, and $\tau_x^{(n_x)}$ denotes the transition time of the $n_x$-th visit. \\ \\

The number of visited traps $K$ and the number of visits $N_x$ at each trap $x$ are both random. Thus, we have to consider the marginal probability density function obtained from summing up the joint probability density function
\begin{equation}\begin{split}\label{fulloccpdf}
p_{\hat{\tau}_x}(t) = \sum_{k=1}^\infty\sum_{n_x=0}^\infty p_{\tau_x^\prime,N_x,K}(t,n_x,k)=\sum_{k=1}^\infty\sum_{n_x=0}^\infty \phi_{N_x,K}(n_x,k) p_{\hat{\tau}_x|N_x,K}(t|n_x,k).
\end{split}\end{equation}
Here, $\phi_{N_x,K}(n_x,k)$ is the probability of having $k$ traps and $n_x$ visits at $x$. The conditional probability density function on the right hand side describes the occupation time for $n_x$ visits, thus we have the $n_x$-fold convolution
\begin{equation}\label{nfoldconv765z}
p_{\hat{\tau}_x|N_x,K}(t|n_x,k)= [p_{\tau_x}\ast\ldots\ast p_{\tau_x}]^{(n_x)}(t).
\end{equation}
Note that the $k$-dependency is implicit because the distribution of $N_x$ depends on $K$. The $2$-fold convolution is $[p_{\tau_x}\ast  p_{\tau_x}]^{(2)}(t) = \int_0^t p_{\tau_x}(t^\prime) p_{\tau_x}(t-t^\prime)\text{d}t^\prime$ and higher orders are obtained subsequently. In Laplace space the $n_x$-fold convolution is the $n_x$-multiple product $\left[ \hat{p}_{\hat{\tau}_x}(s) \right]^{n_x}$, and we can calculate it 
\begin{equation}
\hat{p}_{\hat{\tau}_x|N_x,K}(s|n_x,k)  =\left(\frac{1}{1+\bar{\tau}_x s}\right)^{n_x}.
\end{equation} 
Inverse Laplace transform yields the gamma probability density function
\begin{equation}\label{gammasingle}
p_{\hat{\tau}_x|N_x,K}(t|n_x,k) = \frac{\bar{\tau}_x^{-n_x}t^{n_x-1}\text{exp}(-t/\bar{\tau}_x)}{\Gamma(n_x)},
\end{equation}
which will be used frequently in the following calculations. 

\subsection{Occupation time distribution averaged over disorder}
Here, we calculate the average over disorder for the conditional probability density function Eq.~(\ref{gammasingle}). Generally, this average is performed over the full energy landscape. Since Eq.~(\ref{gammasingle}) describes the statistics at a specific lattice point $x$, the average is performed over this point
\begin{equation}
\langle \circ \rangle_\text{en} = \int\limits_0^\infty \mathrm{d}E_x p_{E}(E_x) \circ,
\end{equation}
while the average over the remaining traps gives one. The index ``en'' refers to ``energy landscape''. From Eq.~(\ref{fulloccpdf}), we have to calculate
\begin{equation}\label{full6354}
\langle p_{\hat{\tau}_x}(t) \rangle_\text{en} = \sum_{k=1}^\infty\sum_{n_x=0}^\infty \langle\phi_{N_x,K}(n_x,k) p_{\hat{\tau}_x|N_x,K}(t|n_x,k) \rangle_\text{en}.
\end{equation}
The average over disorder of the conditional probability density function is
\begin{equation}\begin{split}\label{averagesing5}
\langle p_{\hat{\tau}_x|N_x,K}(t|n_x,k) \rangle_\text{en}=\int\limits_0^\infty \mathrm{d}E_x p_{E}(E_x) p_{\hat{\tau}_x|N_x,K}(t|n_x,k)\sim \alpha (t_0)^\alpha t^{-1-\alpha} \frac{\Gamma(n_x+\alpha)}{\Gamma(n_x)}
\end{split}\end{equation}
with the exponent 
\begin{equation}
\alpha=\frac{T}{T_g}.
\end{equation}
The same asymptotic behavior is derived from Laplace space
\begin{equation}\begin{split}\label{asymlaplaceocc}
\langle \hat{p}_{\hat{\tau}_x|N_x,K}(s|n_x,k) \rangle_\text{en} = \int\limits_0^\infty \mathrm{d}E_xp_{E}(E_x)\left(\frac{1}{1+\bar{\tau}_x s}\right)^{n_x} \sim 1 + \alpha s^\alpha (t_0)^\alpha \Gamma(-\alpha) \frac{\Gamma(n_x+\alpha)}{\Gamma(n_x)}.
\end{split}\end{equation}
Now with Eq.~(\ref{averagesing5}), the asymptotics of Eq.~(\ref{full6354}) are
\begin{equation}\begin{split}\label{mnbvf}
\langle p_{\hat{\tau}_x}(t) \rangle_\text{en} \sim \alpha (t_0)^\alpha t^{-1-\alpha} \left\langle \frac{\Gamma(N_x+\alpha)}{\Gamma(N_x)} \right\rangle
\end{split}\end{equation}
with the mean
\begin{equation}\label{firstapp324}
\left\langle \frac{\Gamma(N_x+\alpha)}{\Gamma(N_x)} \right\rangle=\sum_{k=1}^\infty \sum_{n_x=0}^\infty \phi_{N_x,K}(n_x,k) \frac{\Gamma(n_x+\alpha)}{\Gamma(n_x)}= \sum_{n_x=0}^\infty \phi_{N_x}(n_x) \frac{\Gamma(n_x+\alpha)}{\Gamma(n_x)}
\end{equation}
where we performed the sum over $k$ in the last step. This mean plays an important role in the principle of the single long transition time (see Eq.~(29) in the main text and discussion below).

\subsection{First passage time in the quenched trap model}
The first passage time is
\begin{equation}\label{firstpassqtm2}
t_f = \sum_{x=L+1-k}^L \hat{\tau}_x
\end{equation}
when the number of visited traps is $K=k$. The probability density function is
\begin{equation}\label{fasdgfs}
p_{t_f}(t) = \sum_{k=1}^\infty \sum_{n_x=0}^\infty p_{t_f,N_x,K}(t,n_x,k),
\end{equation}
see also Eq.~(28) in the main text. We write $p_{t_f}(t)$ now in a more convenient way. It is useful to perform the average over $n_x$ in Eq.~(\ref{fasdgfs})
\begin{equation}\label{convpdf}
p_{t_f}(t) = \sum_{k=1}^\infty p_{t_f,K}(t,k) = \sum_{k=1}^\infty \phi_K(k) p_{t_f|K}(t|k)
\end{equation}
because, for given $K=k$, the marginal probability density function on the right hand is the $k$-fold convolution
\begin{equation}\label{kconvo2}
p_{t_f|K}(t|k)=[p_{\hat{\tau}_{L-K+1}|K}\ast\ldots\ast p_{\hat{\tau}_L|K}]^{(k)}(t),
\end{equation}
see Eq.~(\ref{firstpassqtm2}). Here, $\phi_K(k)$ is the marginal probability that a particle visited $K=k$ traps. Thus, in Laplace space Eq.~(\ref{convpdf}) is
\begin{equation}\begin{split}\label{fulltflaplacef}
\hat{p}_{t_f}(s) =  \sum_{k=1}^\infty \phi_K(k) \hat{p}_{t_f|K}(s|k)= \sum_{k=1}^\infty \phi_K(k) \prod_{x=L-k+1}^{L} \hat{p}_{\hat{\tau}_x|K}(s|k)
\end{split}\end{equation}
The convolution in Eq.~(\ref{kconvo2}) depends on the conditional probability density function
\begin{equation}\begin{split}\label{fixedkpdf}
p_{\hat{\tau}_x|K}(t|k) &= \sum_{n_x=0}^\infty p_{\hat{\tau}_x,N_x|K}(t,n_x|k)= \sum_{n_x=0}^\infty \phi_{N_x|K}(n_x,k) p_{\hat{\tau}_x|N_x,K}(t|n_x,k)
\end{split}\end{equation}
where we find $p_{\hat{\tau}_x|N_x,K}(t|n_x,k)$ from Eq.~(\ref{gammasingle}). In Laplace space, this is
\begin{equation}\label{anotherlaplace}
\hat{p}_{\hat{\tau}_x|K}(s|k) = \sum_{n_x=0}^\infty \phi_{N_x|K} (n_x|k) \hat{p}_{\hat{\tau}_x|N_x,K}(s|n_x,k).
\end{equation}
So the Laplace transform of $p_{t_f}(t)$ can be written as
\begin{equation}\label{anfla}
\hat{p}_{t_f}(s) 
= \sum_{k=1}^\infty \phi_K(k) \prod_{x=L-k+1}^{L}
\left( \sum_{n_x=0}^\infty \phi_{N_x|K} (n_x|k)  \hat{p}_{\hat{\tau}_x|N_x,K}(s|n_x,k) \right)
\end{equation}
where we put Eq.~(\ref{anotherlaplace}) into (\ref{fulltflaplacef}). We will use these result next when averaging over disorder.

\subsection{First passage time distribution averaged over disorder}\label{sec:fagd4}
We want to derive the asymptotics of $\langle p_{t_f}(t) \rangle_\text{en}$. The average over disorder $\langle \circ \rangle$ is performed over all traps. We work in Laplace space and take Eq.~(\ref{anfla})
\begin{equation}\begin{split}\label{fsda89d}
\langle \hat{p}_{t_f}(s) \rangle_\text{en}
= \sum_{k=1}^\infty \phi_K(k) \prod_{x=L-k+1}^{L}
\left( \sum_{n_x=0}^\infty \phi_{N_x|K} (n_x|k) \langle \hat{p}_{\hat{\tau}_x|N_x,K}(s|n_x,k) \rangle_\text{en}\right)
\end{split}\end{equation}
where we can see that the average over disorder on the right hand side is performed over $K=k$ traps
\begin{equation}\begin{split}
\langle \circ \rangle_\text{en} = \int\limits_0^\infty \mathrm{d}E_{L-k+1} p_{E}(E_{L-k+1})  \ldots \int\limits_0^\infty  \mathrm{d}E_L p_{E}(E_L) \circ.
\end{split}\end{equation}
We are interested in the asymptotics of Eq.~(\ref{fsda89d}), therefore considering the small $s$ behavior of $\langle \hat{p}_{\hat{\tau}_x|N_x,K}(s|n_x,k) \rangle_\text{en} \sim 1+C_{n_x}s^\alpha$ with $C_{n_x}=\alpha (t_0)^\alpha \Gamma(-\alpha)\Gamma(n_x+\alpha)/\Gamma(n_x)$ as found in Eq.~(\ref{asymlaplaceocc}). So Eq.~(\ref{fsda89d}) behaves as
\begin{equation}\begin{split}\label{fsda89dsdf}
\langle \hat{p}_{t_f}(s) \rangle_\text{en}
&\sim \sum_{k=1}^\infty \phi_K(k) \prod_{x=L-k+1}^{L}
\left( \sum_{n_x=0}^\infty \phi_{N_x|K} (n_x|k) [ 1+C_{n_x}s^\alpha]\right) \\
&= \sum_{k=1}^\infty \phi_K(k) \left(1+ s^\alpha \sum_{x=L-k+1}^L \sum_{n_x=0}^\infty \phi_{N_x|K} (n_x|k) C_{n_x} \right)\\
&=1+ s^\alpha \sum_{k=1}^\infty \phi_K(k)\sum_{x=L-k+1}^L \sum_{n_x=0}^\infty \phi_{N_x|K} (n_x|k) C_{n_x} \\
&=1+ s^\alpha \sum_{k=1}^\infty \sum_{x=L-k+1}^L \sum_{n_x=0}^\infty \phi_{N_x,K} (n_x,k) C_{n_x}.
\end{split}\end{equation}
The triple sum can be written as
\begin{equation}\label{tripplesum}
\sum_{k=1}^\infty \sum_{x=L-k+1}^L \sum_{n_x=0}^\infty \phi_{N_x,K} (n_x,k) C_{n_x} = \sum_{x=-\infty}^L\sum_{k=1}^\infty  \sum_{n_x=0}^\infty \phi_{N_x,K} (n_x,k) C_{n_x}
\end{equation}
because $\phi_{N_x,K}=0$ for $x \le L-k$. So we finally get the asymptotics via inverse Laplace transform of the last step in Eq.~(\ref{fsda89dsdf})
\begin{equation}\label{largetfinal}
\langle p_{t_f}(t) \rangle_\text{en} \sim \alpha (t_0)^\alpha M_\alpha t^{-1-\alpha}
\end{equation}
where we defined the function
\begin{equation}\label{largetfinalvvvv}
M_\alpha = \sum_{x=-\infty}^L\sum_{k=1}^\infty  \sum_{n_x=0}^\infty \phi_{N_x,K} (n_x,k) \frac{\Gamma(n_x+\alpha)}{\Gamma(n_x)}=\sum_{x=-\infty}^L \left\langle \frac{\Gamma(N_x+\alpha)}{\Gamma(N_x)}\right\rangle,
\end{equation}
compare with Eq.~(\ref{firstapp324}). Before we consider the maximum transition time, we provide an alternative derivation of the $\langle p_{t_f}(t) \rangle_\text{en}$ asymptotics, which will be used later for the modified first passage time $t_r$ calculations when removing the deepest trap (see SM Sec.~\ref{sec:app_removal_qtm}).

\subsection{First passage time distribution - Order statistics approach} \label{sec:fptorder23}

A second approach to get the large $t$ behavior of $\langle p_{t_f}(t) \rangle_\text{en}$ uses order statistics. We order the set of the $K=k$ visited traps $\{E_{L-k+1},\ldots,E_{L}\}$ according
\begin{equation}\label{trapoder}
E_{(1)}<\ldots<E_{(k)}.
\end{equation}
The $q$-th order has the probability density function
\begin{equation}\label{enorder}
p_{E_{(q)}}(E)=\frac{k!}{(q-1)!(k-q)!}p_E(E) [P_E(E)]^{q-1}[1-P_E(E)]^{k-q}
\end{equation}
with $q=1,\ldots,k$. The first passage time is
\begin{equation}\label{firstpasorde1223}
t_f = \sum_{q=1}^k \hat{\tau}_{x[E_{(q)}]}
\end{equation}
where $x[E_{(q)}]$ is the lattice point of trap $E_{(q)}$. For the sake of simplicity we write
\begin{equation}\label{simploc}
x[E_{(q)}] = y[q]
\end{equation}
in the following. We repeat Eq.~(\ref{convpdf})
\begin{equation}
p_{t_f}(t)  = \sum_{k=1}^\infty \phi_K(k) p_{t_f|K}(t|k)
\end{equation}
where $p_{t_f|K}(t|k)$ given in Eq.~(\ref{kconvo2}) is now replaced by
\begin{equation}
p_{t_f|K}(t|k)=[p_{\hat{\tau}_{y[1]}|K}\ast\ldots\ast p_{\hat{\tau}_{y[k]}|K}]^{(k)}(t)
\end{equation}
Each single conditional probability density function is given as
\begin{equation}\begin{split}\label{fd89gfsdv}
p_{\hat{\tau}_{y[q]}|K}(t|k) = \sum_{n_{y[q]}=0}^\infty \phi_{N_{y[q]}|K}(n_{y[q]},k) p_{\hat{\tau}_{y[q]}|N_{y[q]},K}(t|n_{y[q]},k).
\end{split}\end{equation}
Therefore, the Laplace of $p_{t_f}(t)$ is
\begin{equation}\label{laplace89423}
\hat{p}_{t_f}(s) = \sum_{k=1}^\infty \phi_K(k) \hat{p}_{t_f|K}(s|k)=
 \sum_{k=1}^\infty \phi_K(k) \prod_{q=1}^{k}
\left( \sum_{n_{y[q]}=0}^\infty \phi_{N_{y[q]}|K} (n_{y[q]}|k)  \hat{p}_{\hat{\tau}_{y[q]}|N_{y[q]},K}(s|n_{y[q]},k) \right),
\end{equation}
compare with Eq.~(\ref{anfla}). We will use this formula for the average over disorder. But first, we need to discuss the conditional probability density function on the right hand side of Eq.~(\ref{fd89gfsdv}), which is the gamma function
\begin{equation}
p_{\hat{\tau}_{y[q]}|N_{y[q]},K}(t|n_{y[q]},k) = \frac{(\bar{\tau}_{y[q]})^{-n_{y[q]}}t^{n_{y[q]}-1}\text{exp}(-t/\bar{\tau}_{y[q]})}{\Gamma(n_{y[q]})}
\end{equation}
with $\bar{\tau}_{y[q]}=t_0 \text{exp} (E_{(q)}/T)$, see Eq.~(\ref{gammasingle}). We perform the average over disorder
\begin{equation}\label{defavtauncc}
\langle p_{\hat{\tau}_{y[q]}|N_{y[q]},K}(t|n_{y[q]},k) \rangle_\text{en}= \int\limits_0^\infty \text{d} E_{(q)} p_{E_{(q)}}(E_{(q)}) \frac{(\bar{\tau}_{y[q]})^{-n_{y[q]}}t^{n_{y[q]}-1}\text{exp}(-t/\bar{\tau}_{y[q]})}{\Gamma(n_{y[q]})}.
\end{equation}
The next steps derive the large $t$ behavior of this integral. With $p_{E_{(q)}}(_{E_{(q)}})$ from Eq.~(\ref{enorder}), we have to calculate
\begin{equation}\begin{split}
&\langle p_{\hat{\tau}_{y[q]}|N_{y[q]},K}(t|n_{y[q]},k) \rangle_\text{en}\\
&= \frac{k!}{(q-1)!(k-q)!} \int\limits_0^\infty \text{d} E_{(q)} p_E(E_{(q)}) [P_E(E_{(q)})]^{q-1}[1-P_E(E_{(q)})]^{k-q} \frac{(\bar{\tau}_{y[q]})^{-n_{y[q]}}t^{n_{y[q]}-1}\text{exp}(-t/\bar{\tau}_{y[q]})}{\Gamma(n_{y[q]})}.
\end{split}\end{equation}
For the term $[P_E(E)]^{q-1}$, we use the Binomial theorem
\begin{equation}
[P_E(E_{(q)})]^{q-1} = [1-\text{exp}(-E_{(q)}/T_g)]^{q-1} = \sum_{j=0}^{q-1} \binom{q-1}{j} (-1)^j \text{exp}(-jE_{(q)}/T_g)
\end{equation}
so that
\begin{equation}\begin{split}
&\langle p_{\hat{\tau}_{y[q]}|N_{y[q]},K}(t|n_{y[q]},k) \rangle_\text{en}\\
&= \frac{k!}{(q-1)!(k-q)!}  \sum_{j=0}^{q-1} \binom{q-1}{j} (-1)^j \underbrace{\int\limits_0^\infty \text{d} E_{(q)}\frac{1}{T_g} \text{exp}[-(1+j+k-q)E_{(q)}/T_g)] \frac{(\bar{\tau}_{y[q]})^{-n_{y[q]}}t^{n_{y[q]}-1}\text{exp}(-t/\bar{\tau}_{y[q]})}{\Gamma(n_{y[q]})}}_{I}.
\end{split}\end{equation}
Using Mathematica finds the integral $I$ as
\begin{equation}
I=\frac{t^{-1+n_{y[q]}}(t_0)^{-n_{y[q]}}\Gamma(\alpha^\prime+n_{y[q]}){}_1F_1(\alpha^\prime+n_{y[q]},\alpha^\prime+n_{y[q]}+1,-t/t_0)}{\Gamma(n_{y[q]})\Gamma(\alpha^\prime+n_{y[q]}+1)}
\end{equation}
with $\alpha^\prime=\alpha(1+j+k-q)$ and the Kummer confluent hypergeometric function ${}_1F_1$. The asymptotic behavior is
\begin{equation}
I \sim \alpha (t_0)^{\alpha^\prime}t^{-1-\alpha^\prime} \frac{\Gamma(\alpha^\prime+n_{y[q]})}{\Gamma(n_{y[q]})}
\end{equation}
where we see that we only need $j=0$ for the large $t$ behavior. Therefore, the large $t$ behavior is found
\begin{equation}\label{fdsk6432}
\langle p_{\hat{\tau}_{y[q]}|N_{y[q]},K}(t|n_{y[q]},k) \rangle_\text{en} \sim J_q t^{-1-\alpha(k+1-q)}.
\end{equation}
The prefactor is
\begin{equation}
J_q = \alpha(t_0)^{\alpha(k+1-q)} \frac{k!}{(k-q)!(q-1)!}\frac{\Gamma(\alpha(k+1-q)+n_{y[q]})}{\Gamma(n_{y[q]})}.
\end{equation}
The small $s$ behavior in Laplace space of Eq.~(\ref{fdsk6432}) is
\begin{equation}
\langle \hat{p}_{\hat{\tau}_{y[q]}|N_{y[q]},K}(s|n_{y[q]},k) \rangle_\text{en} \sim 1 +J_q \Gamma(-\alpha(k+1-q))s^{\alpha(k+1-q)}.
\end{equation}
which we use to calculate $\langle \hat{p}_{t_f}(s) \rangle_\text{en}$ with $\hat{p}_{t_f}(s)$ taken from Eq.~(\ref{laplace89423})
\begin{equation}\label{gdfavsdfhnnn234}
\langle \hat{p}_{t_f}(s) \rangle_\text{en} \sim
 \sum_{k=1}^\infty \phi_K(k) \prod_{q=1}^{k}
\left( \sum_{n_{y[q]}=0}^\infty \phi_{N_{y[q]}|K} (n_{y[q]}|k)  \left(1+J_q \Gamma(-\alpha(k+1-q))s^{\alpha(k+1-q)}\right) \right).
\end{equation}
We are interested in the large $t$ behavior of $\langle p_{t_f}(t) \rangle_\text{en}$ so that only $q=k$ is required. That is basically the principle of the single long transition time because $q=k$ refers to the deepest trap $E_{(k)}$, see discussion in SM Sec.~\ref{sec:fagd45}. Hence, it follows
\begin{equation}\begin{split}
\langle \hat{p}_{t_f}(s) \rangle_\text{en} &\sim
 \sum_{k=1}^\infty \phi_K(k) 
 \sum_{n_{y[k]}=0}^\infty \phi_{N_{y[k]}|K} (n_{y[k]}|k)  \left(1+J_k \Gamma(-\alpha)s^{\alpha}\right) \\
 &=1+ s^\alpha \Gamma(-\alpha) \sum_{k=1}^\infty \sum_{n_{y[k]}=0}^\infty \phi_{N_{y[k]}|K}(n_{y[k]}|k) J_k
\end{split}\end{equation}
with
\begin{equation}
J_k=\frac{\alpha(t_0)^{\alpha}\Gamma(\alpha+n_{y[k]})}{\Gamma(n_{y[k]})} k.
\end{equation}
We finally get the asymptotic behavior
\begin{equation}\label{fsdsssag3}
\langle p_{t_f}(t) \rangle_\text{en} \sim \alpha (t_0)^\alpha t^{-1-\alpha} \sum_{k=1}^\infty \sum_{n_{y[k]}=0}^\infty \phi_{N_{y[k]},K}(n_{y[k]},k) \frac{\Gamma(\alpha+n_{y[k]})}{\Gamma(n_{y[k]})} k.
\end{equation}
The average on the right hand side picks a specific location of the deepest trap $y[k]=x[E_{(k)}]$ (see Eq.~(\ref{simploc})). However, this location is random, moreover, uniformly distributed over the span $\{L-k+1,\ldots,L\}$ with probability $1/k$. When we take out the average over this location $\sum_{x=L-k+1}^L 1/k $, we obtain
\begin{equation}\begin{split}\label{fsdsssag33}
\langle p_{t_f}(t) \rangle_\text{en} &\sim \alpha (t_0)^\alpha t^{-1-\alpha} \sum_{k=1}^\infty \sum_{x=L-k+1}^L \sum_{n_x=0}^\infty \phi_{N_x,K}(n_x,k) \frac{\Gamma(\alpha+n_x)}{\Gamma(n_x)} \\
&=\alpha (t_0)^\alpha t^{-1-\alpha} \sum_{x=-\infty}^L \sum_{k=1}^\infty  \sum_{n_x=0}^\infty \phi_{N_x,K}(n_x,k) \frac{\Gamma(\alpha+n_x)}{\Gamma(n_x)}
\end{split}\end{equation}
where we used Eq.~(\ref{tripplesum}) to rewrite the triple sum. The result is exactly Eq.~(\ref{largetfinal}). The purpose of this second approach here is that we will need it below for the large $t$ behavior of $\langle p_{t_r}(t) \rangle_\text{en}$ and now for the occupation time in the deepest trap.

\subsection{Occupation time in the deepest trap}\label{sec:fagd45}

The occupation time in the deepest trap $E_\text{max}=E_{(k)}=\text{max}(E_{L-k+1},\ldots,E_L)$ is
\begin{equation}\label{occdeep324dd}
\hat{\tau}_\text{max} = \hat{\tau}_{y[k]}
\end{equation}
when the number of visited traps is $K=k$. Here, $y[q]$ is the location of the deepest trap Eq.~(\ref{simploc}), and $E_{(k)}$ can be found in Eq.~(\ref{trapoder}). Since $K$ is random, we have to consider the marginal probability density function
\begin{equation}\begin{split}\label{gggbvbb}
p_{\hat{\tau}_\text{max}}(t) &= \sum_{k=1}^\infty \phi_K(k) p_{\hat{\tau}_{y[k]}|K}(t|k) \\
 &= \sum_{k=1}^\infty \phi_K(k) \sum_{n_{y[q]}=0}^\infty \phi_{N_{y[q]}|K}(n_{y[q]},k) p_{\hat{\tau}_{y[q]}|N_{y[k]},K}(t|n_{y[q]},k),
\end{split}\end{equation}
see Eq.~(28) in the main text. Note that in Eq.~(28) we implicitly suppose the location of the deepest trap $x=y[q]$. The average over the disorder of Eq.~(\ref{gggbvbb}) is
\begin{equation}\begin{split}\label{johnnnn}
\langle p_{\hat{\tau}_\text{max}}(t) \rangle_\text{en}= \sum_{k=1}^\infty \phi_K(k) \sum_{n_{y[k]}=0}^\infty \phi_{N_{y[k]}|K}(n_{y[k]},k) \langle p_{\hat{\tau}_{y[k]}|N_{y[k]},K}(t|n_{y[k]},k) \rangle_\text{en}.
\end{split}\end{equation}
In Eq.~(\ref{fdsk6432}), we already calculated the large $t$ behavior of $\langle p_{\hat{\tau}_{y[q]}|N_{y[q]},K}(t|n_{y[q]},k) \rangle_\text{en}$ for any $q\le k$. Using $q=k$, we obtain
\begin{equation}\begin{split}
\langle p_{\hat{\tau}_\text{max}}(t) \rangle_\text{en} \sim \alpha (t_0)^\alpha t^{-1-\alpha} \sum_{k=1}^\infty \sum_{n_{y[k]}=0}^\infty \phi_{N_{y[k]},K}(n_{y[k]},k) \frac{\Gamma(\alpha+n_{y[k]})}{\Gamma(n_{y[k]})},
\end{split}\end{equation}
which is exactly the same asymptotic behavior as $\langle p_{t_f}(t) \rangle_\text{en}$ presented in Eq.~(\ref{fsdsssag3}). Following the discussion around Eq.~(\ref{fsdsssag33}) of rewriting the double some, we can also use
\begin{equation}\label{dgvds4t34ewf}
\langle p_{\hat{\tau}_\text{max}}(t) \rangle_\text{en} \sim \alpha (t_0)^\alpha t^{-1-\alpha} \sum_{x=-\infty}^L \sum_{k=1}^\infty  \sum_{n_x=0}^\infty \phi_{N_x,K}(n_x,k) \frac{\Gamma(\alpha+n_x)}{\Gamma(n_x)}
\end{equation}
which is the same expression found for $\langle p_{t_f}(t) \rangle_\text{en}$ in Eq.~(\ref{largetfinal}). We summarize this asymptotic equivalence in the next subsection. 
 
\subsection{Principle of the single long transition time for weak bias $1/2<p<1$}

The asymptotic equivalence between $\langle p_{t_f}(t) \rangle_\text{en}$ Eq.~(\ref{largetfinal}) and $\langle p_{\hat{\tau}_\text{max}}(t) \rangle_\text{en}$ Eq.~(\ref{dgvds4t34ewf}) is the principle of the single long transition time
\begin{equation}\label{principle345435}
\langle p_{t_f}(t) \rangle_\text{en} \sim \langle p_{\hat{\tau}_\text{max}}(t) \rangle_\text{en}
\end{equation}
for the quenched trap model with weak bias $1/2<p<1$, which we presented in the main text in Eq.~(29). With Eq.~(\ref{mnbvf}), we can write the principle also as
\begin{equation}
\langle p_{t_f}(t) \rangle_\text{en} \sim \sum_{x=-\infty}^L \langle p_{\hat{\tau}_x}(t) \rangle_\text{en},
\end{equation}
which is the analog to the reference case of the unidirectional transport $p_{t_f}(t) \sim N p_{\tau_n}(t)$ in Eq.~(\ref{bbbvcbb}).

\subsection{The strongly biased case $p=1$}\label{sec:strongqtm}
For the one-dimensional environment, the principle of the single long transition time and related formulas are presented for the strongly biased $p=1$ quenched trap model in the main text in Eq.~(25), (26) and (27). We discuss these equations here. \\ \\

When the bias is strong $p=1$, the particles move directly from $x=1$ to the right until being absorbed at $x=L+1$. The traps are $\{(E_1,\ldots,E_L)\}$. A particle visits each trap at $x$ only once $N_x=1$, and the total number of traps is $K=L$. Obivously, the occupation time at $x$ Eq.~(\ref{occdefnvdsfa4}) is the only transition time 
\begin{equation}
\hat{\tau}_x = \tau_x,
\end{equation}
the first passage time Eq.~(\ref{firstpassqtm2}) is
\begin{equation}
t_f = \sum_{x=1}^L \tau_x,
\end{equation}
and the occupation time in the deepest trap Eq.~(\ref{occdeep324dd}) is simply the only transition time in the deepest trap
\begin{equation}
\hat{\tau}_\text{max}= \tau_\text{max} = \tau_{y[k]}
\end{equation}
with $y[k]$ as  the location of the deepest trap $E_\text{max}=E_{(k)}$. The probability functions of the latter two are
\begin{equation}\begin{split}
p_{t_f}(t) &= [p_{\tau_1}\ast\ldots\ast p_{\tau_L}]^{(L)} (t), \\
p_{\tau_\text{max}}(t) &= p_{\tau_{y[q]}}(t).
\end{split}\end{equation}
Compare the first equation with Eq.~(\ref{fasdgfs}) to (\ref{kconvo2}) while
\begin{equation}\label{phiphiphi}
\phi_{N_x,K}(n_x,k) = \delta_{n_x,1} \delta_{k,L}
\end{equation}
because $K=L$ and $N_x=1$ are fixed. The first equation in Laplace space and the second equation are
\begin{equation}\begin{split}
\hat{p}_{t_f}(s) &= \prod_{x=1}^L (1+\bar{\tau}_x s)^{-1}, \\
p_{\tau_\text{max}}(t) &= \frac{1}{\bar{\tau}_{y[q]}}\text{exp}(-t/\bar{\tau}_{y[q]}),
\end{split}\end{equation}
which is Eq.~(25) in the main text. The solution (and consequently the large $t$ behavior) of the first equation is derived in the main text. We get the principle of the single long transition time
\begin{equation}\begin{split}
\text{Prob}(t_f > t) &\sim Q_m \text{Prob}(\tau_\text{max}>t)= Q_m  \text{exp} \left(-\frac{t}{\bar{\tau}_m} \right)
\end{split}\end{equation}
with the prefactor 
\begin{equation}
Q_{y[k]}= (\bar{\tau}_{y[k]})^{L-1} \prod_{x=1,x\neq {y[k]}}^L [\bar{\tau}_{y[k]}-\bar{\tau}_x]^{-1},
\end{equation}
compare with Eq.~(26) in the main text. \\ \\

When averaging over the disorder, we already found the principle of the single long transition time Eq.~(\ref{principle345435}) with the exact expression Eq.~(\ref{largetfinal}). So we have here 
\begin{equation}
\langle p_{t_f}(t) \rangle_\text{en} \sim \langle p_{\tau_\text{max}}(t) \rangle_\text{en} \sim \alpha (t_0)^\alpha \Gamma(1+\alpha) L t^{-1-\alpha}
\end{equation}
where $\phi_{N_x,K}(n_x,k) = \delta_{n_x,1} \delta_{k,L}$ Eq.~(\ref{phiphiphi}) was used to calculate the function $M_\alpha$ Eq.(\ref{largetfinalvvvv})
\begin{equation}
M_\alpha = \sum_{x=-\infty}^L \left\langle \frac{\Gamma(N_x+\alpha)}{\Gamma(N_x)}\right\rangle = \sum_{x=1}^L \frac{\Gamma(1+\alpha)}{\Gamma(1)} = L \Gamma(1+\alpha).
\end{equation}
This principle is Eq.~(26) in the main text.

\subsection{Principle of the single long transition time for deterministic transition times}
The content of this subsection is not presented in the main text. However, one often uses deterministic transition times
\begin{equation}
p_{\tau_x}(t)=\delta(t-\bar{\tau}_x)
\end{equation}
with $\bar{\tau}_x=t_0\mathrm{exp}(E_x/T)$ instead of the exponential distribution Eq.~(\ref{randomtrans}). For that case, we get the conditional probability density function Eq.~(\ref{nfoldconv765z}) as
\begin{equation}
p_{\hat{\tau}_x|N_x,K}(t|n_x,k)=\delta(t-n_x\mu_x).
\end{equation}
Now we proceed as in Eq.~(\ref{averagesing5}), i.e. we calculate the average over the disorder
\begin{equation}\begin{split}\label{fasbcxbhh}
\langle p_{\hat{\tau}_x|N_x,K}(t|n_x,k) \rangle_{E_x}&=\int\limits_0^\infty \mathrm{d}E_x p_{E}(E_x) p_{\hat{\tau}_x|N_x,K}(t|n_x,k)=\int\limits_0^\infty \mathrm{d}E_x p_{E_x}(E_x) \delta(t-n_x \mu_x)= \frac{p_{E_x}(\tilde{E})}{|g^\prime(\tilde{E})|}\\
&= \alpha (t_0)^\alpha (n_x)^\alpha t^{1-\alpha} .
\end{split}\end{equation}
Here, $\tilde{E}$ is the only zero of the function $g(E)=t-n_x \bar{\tau}_x$. We repeat the previous calculations, and find the principle of the single long transition time
\begin{equation}
\langle p_{t_f}(t) \rangle_\text{en} \sim \langle p_{\hat{\tau}_\text{max}}(t)\rangle_\text{en} \sim \alpha (t_0)^\alpha t^{-1-\alpha} \sum_{x=-\infty}^L\left\langle (N_x)^\alpha\right\rangle.
\end{equation}
In contrast to the principle Eq.~(\ref{principle345435}) where the transition times are exponentially distributed, we perform here the sum on $\left\langle (N_x)^\alpha\right\rangle$ and not on $\left\langle \Gamma(N_x+\alpha)/\Gamma(N_x)\right\rangle$.

\section{Elimination of the deepest trap in the quenched trap model}\label{sec:app_removal_qtm}
For the quenched trap model, we presented in the main text the results Eq.~(31) to (35), which consider the statistics after removing the deepest trap. We derive these results here. Generally, the following calculations assume any bias $p>1/2$. The special case of strong bias $p=1$ can be found below in SM Sec.~\ref{sec:strongelim}. 

\subsection{Modified first passage time in the quenched trap model}

A particle visits $K=k$ traps. We remove the maximum of these visited traps $E_\text{max}=E_{(k)}=\text{max}(E_{L-k+1},\ldots,E_{L})$ and deal with $E_{(1)}<\ldots < E_{(k-1)}$. The order statistics of the traps were defined in Eq.~(\ref{trapoder}) and (\ref{enorder}). The modified first passage time is
\begin{equation}
t_r = t_f - \hat{\tau}_\text{max} = \sum_{q=1}^{k-1} \hat{\tau}_{x[E_{(q)}]},
\end{equation}
compare with Eq.~(\ref{firstpasorde1223}). In the following, we basically repeat the steps of SM Sec.~\ref{sec:fptorder23}. We are looking for the marginal probability density function
\begin{equation}
p_{t_r}(t)  = \sum_{k=1}^\infty \phi_K(k) p_{t_r|K}(t|k)
\end{equation}
and its average over disorder. The conditional probability density function is the $(n-1)$-fold convolution
\begin{equation}
p_{t_r|K}(t|k)=[p_{\hat{\tau}_{y[1]}|K}\ast\ldots\ast p_{\hat{\tau}_{y[k-1]}|K}]^{(k-1)}(t).
\end{equation}
Each single conditional probability density function is given as
\begin{equation}\begin{split}
p_{\hat{\tau}_{y[q]}|K}(t|k) = \sum_{n_{y[q]}=0}^\infty \phi_{N_{y[q]}|K}(n_{y[q]},k) p_{\hat{\tau}_{y[q]}|N_{y[q]},K}(t|n_{y[q]},k),
\end{split}\end{equation}
which was already shown in Eq.~(\ref{fd89gfsdv}). Following the steps after Eq.~(\ref{fd89gfsdv}), we get again the small $s$ behavior of the Laplace transform Eq.~(\ref{gdfavsdfhnnn234}) 
\begin{equation}
\langle \hat{p}_{t_r}(s) \rangle_\text{en} \sim
 \sum_{k=1}^\infty \phi_K(k) \prod_{q=1}^{k-1}
\left( \sum_{n_{y[q]}=0}^\infty \phi_{N_{y[q]}|K} (n_{y[q]}|k)  \left(1+J_q \Gamma(-\alpha(k+1-q))s^{\alpha(k+1-q)}\right) \right).
\end{equation}
but the product ends at $k-1$ instead of $k$. We are interested in the large $t$ behavior of $\langle p_{t_r}(t) \rangle_\text{en}$ so that only $q=k-1$ is required. Hence, it follows
\begin{equation}\begin{split}
\langle \hat{p}_{t_r}(s) \rangle_\text{en} \sim
1+ s^{2\alpha} \Gamma(-2\alpha) \sum_{k=1}^\infty \sum_{n_{y[k-1]}=0}^\infty \phi_{N_{y[k-1]}|K}(n_{y[k-1]}|k) J_{k-1}
\end{split}\end{equation}
with
\begin{equation}
J_{k-1}=\frac{\alpha(t_0)^{2\alpha}\Gamma(2\alpha+n_{y[k-1]})}{\Gamma(n_{y[k-1]})} k(k-1).
\end{equation}
We finally get the asymptotic behavior
\begin{equation}\label{fsdsssag3vcx}
\langle p_{t_r}(t) \rangle_\text{en} \sim \alpha (t_0)^{2\alpha} t^{-1-2\alpha} \sum_{k=1}^\infty \sum_{n_{y[k-1]}=0}^\infty \phi_{N_{y[k-1]},K}(n_{y[k-1]},k) \frac{\Gamma(2\alpha+n_{y[k-1]})}{\Gamma(n_{y[k-1]})} k(k-1).
\end{equation}
The average on the right hand side picks a specific location of the second deepest trap $y[k-1]=x[E_{(k-1)}]$ (see Eq.~(\ref{simploc})). However, this location is random, moreover, uniformly distributed over the span $\{L-k+1,\ldots,L\}$ with probability $1/k$. When we take out the average over this location $\sum_{x=L-k+1}^L 1/k $, we obtain
\begin{equation}\begin{split}\label{fsdsssag33vcxy}
\langle p_{t_r}(t) \rangle_\text{en} &\sim \alpha (t_0)^{2\alpha} t^{-1-2\alpha} \sum_{k=1}^\infty \sum_{x=L-k+1}^L \sum_{n_x=0}^\infty \phi_{N_x,K}(n_x,k) \frac{\Gamma(2\alpha+n_x)}{\Gamma(n_x)} (k-1) \\
&=\alpha (t_0)^{2\alpha} t^{-1-2\alpha} \sum_{x=-\infty}^L \sum_{k=1}^\infty  \sum_{n_x=0}^\infty \phi_{N_x,K}(n_x,k) \frac{\Gamma(2\alpha+n_x)}{\Gamma(n_x)}(k-1)
\end{split}\end{equation}
where we used Eq.~(\ref{tripplesum}) to rewrite the triple sum. Next, we derive the occupation time in the second deepest trap.

\subsection{Occupation time in the second deepest trap}
After removing $E_\text{max}=E_{(k)}$ from a given number of visited traps $K=k$, the occupation time in the newly deepest trap $E_{(k-1)}$ is
\begin{equation}
\hat{\tau}_\text{max}^\star= \hat{\tau}_{y[k-1]}.
\end{equation}
We repeat the steps of SM Sec.~\ref{sec:fagd45} to obtain the asymptotics of $\langle p_{\hat{\tau}_\text{max}^\star} (t) \rangle_\text{en}$. From Eq.~(\ref{johnnnn}) we learn that
\begin{equation}\begin{split}
\langle p_{\hat{\tau}_\text{max}^\star}(t) \rangle_\text{en}= \sum_{k=1}^\infty \phi_K(k) \sum_{n_{y[k-1]}=0}^\infty \phi_{N_{y[k-1]}|K}(n_{y[k-1]},k) \langle p_{\hat{\tau}_{y[k-1]}|N_{y[k-1]},K}(t|n_{y[k-1]},k) \rangle_\text{en}.
\end{split}\end{equation}
where we replaced the location of the deepest trap $y[k]$ with $y[k-1]$. In Eq.~(\ref{fdsk6432}), we already calculated the large $t$ behavior of $\langle p_{\hat{\tau}_{y[q]}|N_{y[q]},K}(t|n_{y[q]},k) \rangle_\text{en}$ for any $q\le k$. Using $q=k-1$, we obtain
\begin{equation}\begin{split}
\langle p_{\hat{\tau}_\text{max}^\star}(t) \rangle_\text{en} \sim \alpha (t_0)^{2\alpha} t^{-1-2\alpha} \sum_{k=1}^\infty \sum_{n_{y[k-1]}=0}^\infty \phi_{N_{y[k-1]},K}(n_{y[k-1]},k) \frac{\Gamma(2\alpha+n_{y[k-1]})}{\Gamma(n_{y[k-1]})} k(k-1),
\end{split}\end{equation}
which is exactly the same asymptotic behavior as $\langle p_{t_r}(t) \rangle_\text{en}$ presented in Eq.~(\ref{fsdsssag3}). As explained above, we can also write
\begin{equation}\label{dgvds4t34ewf2222}
\langle p_{\hat{\tau}_\text{max}^\star}(t) \rangle_\text{en} \sim\alpha (t_0)^{2\alpha} t^{-1-2\alpha} \sum_{x=-\infty}^L \sum_{k=1}^\infty  \sum_{n_x=0}^\infty \phi_{N_x,K}(n_x,k) \frac{\Gamma(2\alpha+n_x)}{\Gamma(n_x)}(k-1)
\end{equation}
which is the same expression used for $\langle p_{t_r}(t) \rangle_\text{en}$ in Eq.~(\ref{fsdsssag33vcxy}). We summarize this asymptotic equivalence in the next subsection.

\subsection{Relationship between $t_r$ and $\hat{\tau}_\text{max}^\star$ for the bias $p>1/2$}
From the equivalence of the asymptotics in Eq.~(\ref{fsdsssag33vcxy}) and (\ref{dgvds4t34ewf2222}), we see the relationship between the modified first passage time and the occupation time in the second deepest trap
\begin{equation}\label{almostlast}
\langle p_{t_r}(t) \rangle_\text{en} \sim \langle p_{\hat{\tau}_\text{max}^\star}(t) \sim \alpha (t_0)^ {2\alpha} t^{-1-2\alpha}M_\alpha^\star
\end{equation}
with the function
\begin{equation}\label{mstar}
M_\alpha^\star = \sum_{x=-\infty}^L \sum_{k=1}^\infty  \sum_{n_x=0}^\infty \phi_{N_x,K}(n_x,k) \frac{\Gamma(2\alpha+n_x)}{\Gamma(n_x)}(k-1).
\end{equation}
This result is shown in Eq.~(35) in the main text.

\subsection{Asymptotics in the strongly biased case $p=1$}\label{sec:strongelim}
When the bias is strong $p=1$, the transport in a one-dimensional environment is fully discussed in the main text, see Eq.~(31) and (32). We concentrate here on the derivation of Eq.~(33) considering the average over the disorder. \\ \\

The bias $p=1$ implies that a particle only jumps from left to right through the $K=L$ traps $\{E_1,\ldots,E_L\}$ until absorption at $L+1$. Each trap is visited only once. Therefore, the function Eq.~(\ref{mstar}) is
\begin{equation}
M_\alpha^\star = \sum_{x=1}^L  \sum_{k=1}^\infty  \sum_{n_x=0}^\infty \delta_{k,L} \delta_{n_x,1} \Gamma(2\alpha +1)(k-1) = L(L-1) \Gamma(2\alpha+1).
\end{equation}
So the relationship in Eq.~(\ref{almostlast}) becomes
\begin{equation}
\langle p_{t_r}(t) \rangle_\text{en} \sim \langle p_{\hat{\tau}_\text{max}^\star}(t) \sim \alpha (t_0)^{2\alpha} L(L-1) \Gamma(2\alpha+1)t^{-1-2\alpha},
\end{equation}
see Eq.~(33) in the main text.

\subsection{Measure of gain $G$ in the strongly biased case $p=1$}

Here, we derive $G=\langle \bar{t}_r \rangle_\text{en} / \langle \bar{t}_f \rangle_\text{en}$ from Eq.~(34) in the main text. The bias is strong $p=1$ so that the particles move from left to right through the traps $\{E_1,\ldots,E_L\}$. First, we calculate the mean first passage time in this one-dimensional environment 
\begin{equation}
\bar{t}_f = \sum_{x=1}^L \bar{\tau}_x =\sum_{x=1}^L t_0 \text{exp}(E_x/T).
\end{equation}
The average over disorder yields
\begin{equation}\label{fsdav44tqqqc}
\langle \bar{t}_f \rangle_\text{en}=\sum_{x=1}^L \langle\bar{\tau}_x \rangle_\text{en}=\sum_{x=1}^L  \underbrace{\int\limits_0^\infty \text{d}E_x p_E(E_x) t_0\text{exp}(E_x/T)}_{=\alpha/(\alpha-1)} = L t_0   \frac{\alpha}{\alpha-1}
\end{equation}
under the assumption $\alpha=T/T_g>1$. \\ \\

Secondly, the mean of the modified first passage time in the one-dimensional environment is
\begin{equation}
\bar{t}_r = \sum_{q=1}^{L-1} \bar{\tau}_{y[q]} = \sum_{q=1}^{L-1} t_0\text{exp}(E_{(q)}/T).
\end{equation}
The average over disorder is
\begin{equation}\begin{split}\label{almostlast}
\langle \bar{t}_r\rangle_\text{en} &=  \sum_{q=1}^{L-1}\langle \bar{\tau}_{y[q]}\rangle_\text{en} \\
&= \sum_{q=1}^{L-1} \int\limits_0^\infty \mathrm{d}E_{(q)} p_{E_{(q)}}(E_{(q)}) t_0 \text{exp}(E_{(q)}/T)\\
&=\sum_{q=1}^{L-1}  \frac{L!}{(q-1)!(L-q)!} \int\limits_0^\infty \mathrm{d}E_{(q)}  p_{E}(E_{(q)})[P_{E}(E_{(q)})]^{q-1}[1-P_{E}(E_{(q)})]^{L-q} t_0\text{exp}(E_{(q)}/T). 
\end{split}\end{equation}
We use the Binomial theorem for the term
\begin{equation}\begin{split}
[P_{E}(E_{(q)})]^{q-1} =\sum_{j=0}^{q-1} \binom{q-1}{j} (-1)^j \text{exp}(-jE_{(q)}/T_g).
\end{split}\end{equation}
Thus, we get
\begin{equation}\begin{split}
\langle \bar{t}_r\rangle_\text{en}
&=\sum_{q=1}^{L-1} \frac{L!}{(q-1)!(L-q)!} t_0 \sum_{j=0}^{q-1}\binom{q-1}{j} (-1)^j \frac{\alpha}{\alpha(1+L-q+j)-1}\\
&=\sum_{q=1}^{L-1} \frac{L!}{(q-1)!(L-q)!} t_0  (-1)^q (q-1)! \frac{\Gamma(-L+1/\alpha)}{\Gamma(-L+q+1/\alpha)}
\end{split}\end{equation}
under the assumption $\alpha>1/2$.  This assumptions comes from  Eq.~(\ref{almostlast}) where the integral exist for $\alpha>1/(1+L-q)$. Finally, we obtain
\begin{equation}\begin{split}\label{fsdav44tqqq}
\langle \bar{t}_r\rangle_\text{en} = L t_0 \frac{\alpha}{\alpha -1 } + \frac{(-1)^{L+1}t_0 L! \Gamma(-L+1/\alpha)}{\Gamma(1/\alpha)}.
\end{split}\end{equation}
The ratio between Eq.~(\ref{fsdav44tqqq}) and (\ref{fsdav44tqqqc}) gives the measure $G$ presented in the main text Eq.~(34).
under the condition $\alpha>1/2$. So we found the formula Eq.~(34) in the main text.

\end{appendix}

\bibliographystyle{apsrev4-2}

\bibliography{references}